\documentclass[preprint, aps, prmaterials, onecolumn, superscriptaddress, floatfix, altaffilletter, longbibliography]{revtex4-2}
\usepackage{graphicx} 
\usepackage{dcolumn}
\usepackage{bm}
\usepackage{booktabs}
\usepackage{textcomp}
\usepackage{makecell}
\usepackage{hhline}
\usepackage[usestackEOL]{stackengine}
\usepackage[version=3]{mhchem}
\usepackage{amsmath}
\usepackage{braket}
\usepackage{tikz}
\usepackage{multirow}
\usepackage[T1]{fontenc}
\usepackage[utf8]{inputenc}

\begin{document}

\title{Enhanced interlayer interactions in Ni-doped MoS$_2$, and structural and electronic signatures of doping site}

\author{Rijan Karkee} 
\email{rkarkee@ucmerced.edu}
\affiliation{Department of Physics, University of California, Merced, Merced, CA 95343}
\author{Enrique Guerrero} 
\email{eguerrero23@ucmerced.edu}
\affiliation{Department of Physics, University of California, Merced, Merced, CA 95343}
\author{David A. Strubbe} 
\email{dstrubbe@ucmerced.edu}
\affiliation{Department of Physics, University of California, Merced, Merced, CA 95343}

\begin{abstract}\label{Abstract}
The crystal structure of MoS$_2$ with strong covalent bonds in plane and weak Van der Waals interactions out of plane gives rise to interesting properties for applications such as solid lubrication, optoelectronics, and catalysis, which can be enhanced by transition-metal doping. However, the mechanisms for improvement and even the structure of the doped material can be unclear, which we address with theoretical calculations. Building on our previous work on Ni-doping of the bulk 2H phase, now we compare to polytypes (1H monolayer and 3R bulk), to determine favorable sites for Ni and the doping effect on structure, electronic properties, and the layer dissociation energy. The most favorable intercalation/adatom sites are tetrahedral intercalation for 3R (like 2H) and Mo-atop for 1H. The relative energies indicate a possibility of phase change from 2H to 3R with substitution of Mo or S. We find structural and electronic properties that can be used to identify the doping sites, including metallic behavior in Mo-substituted 3R and 2H, and in-gap states for  Mo- and S-substituted 1H, which could have interesting optoelectronic applications. We observe a large enhancement in the interlayer interactions of Ni-doped MoS$_2$, opposite to the effect of other transition metals. For lubrication applications, this increased layer dissociation energy could be the mechanism of low wear. Our systematic study shows the effect of doping concentration and we extrapolate to the low-doping limit. This work gives insight into the previously unclear structure of Ni-doped MoS$_2$ and how it can be detected experimentally, the relation of energy and structures of doped monolayers and bulk systems, the electronic properties under doping, and the effect of doping on interlayer interactions.
\end{abstract}

\date{\today}
\maketitle

\section{Introduction}
There has been great interest in transition metal dichalcogenides, particularly in MoS$_2$ which has been used in a wide variety of applications because of its unique physical, optical and electrical properties \cite{appl_opticL,gap_variation,appl_magnetic}. Tunability of the band gap from direct to indirect \cite{appl_opticL,gap_variation},  good electron mobility \cite{mobility},  the possibility of defect engineering  including creating quantum emitters \cite{defect_engg}, and high current-carrying capabilities \cite{high_current} open great potential in optics and electronics. MoS$_2$ has
already shown promising applications in lubrication, hydrodesulfurization, and optoelectronics \cite{Mos2_application1,Mos2_application2,Mos2_application3,Mos2_application4}.
Theoretical and experimental work has shown that transition-metal-doping in MoS$_2$ can improve catalytic reactivity \cite{doped_apl1,doped_apl2,solar_hydro_pro}, lubrication \cite{appl_tribology} and gas sensing \cite{CO_and_NO_ads,doped_apl4}. However, the detailed  mechanism is still unclear for improved catalysis and lubrication with doping, and in general the exact location of dopants in MoS$_2$ is also uncertain. Our previous work addressed the most favorable sites for Ni in 2H-MoS$_2$ as a function of chemical potential and investigated signatures in Raman and IR spectroscopy \cite{enrique}. We now extend this study to a systematic exploration of other polytypes, with consideration of the relation between monolayer and bulk structure and energetics, and study of doping concentration-dependence. This work will look specifically at implications for solid lubrication, but also give insight into optoelectronic and other fundamental properties.

MoS$_2$ is a layered structure with weak Van der Waals forces between layers and strong covalent forces within the layers giving rise to interesting applications in solid lubricants because the weakly bound layers can slide against each other 
easily. Graphite is a commonly used solid lubricant, which also has a layered Van der Waals structure, but it needs to adsorb oxygen and moisture to develop the low shear strength necessary for lubrication \cite{graphite}. MoS$_2$, on the other hand, works well under vacuum but shows degradation in lubrication under humidity and high temperature, for unclear reasons \cite{lubrication_in_mos2_in_ambient}. Therefore MoS$_2$ is preferred for solid lubrication in space applications such as  NASA's  James Webb Space Telescope \cite{Mos2_application4} where the wide range of temperatures precludes the use of oils or greases for lubrication of moving parts. 

A clear understanding of the mechanisms by which Ni-doped MoS$_2$ enhances lubrication performance of pristine MoS$_2$ is still lacking. A recent review \cite{lubricant_new} discusses low friction mechanisms based on ordering of crystallites in adjacent platelets. The early experimental work of Stupp  \cite{mos2_experiment} studied the friction properties in MoS$_2$  doped with various transition metals. MoS$_2$ doped with Ni or Ta showed the best results in terms of stable friction, excellent endurance, easy control, good effect on aging, and lower coefficient of friction than pristine MoS$_2$. Later, other researchers also confirmed the lower coefficient of friction for Ni-doping \cite{mos2_experiment_2}. Ni is more abundant and less expensive than Ta \cite{Ni_abundance,Ta_abundance}, making it more attractive. Most of the literature on doped MoS$_2$ is focused on electronic, optical, and catalytic applications and there have been only a small number of studies on tribology. There are already some studies on Ni-doped MoS$_2$ monolayers and nanosheets, for catalysis in solar hydrogen production \cite{solar_hydro_pro}, adsorption of CO or NO for potential gas sensors \cite{CO_and_NO_ads}, or adsorption of O$_2$ molecules \cite{O2_ads}, but tribological applications have not been investigated. Different dopants such as Ni, Cr, Ti, Au, Zr and Sb$_2$O$_3$ have been studied for the improvement of tribological properties \cite{lubricant_by_David}. One popular proposed mechanism is increase in hardness as a result of distortion of the MoS$_2$ crystal structure, and the resultant hardness reduces wear which is crucial to coating life (endurance) \cite{fric_proposed_mechanism1,fric_proposed_mechanism2}. A recent experiment specifically on Ni showed that wear is reduced by doping, extending the lifetime of MoS$_2$ films \cite{low_wear_rate}. This work aims to understand mechanisms of wear reduction in those experimental results through theoretical and atomistic study of Ni-doped MoS$_2$.

We investigated Ni-doped MoS$_2$ structures and properties by Density Functional Theory (DFT). MoS$_2$ can exist in different phases \cite{lubricant_by_David}: 2H and 3R  bulk, and 1T and 1H monolayers. Friction in MoS$_2$ involves sliding of layers, and 2H and 3R phases only differ by stacking, \textit{i.e.} sliding followed by rotation. While 2H is the most studied bulk polytype and the predominant naturally occurring one, 3R is also found naturally and has a very similar energy to 2H \cite{lubricant_by_David}. Also, it is found that doping may alter the phase: for example, Li intercalation in MoS$_2$ causes a change from 2H to 1T phase \cite{phase_change_Li} and Nb substitution on Mo favors 3R over 2H \cite{phase_change_Nb}. The possibility of alteration in phase due to doping, layer sliding, or the effect of mechanical load in tribology application is a motivation for considering other phases in MoS$_2$, as well as to assess whether deliberate production of different polytypes could  improve performance.

In this paper, we will  consider the effect of Ni-doping in MoS$_2$  polytypes, and investigate changes in local structure, the energetics of dopants at different sites,  the possibility of phase changes, and the effect on interlayer interactions. Section \ref{Methods} details our computational approach. Section \ref{structure} discusses the  choice of dopant structures of different polytypes in MoS$_2$. Section \ref{results} presents results on: (a)  doping formation energy, (b) effect of doping on crystal structure and local structure, (c) effect on electronic properties (including oxidation states) and (d) effect on the layer dissociation energy. Finally, Section \ref{conclusion}  concludes to discuss the impact of Ni-doping, experimentally measurable signatures of different doped structures, and a possible explanation for low wear in MoS$_2$ lubrication. 

\section{Methods}
\label{Methods}
Our calculations use plane-wave density functional theory (DFT)  implemented in the code Quantum ESPRESSO, version 6.1 \cite{QE-2009,QE_2}. We used the Perdew-Burke-Ernzerhof (PBE) generalized gradient approximation \cite{original_pbe}  for all of our analysis
except for the doping formation energy which is calculated with Perdew-Wang \cite{lda_perdew} local density approximation (LDA)   functional. 
With PBE, we used the semi-empirical  Grimme-D2 (GD2) \cite{Grimme} Van der Waals correction to the total energy, which gives lattice  parameters  and other properties considerably closer to experimental results for MoS$_2$ \cite{Mos2_exp_fit,enrique}. Calculation with LDA has also been shown to give accurate lattice parameters \cite{enrique}. The doping formation energy requires the energies of metal solids such as Ni and Mo which are not well described by PBE+GD2 \cite{GD2_metal}, and therefore we used LDA for these calculations to obtain accurate energies for both metals and doped MoS$_2$ as in our previous work on 2H-MoS$_2$ \cite{enrique}. We used Optimized Norm-Conserving Vanderbilt pseudopotentials \cite{ONCV} parametrized by Schlipf and Gygi\cite{Gygi} from the SG15 set \cite{web_sg15} generally, except we used the Pseudodojo set \cite{web_pseudojo} for partial density of states (PDOS), to have available pseudo-wavefunctions; and for LDA, to be consistent with our previous calculations \cite{enrique}. Kinetic energy cutoffs of 816 eV (60 Ry) for PBE and 1088 eV (80 Ry) for LDA were used.  Half-shifted $k$-point grids of $6\times6\times2$ for bulk and $12\times12\times1$ for monolayers  (in the pristine case) were chosen to converge the total energies within 0.001 eV/atom. Atomic coordinates were relaxed using a force threshold of $1.0\times 10^{-4}$ Ry/bohr and the stresses were relaxed below $0.1$ kbar. We used a Gaussian smearing of 0.05 eV for the bulk metals or whenever metallic cases arose. For doped structures, we decreased the $k$-grid in  $x$- , $y$- or $z$-directions in proportion to the supercell size, to maintain consistent Brillouin zone sampling. For density of states calculations, a broadening of  0.02 eV and $k$-grid  sampling of $20 \times20\times1$ for $4\times4\times1$ monolayers and $20\times20\times10$ for $2\times2\times1$ bulk was used.

The doping concentration is varied by putting a single Ni atom in increasing supercells such as $2\times2\times1$, $3\times3\times1$, $3\times3\times2$, etc. All supercells are taken as neutral, as appropriate for neutral impurities or high concentration, though charged impurities are also possible \cite{lattice}. In the 2H structure, we also considered tetrahedrally intercalated structures with two Ni atoms per cell, one between each pair of layers, vertically above one another. Our supercells represent not only approximations to a disordered low-doping limit, but also possible ordered phases, as are found for Li intercalation \cite{phase_change_Li}. The Ni atom fraction ranges from a minimum of 0.5\% to maximum of 17\%.  The monolayers contain vacuum of 15 {\rm\AA} between layers to reduce the effect of spurious periodicity. We have tested the monolayers with dipole corrections \cite{dipole_corr} in case there would be an out-of-plane dipole moment in the doped phases, but the change in total energy was negligible ($\sim10^{-4}$ eV) and structural parameters were unaffected. Our calculations in general are non-spin-polarized, as use of spin polarization did not significantly change the  energy (less than 0.1 meV per atom). However, we used spin-polarized calculations for ferromagnetic bulk Ni, and also for test calculations on each doping case to check for magnetic moments.

Before calculations on doped systems, the structure of 2H-MoS$_2$ was bench-marked with PBE+GD2  and lattice parameters were calculated as $a=b=3.19$ $\pm$ 0.02 {\rm\AA} and $c=12.4$ $\pm$ 0.2 {\rm\AA}. The uncertainties come from a cubic fit to the total energy as a function of the lattice parameters.  These results are in good agreement with previous theoretical \cite{lattice} and experimental work \cite{lattice_exp}. For the energies of reference elemental phases, bulk Mo and Ni are calculated in  body-centered and face-centered cubic lattice of 3.19 \rm\AA\ and 3.52 \rm\AA, respectively, with a $12\times12\times12$ half-shifted $k$-grid. For  bulk S, a unit cell containing 48 atoms \cite{web_S} is calculated with a $3\times3\times3$ $k$-grid. The energies of these reference phases, as in our previous work \cite{enrique}, are used to calculate the doping formation energy.

\section{Structure and doping site selection}
\label{structure}	

\begin{figure}[h]
		\includegraphics[width=0.6\linewidth]{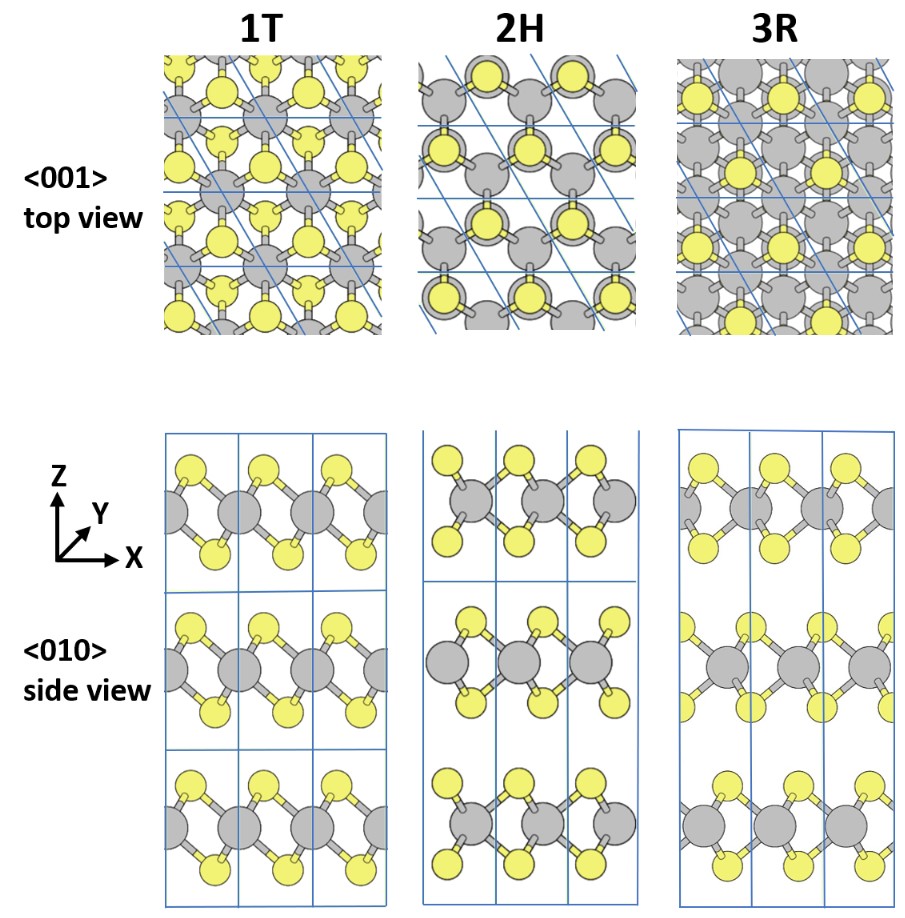}
		\caption{Structure of MoS$_2$ polytypes with conventional unit cell marked by blue lines. Mo is gray and S is yellow. 1H is a single monolayer of MoS$_2$ containing 3 atoms per unit cell, as in each layer of 2H or 3R. This work studies 1H, 2H, and 3R.}
		\label{fig:polytypes}
\end{figure}

MoS$_2$ can exist in different possible structures in bulk and monolayer phases (Fig. \ref{fig:polytypes}) \cite{lubricant_by_David}. The known phases of MoS$_2$ are  1H (hexagonal monolayer),  1T (trigonal monolayer),  2H (hexagonal bulk) and   3R (rhombohedral bulk). For our interest in lubrication applications, we will not consider the 1T phase further due to its metastability and likelihood of converting to 1H. The 1H, 2H and 3R phases contain 3, 6, and 9 atoms in a conventional unit cell respectively, and they differ by stacking. In 1H, each layer of MoS$_2$ is stacked exactly above each other (AA stacking). In 2H, the Mo atom in one layer is above an S atom of the preceding one, forming AB stacking, and 3R has ABC stacking with 3 layers per conventional unit cell. In 3R, all layers have the same orientation, but 2H has inversion symmetry between layers. Our DFT calculation (Table \ref{tab:formation_energies})
shows only a small difference in total energy of the 2H and 3R phases 
in agreement with previous literature \cite{2H&3R_agreement}.

 		\begin{table}[htbp]
 	\centering

 		\begin{tabular}{|c| r| r| }

 			\hline
 		\textbf{$E_{\rm f}-E_{\rm f,2H}$ (eV)} & \multicolumn{1}{c|}{\textbf{LDA}} & \multicolumn{1}{c|}{\textbf{PBE+GD2}} \\
 		\hline
 		\hline
 		2H & 0.00 & 0.00 \\
 		\hline
 		1H & 0.11 & 0.15 \\
 		\hline
 		3R & $1.3 \times 10^{-3}$ & $1.2 \times 10^{-4}$ \\
 		\hline

 		\end{tabular}
		
 		\caption{DFT calculation of formation energy per unit of MoS$_2$ in MoS$_2$ polytypes, with respect to 2H-MoS$_2$. The formation energy of 2H with respect to bulk Mo and S is-3.05 eV with LDA (expected to be more accurate) and -2.59 eV with PBE+GD2.}
 		\label{tab:formation_energies}
 	\end{table}

\begin{figure}[h]
	\includegraphics[scale=0.45]{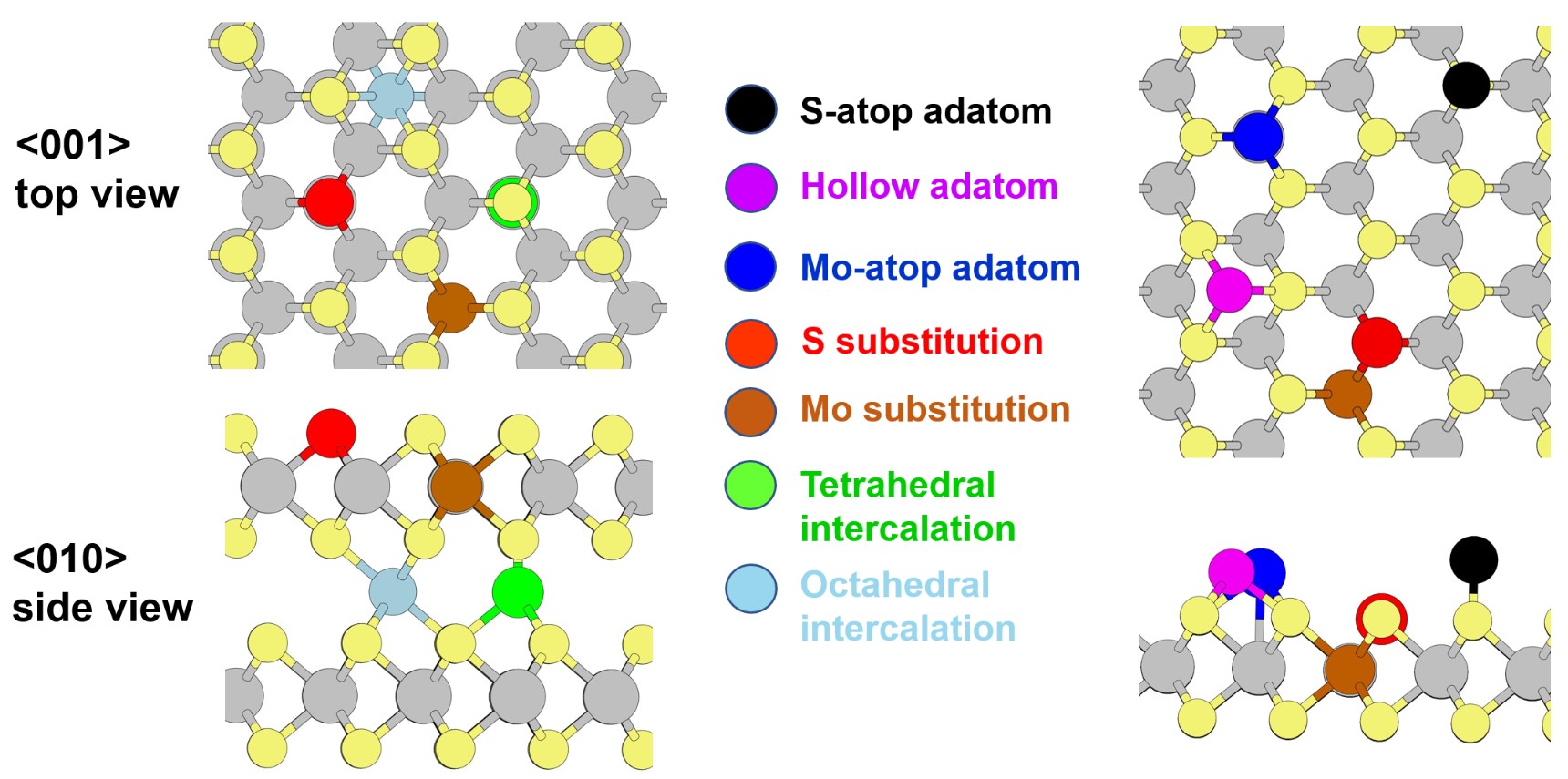}
	\caption{\small{Different dopant sites for Ni in 2H (left) and 1H (right) MoS$_2$ with top view (upper) and side view (lower). Sites include substitution in the  Mo site or S site, tetrahedral or octahedral intercalation for  bulk phases (2H and 3R), and Mo-atop, S-atop and hollow adatoms for 1H (monolayer). 3R sites are similar to 2H (see detail in Fig. \ref{fig:bulk_intercal}).}}
	\label{fig:doping sites}
\end{figure}

  The  most probable sites for dopants \cite{enrique}  in MoS$_2$  include substitutions and intercalations for bulk or adatoms for monolayers, and are shown in Fig. \ref{fig:doping sites}. We have also considered the bridge site along the Mo-S bond but found that intercalation/adatom in this site was unstable and relaxed to tetrahedral intercalation in 2H and 3R, and Mo-atop in 1H. Therefore the bridge site was not considered in further calculations. We will use these sites to study the various structural and electronic properties of Ni-doped MoS$_2$.

\section{Results and discussion}
\label{results}

	\subsection{Doping Formation Energy}
	We begin by establishing which doped structures are energetically accessible, via the doping formation energy, which tells us the energy needed to incorporate the dopant (Ni) into the host system (MoS$_2$). As noted in our previous work \cite{enrique}, the doping formation energy depends on the chemical potentials during the growth process, and so the relative favorability between different stoichiometries (\textit{e.g.} substitution \textit{vs.} intercalation) depends on the conditions, which can be depicted in a phase diagram (Fig. S3 \cite{SM}). The doping formation energy is calculated as:
	\begin{equation}
	E_{\rm formation} = E_{\rm system} - N E_{\rm pristine} - \mu_{\rm Ni} + \mu_{\rm removed}
	\end{equation}
	 for substitution cases, and
	\begin{equation}
	E_{\rm formation} = E_{\rm system} - N E_{\rm pristine} - \mu_{\rm Ni}
	\end{equation}
	 for intercalation cases, where $E_{\rm system}$ is the energy of the doped system, $N$ is the number of cells in the supercell, $E_{\rm pristine}$  is the energy per unit cell of the corresponding phase of MoS$_2$, and $\mu_{\rm Ni}$ and $\mu_{\rm removed}$ are the chemical potential of the dopant (Ni) and removed atom, respectively. We consider $\mu_{\rm Ni}$ equal to the energy per atom of bulk Ni (``Ni-rich'' conditions). For substitution cases, we consider equilibrium conditions for the given phase where $\mu_{\rm Mo} + 2 \mu_{\rm S} = E_{\rm pristine}/N_{\rm f.u.}$ and $N_{\rm f.u.}$ is the number of MoS$_2$ units per pristine cell; then we look at conditions where either $\mu_{\rm Mo} = E_{\rm Mo}$ (``Mo-rich,'' which favors S substitution) or $\mu_{\rm S} = E_{\rm S}$ (``S-rich'', which favors Mo substitution). These energies of Mo and S are calculated from their bulk phases, as in previous literature  \cite{E_form_bulk_ref1,E_form_bulk_ref2,enrique}. For example, in $2\times2\times1$ Mo substitution by Ni in 2H, $E_{\rm system}$ is the  energy of NiMo$_7$S$_{16}$, $N$ is 4, $\mu_{\rm removed}$ is the chemical potential of bulk Mo, and $E_{\rm pristine}$ is the energy of Mo$_2$S$_4$. Similarly in $4\times4\times1$ Ni intercalation in 2H, $E_{\rm system}$ is the energy of NiMo$_{32}$S$_{64}$ and $N$ is 16.
	 
	\subsubsection{Monolayer: 1H}
	\label{monolayer}
	
\begin{figure}[h]
			\centering{
			\begin{tikzpicture}
			\node [anchor=north west] (imgA) at (0.000\linewidth,.58\linewidth){\includegraphics[width=0.49\linewidth]{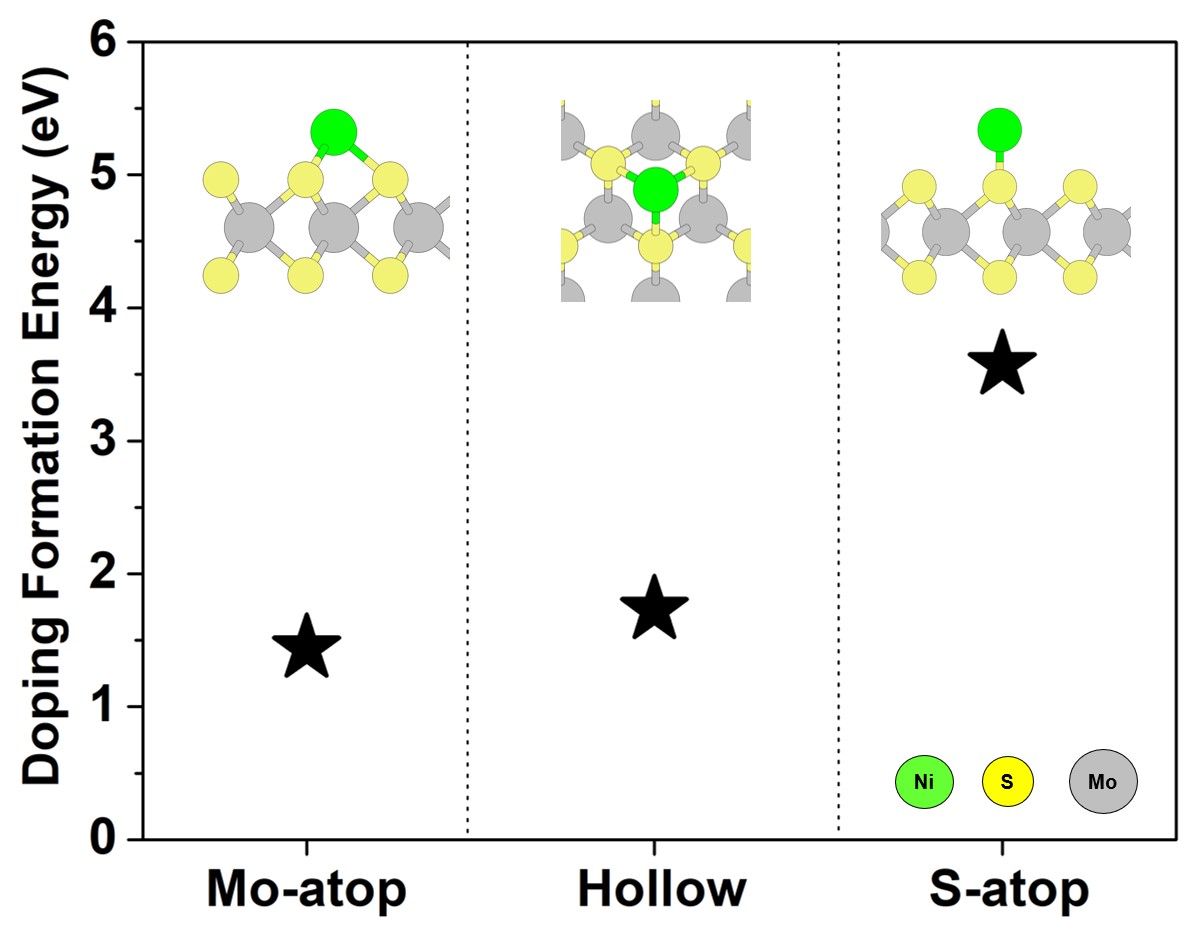}};
            \node [anchor=north west] (imgB) at (0.52\linewidth,.578\linewidth){\includegraphics[width=0.398\linewidth]{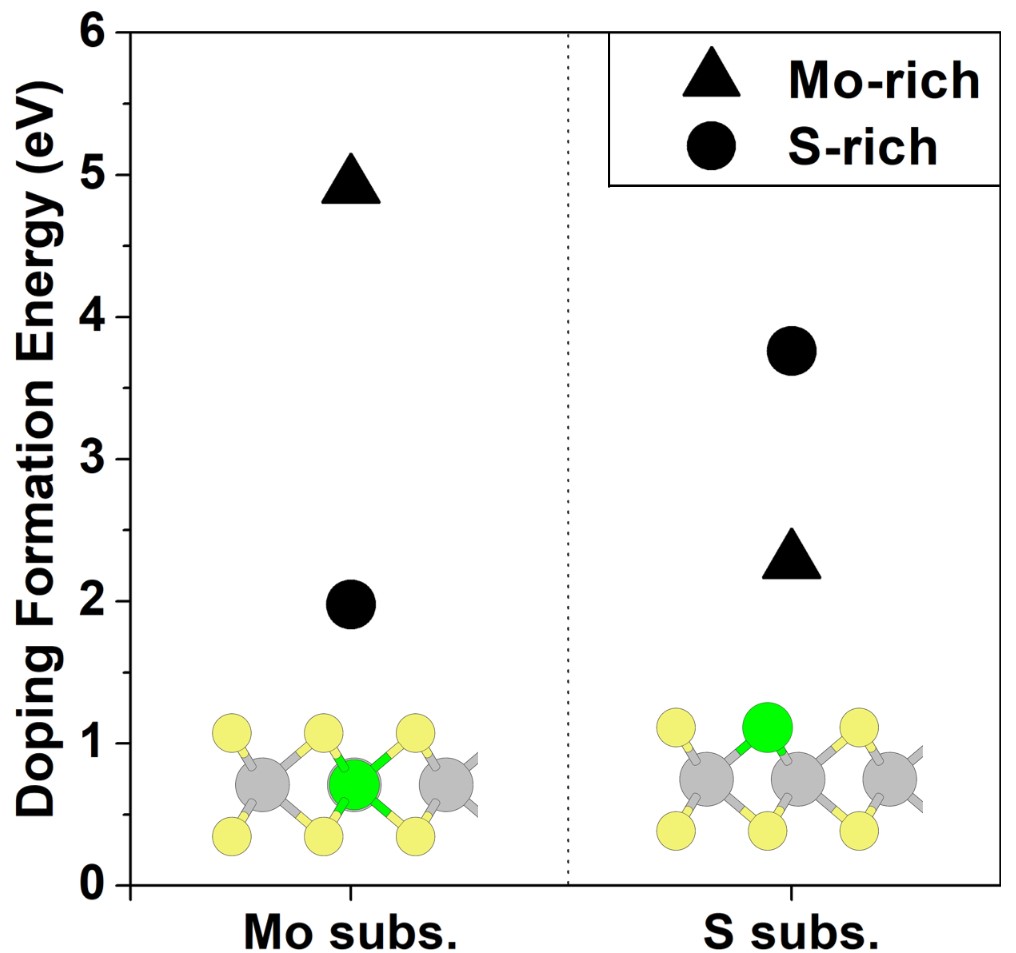}};
          
            \draw [anchor=north west] (0.000\linewidth, .62\linewidth) node {(a)};
            \draw [anchor=north west] (0.520\linewidth, .62\linewidth) node {(b)};
         
            \end{tikzpicture}
			}
		
			\caption{Doping formation energies from LDA for a $4\times4\times1$ supercell of monolayer 1H: a) adatoms, and b) substitution, under Mo-rich and S-rich conditions (which do not affect the value for adatoms). Insets show corresponding structures (side or top view).}
			\label{fig:mono_atop}
			
		\end{figure}

		The doping formation energies for Ni-doped MoS$_2$ in the 1H  phase with Mo or S substitution and different adatom sites (Fig. \ref{fig:doping sites}) were computed with  supercells from $2\times2\times1$  to $4\times4\times1$. Fig. \ref{fig:mono_atop}(a) shows results for adatoms in  $4\times4\times1$ supercell. The Mo-atop adatom is the energetically favored  adatom site and this result, along with the bond lengths, agrees with previous literature \cite{Mo-atop1}. Fig. \ref{fig:mono_atop}(b) shows the substitutional doping formation energy under Mo-rich and S-rich conditions.  The doping formation energy for substitutional doping remains positive even in the most favorable conditions, indicating that substitutional doping may not be achieved at equilibrium conditions. At higher doping concentrations, the doping formation energy is reduced slightly but remains positive (Fig. S2(b-c)).
		
	\subsubsection{Bulk}
	\label{Bulk: 2H and 3R}
\begin{figure}[h]
			\includegraphics[width=1\linewidth]{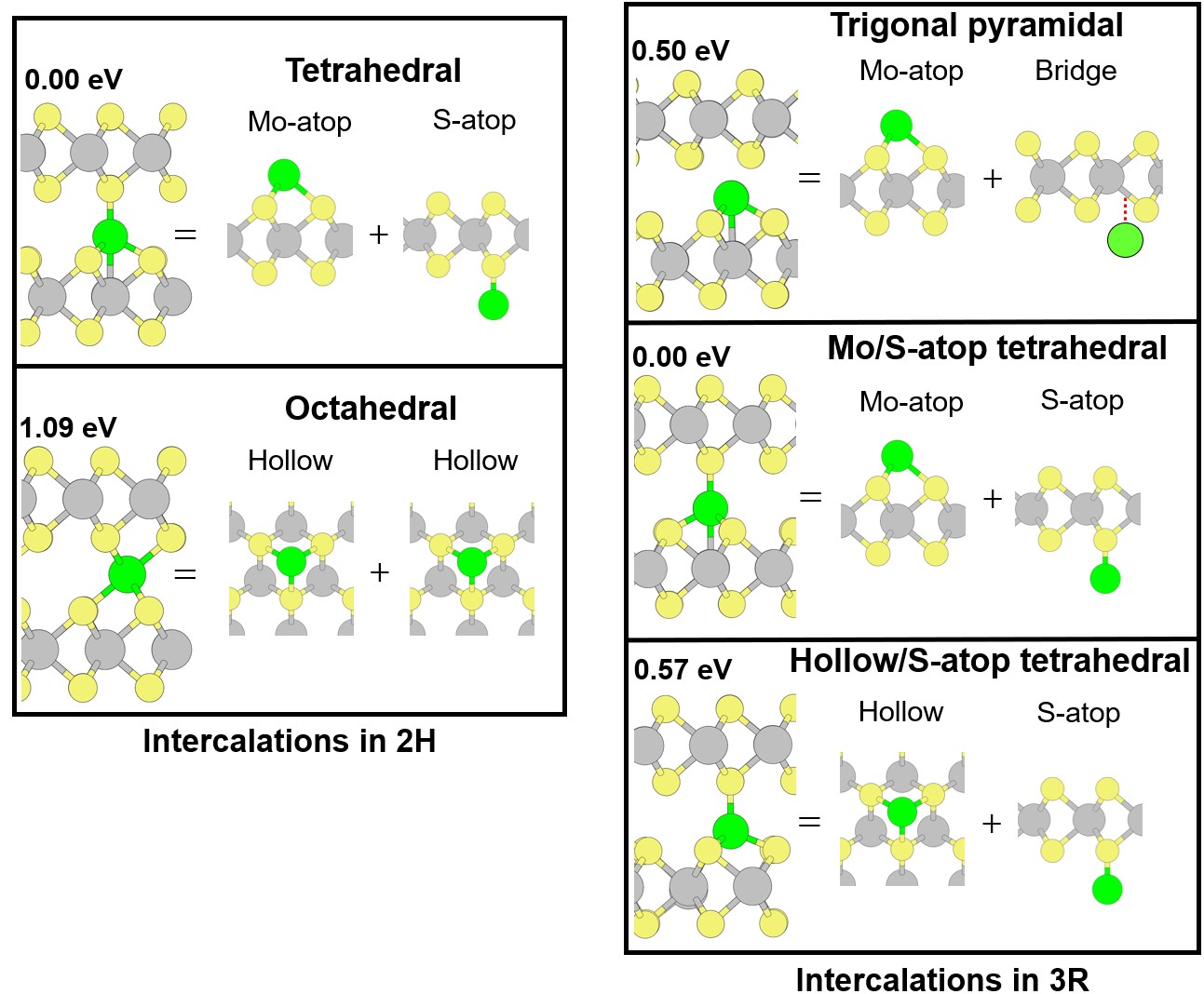}
				\caption{Intercalation structures in 2H and 3R $4\times4\times1$ supercells, classified by bonding geometry and compared to adatom sites with respect to each layer. Mo/S-atop tetrahedral sites are the most favorable for both 2H and 3R, consistent with Mo-atop as the lowest-energy adatom site on 1H. Doping formation energies from LDA are shown relative to the most stable structure.}
				\label{fig:bulk_intercal}
		\end{figure}

		Bulk systems have intercalation rather than adatoms. In 2H, tetrahedral intercalation is energetically favored over octahedral intercalation at all concentrations. The tetrahedral intercalation site is Mo-atop and S-atop with respect to the layers on the either side as shown in Fig. \ref{fig:bulk_intercal}, whereas octahedral intercalation  consists of hollow sites  on each layer. In 3R, we find three metastable intercalation cases. There are two sites with tetrahedral geometry: in one, the Ni atom is Mo-atop for one layer and S atop for the other (``Mo/S-atop''); in the second, the Ni atom is at the hollow site for one layer and Mo-atop for the other (``hollow/Mo-atop''). The third intercalation site has trigonal pyramidal geometry and is Mo-atop for one layer and  bridge site for the other; this site is only stable for supercells larger than $1\times1\times1$. Octahedral intercalation is unstable in 3R and generally relaxes to the trigonal pyramidal structure. The Mo/S-atop tetrahedral intercalation (similar to 2H tetrahedral intercalation) is the most favored intercalation site (Fig. \ref{fig:bulk_intercal} and Table S4 \cite{SM}). These most favorable sites in bulk phases are clearly related to energetics in the 1H phase. As discussed in Section \ref{monolayer}, adatom on Mo is the most favorable site in 1H and all the most favored sites in bulk (2H and 3R) involve Mo-atop. The intercalant acts as an adatom for the neighboring layers and lower doping formation energy is achieved if the intercalant site resembles the monolayers' lowest-energy adatom sites.
        Doping formation energies for substitutions in 2H and 3R are similar to 1H, except that Mo substitution can be favorable under S-rich conditions for 2H and 3R in some cases (Table S1), such as $1\times1\times1$ (which would mean NiS$_2$ for 1H) or other $1\times1\times N$ supercells of 1H. 
		
		\subsubsection{Concentration Dependence of Doping Formation Energy}
		\label{doping_conc}
		
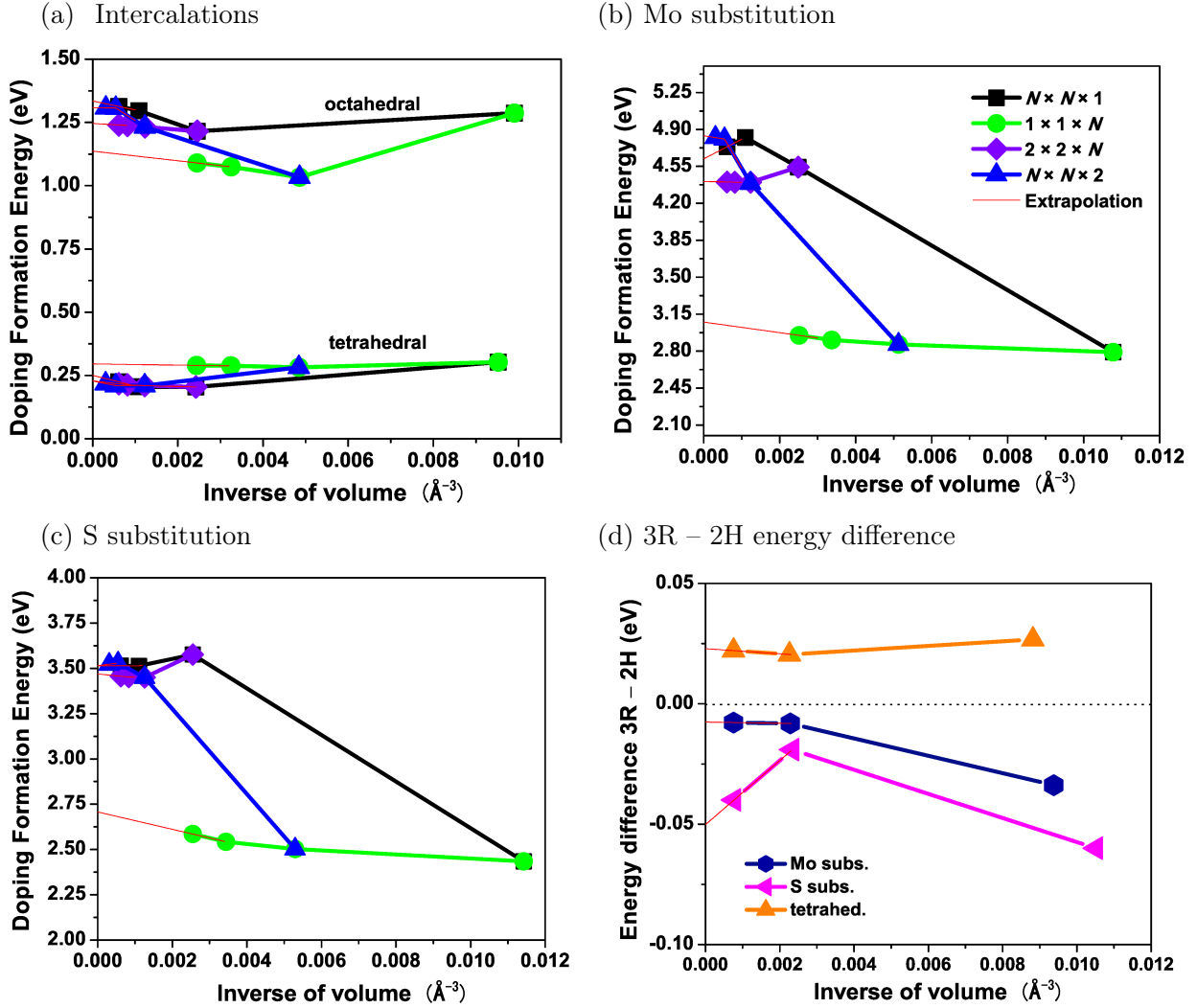
\begin{figure}[h]
			
			\centering{
			\begin{tikzpicture}
			\node [anchor=north west] (imgA) at (-0.15\linewidth,.58\linewidth){\includegraphics[width=0.48\linewidth]{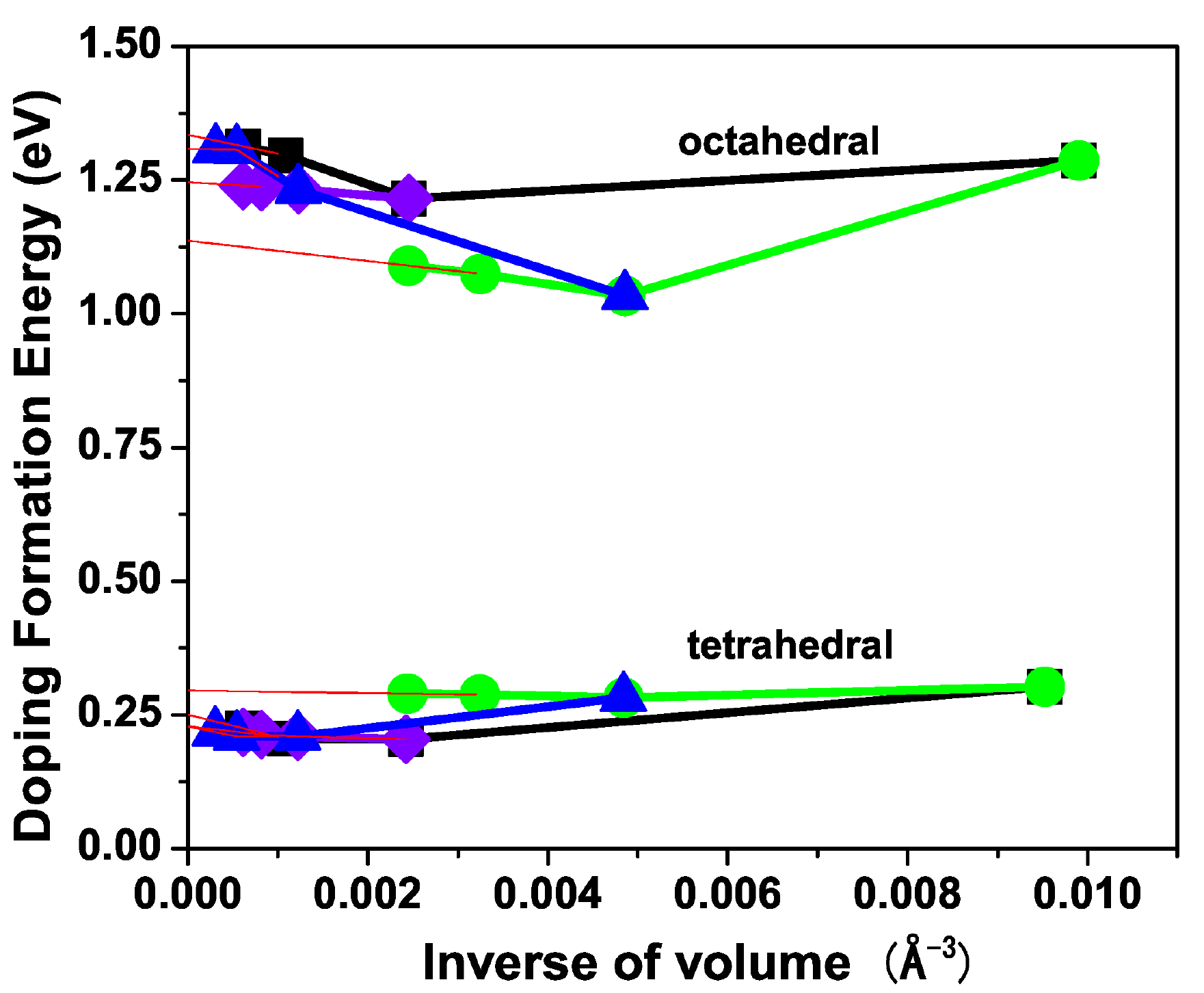}};
            \node [anchor=north west] (imgB) at (0.36\linewidth,.57\linewidth){\includegraphics[width=0.49\linewidth]{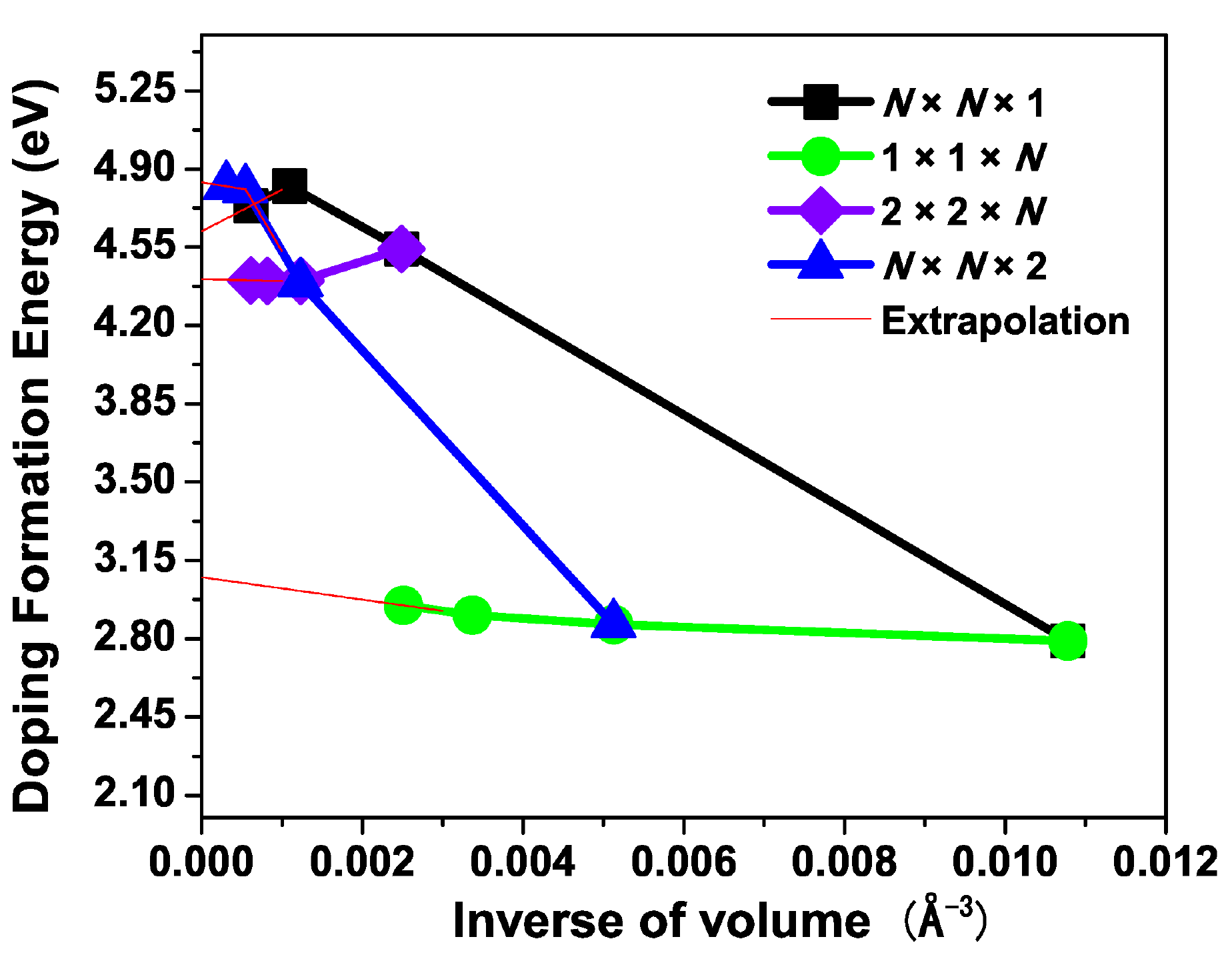}};
          	\node [anchor=north west] (imgC) at (-0.15\linewidth,.14\linewidth){\includegraphics[width=0.48\linewidth]{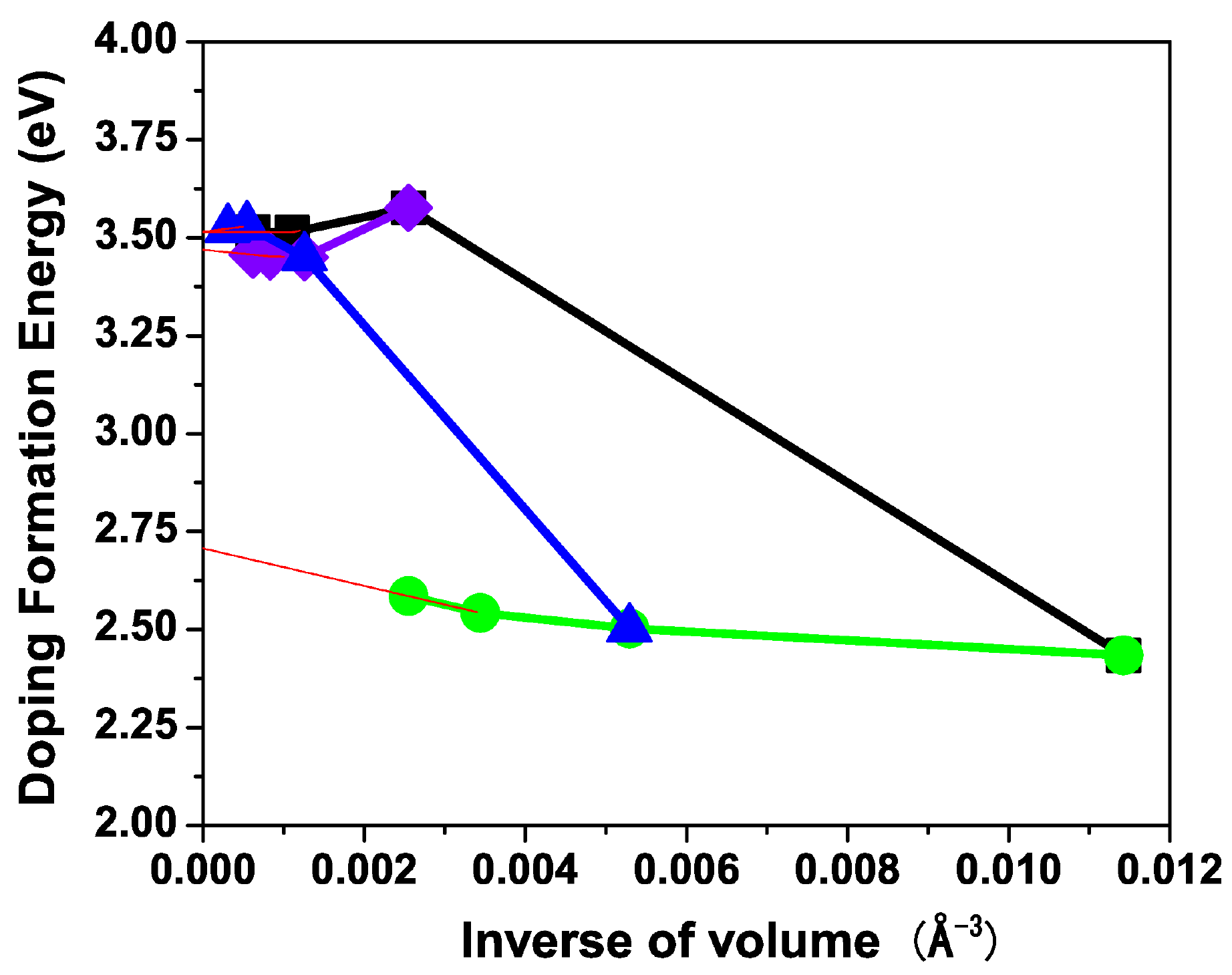}};
            \node [anchor=north west] (imgD) at (0.36\linewidth,.14\linewidth){\includegraphics[width=0.48\linewidth]{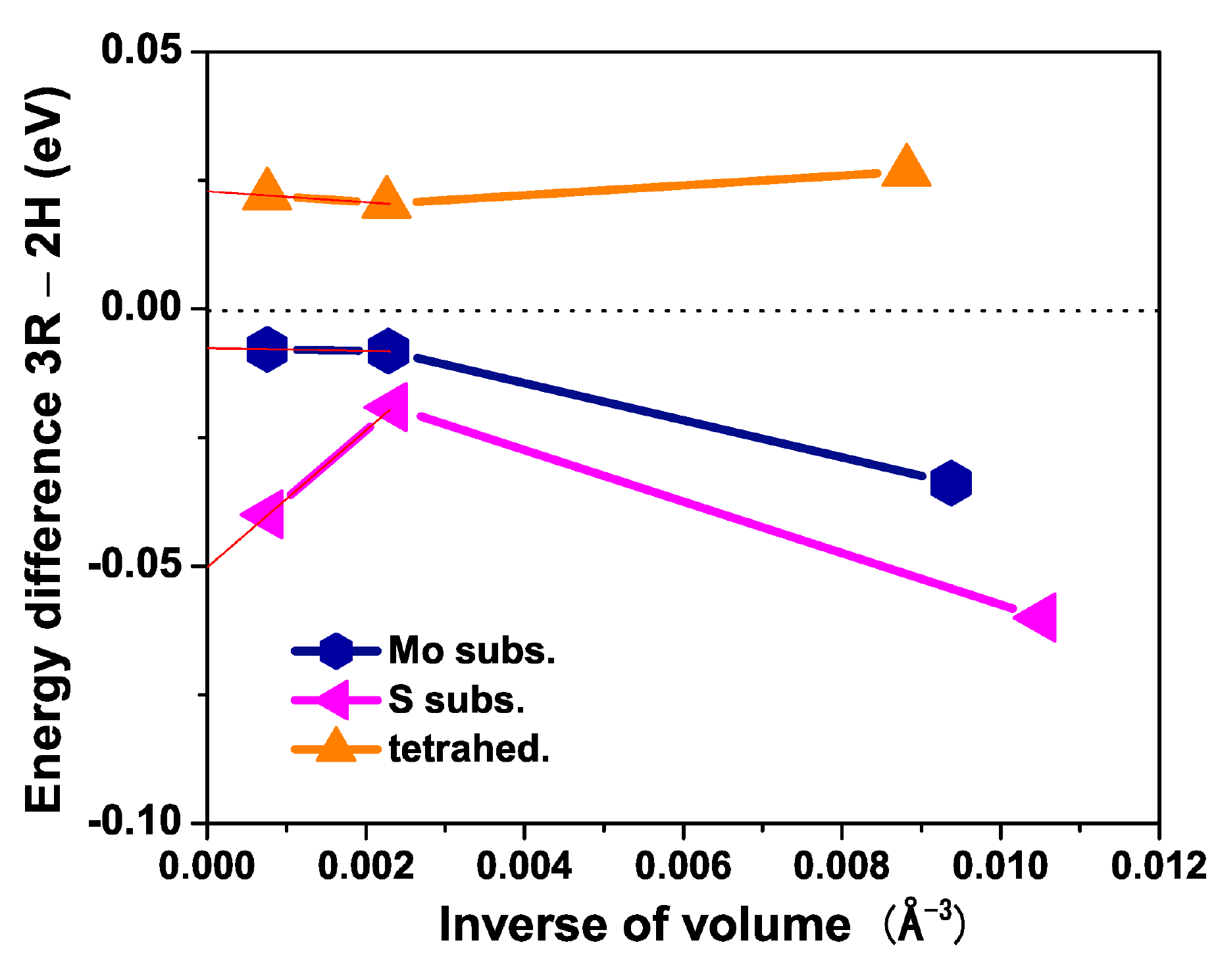}};

            \draw [anchor=north west] (-0.12\linewidth, .61\linewidth) node {(a) \textbf{ {\fontfamily{Arial}\selectfont Intercalations}}};
            \draw [anchor=north west] (0.350\linewidth, .61\linewidth) node {(b) {\fontfamily{Arial}\selectfont \textbf{Mo substitution}}};
            \draw [anchor=north west] (-0.12\linewidth, .17\linewidth) node {(c) {\fontfamily{Arial}\selectfont \textbf{S substitution}}};
            \draw [anchor=north west] (0.350\linewidth, .17\linewidth) node {(d) {\fontfamily{Arial}\selectfont \textbf{3R -- 2H energy difference} }};
            \end{tikzpicture}
			}
				\caption{Doping formation energy from LDA in Ni-doped 2H-MoS$_2$ with respect to volume per Ni atom, in a) intercalations, b) Mo substitution (under Mo-rich conditions)  and c) S substitution (under S-rich conditions). For S-rich conditions, (b) is shifted down by 3.06 eV; for Mo-rich conditions, (c) is shifted down by 1.53 eV. $N$ is the parameter for the supercell size. Tetrahedral intercalation is favored over octahedral by a fairly constant amount. Red lines show extrapolation from the last two points (not always well converged yet) to the low-doping limit, achieved much more rapidly for the $z$-varying supercell series. d) Energy difference per Ni atom between Ni-doped 2H and 3R structures.}
				\label{fig:2H_all_intercal}
		\end{figure}

			 We can extrapolate  defect energies to the low concentration limit  as a function of inverse  volume ($1/V$) according to the scheme of Makov and Payne (based on behavior of both local strain and electrostatics) \cite{extrapolation}. We use this method for the doping formation energy and show results for the 2H phase in Fig. \ref{fig:2H_all_intercal}. The doping formation energies for each polytype and concentration are tabulated in Tables S1-S5 \cite{SM}. Variation in doping formation energy is small along the $z$-axis ($1\times1\times N$ and $2\times2\times N$) in all cases, expected due to weak interaction between layers. This finding indicates the energetics of doped bilayers are similar to bulk.  The  doping formation energy for in-plane supercells  depends significantly on the size for small supercells ($N\times N\times1$ and $N\times N\times2$), but converges at high supercell size. The limit of low doping concentration can be estimated by extrapolation to $1/V=0$ from the last two points (red line in Fig. \ref{fig:2H_all_intercal}) as tabulated in Table S6 \cite{SM}. Low doping concentration calculations involve increasingly  large number of atoms and it is more efficient to extrapolate. 
			 
			 The energy changes with concentration are due to interaction between dopants, and so the results demonstrate that that in both 2H (Fig. \ref{fig:2H_all_intercal}) and 3R (Fig. S1 \cite{SM}), 
			  Mo substitution has the most interaction, S substitution less, and intercalations very little. The negligible Ni-Ni interaction in intercalations is also seen by the fact that doubly intercalated 2H structures (one Ni between each pair of layers) have almost the same doping formation energy as the ones with one Ni per cell (Table S3 \cite{SM}). In the monolayer, the difference in doping formation energy with increase in supercell sizes is considerably smaller (a few tenths of an eV) than in bulk, except for the case of the S-atop adatom (Fig. S2 and Table S5 \cite{SM}). 
			 
			 When we compared the total energy with different supercell sizes but the same number of MoS$_2$ units, we found a significant difference ($\sim$1 eV) for smaller supercells ($1\times1\times4$ vs $2\times2\times1$) but much less ($\sim$0.2 eV) with larger supercells ($4\times4\times1$ vs $2\times2\times4$). This variation arises from the spatial arrangement of dopant atoms and can be thought as micro-clustering of dopants or the effect of local ordering of dopants. The reduction in differences at low doping can be clearly observed in the extrapolated lines in  Fig. \ref{fig:2H_all_intercal} where the doping formation energies are closer for different supercells of the same volume.  
	
		Next, we compared the energy difference between doped 2H and 3R to look for the possibility of phase changes. To compare directly the energy of doped 2H and 3R structures, we compared  $2\times2\times2$ for 3R and $2\times2\times3$ for 2H, supercells with 24 units of MoS$_2$. We also compared supercells from the primitive cell of  3R that contains only one MoS$_2$ unit, $1\times1\times2$ and $2\times2\times2$  \textit{vs.} $1\times1\times1$ and $2\times2\times1$  of 2H. Energy differences are shown in Fig. \ref{fig:2H_all_intercal}(d). Recall that pristine 3R is slightly higher in energy than 2H. We find Mo-substituted and S-substituted 3R are lower in energy  than 2H at all considered doping concentrations, suggesting a possibility of phase change from 2H to 3R with substitutional doping.  In experiment, Mo substitution by Ni was reported to modify the stacking sequence of  MoS$_2$ layers forming 3R \cite{Ni_2H_3R}, consistent with our finding. Similarly, a phase change from 2H to 3R was reported  for Nb substitution of Mo in MoS$_2$ \cite{phase_change_Nb}. By contrast, for tetrahedral intercalation, doped 2H is lower than doped 3R (Mo/S-atop)  at all considered doping concentrations, further strengthening the stability of 2H over 3R.
		
		We additionally constructed phase diagrams of the stable doping site as a function of chemical potentials $\mu_{\rm Mo}$ and $\mu_{\rm S}$, for each polytype, as we did previously for 2H-MoS$_2$ \cite{enrique}. These phase diagrams (Fig. S3 \cite{SM}) allow us to identify which doped structures can form in equilibrium and which are compatible with stability of the corresponding pristine structure. The 1H diagram is similar to 2H, with a shift to higher chemical potentials, and the Mo-atop adatom site is favored in the pristine stability region. The 3R phase diagram also closely resembles the 2H one \cite{enrique}, and only the Mo/S-atop intercalation is compatible with pristine stability, except at the highest doping levels. The other doping sites are predicted to be accessible only out of equilibrium.
		
	\subsection{Structural parameters}
	\label{structural_parameters}
	Incorporating Ni into MoS$_2$ not only changes the energy, but also can cause local changes in structure (bond lengths and lattice parameters), which can be useful for experimental identification of the structure. Studying bonding also helps understand interlayer interactions and sliding. The maximum difference in Ni-S and Ni-Mo bond lengths with respect to the pristine Mo-S bond length  is around 8\% for the highest doping concentration in substitution sites. Mo-S bond lengths away from the doping site were very close to the pristine Mo-S bond (differing  by less than 0.04 \rm\AA), showing only a local effect on the structure. For the $1\times1\times1$ supercell of 2H, the Ni-S bond length decreased \textit{vs.} the pristine Mo-S bond length, and the Ni-Mo bond in S substitution increased, as shown in Table \ref{tab:Bond_length}. The bond lengths for substitutional doping in all 2H and 3R structures are consistent with a covalent bond, judging by the covalent radii (Mo = 1.54 \rm\AA, Ni = 1.24 \rm\AA, S = 1.05 \rm\AA ) \cite{covalent_radii}. Tetrahedral intercalation also shows covalent bonds, in agreement with our previous results based on electron density \cite{enrique}: the Ni-S, and even Ni-Mo, bond lengths are far lower than the sum of Van der Waals radii  (S = 1.81 \rm\AA\ \cite{Van_radii_S}, Ni = 1.97 \rm\AA\  \cite{Van_radii_MoNi}  \space and Mo = 2.16\rm\AA \space \cite{Van_radii_MoNi}). For comparison, the  S-S distance between pristine MoS$_2$ layers is 3.59 \rm\AA \space which  is close to the sum of the Van der Waals radii of two S atoms. For a concentration less than  1\% of Mo substitution in both 2H and 3R, we found only 4 Ni-S bonds instead of 6, breaking the symmetry in a pseudo-Jahn-Teller distortion. A similar five-fold coordination was calculated \cite{doped_apl2} for Ni substituting Mo at a 2H basal plane surface, attributed to population of anti-bonding $d$-orbitals by extra charge from Ni. Four-fold coordination like our result was reported in a previous theoretical calculation for Mo-substituted 2H-MoS$_2$ in few-layer nanosheets \cite{four_coodination}.
	
		\begin{table}
	
	   \footnotesize

		\begin{tabular}{c| r| r| r| r| r| r}

			\Longstack{ \textbf{Pristine} \\ \textbf{Mo-S=2.41 \rm\AA}} & \Longstack{\textbf{2H} \\ \textbf{($1\times1\times1$)}}  & \Longstack{\textbf{2H} \\  \textbf{ ($4\times4\times1$) }} & \Longstack{\textbf{3R} \\ \textbf{($1\times1\times1$)}} & \Longstack{\textbf{ 3R} \\  \textbf{($4\times4\times1$)}} & \Longstack{\textbf{1H} \\ \textbf{($4\times4\times1$)}} & \Longstack{\textbf{Covalent}  \\ \textbf{radii sum}} \\
			\hline
			\hline

			Ni-S (Mo subs.) & 2.23 {\rm\AA} &  2.23 {\rm\AA} & {2.25 \rm\AA}  & {2.24 \rm\AA} & 2.34 \rm\AA &   2.29 \rm\AA \\
			\hline
			Ni-Mo (S subs.) & 2.67 {\rm\AA}  & 2.55 {\rm\AA}  & {2.58  \rm\AA}  & {2.55 \rm\AA} & 2.51 \rm\AA &2.78 \rm\AA \\
			\hline
			Ni-S (Intercal./adatom) & 2.15 (2.17) {\rm\AA} &  2.16 (2.10) {\rm\AA} & 2.18 (2.16) {\rm\AA}  & 2.16 (2.10) {\rm\AA} & 2.12 \rm\AA &   2.29 \rm\AA \\
			\hline
		\end{tabular}
		
		\caption{Comparison of Ni bond lengths in Ni-doped MoS$_2$ polytypes, from PBE+GD2. The lowest-energy intercalation and adatom structures are considered. For 2H/3R tetrahedral intercalation, the 3 equivalent Ni-S bonds' length is listed first, followed by the 4th one (see Fig. \ref{fig:bulk_intercal}).
		\label{tab:Bond_length}}
	\end{table}
	
		The 1H monolayer showed structural effects similar to the 2H phase. In Mo substitution, the Ni-S bond is decreased  and in S substitution the Ni-Mo bond is increased, for all supercells considered. The bond lengths  are close to the sums of  covalent radii except for the S-atop case (1.96 \rm\AA), which is considerably shorter, presumably due to the fact that Ni is forming only one bond.
	
	Results on lattice parameters are summarized in Table \ref{tab:cparameter} for $2\times2\times1$ supercells. The change in plane ($a$ and $b$) was small (less than 2\%) for all doping sites and polytypes. However, the $c$-parameter out of plane decreases slightly in the 2H and 3R phases in substitutional cases. In intercalations, the $c$-parameter was reduced slightly, though perhaps not significantly (within error bars). This is notable because intercalations are usually expected to increase the $c$-parameter due to Van der Waals interaction with the intercalant \cite{Intercalation_inc_layers,phase_change_Li}. However, as we previously showed, intercalated Ni forms covalent bonds \cite{enrique}, resulting in little or no change in interlayer spacing on intercalation. We present an error bar in $c$-parameters  in Table \ref{tab:cparameter}  due to the insensitivity in total energy change in the $c$-direction with weak Van der Waals interactions (not observed along the stronger-bonded $a$- and $b$-directions). This error bar is calculated by parabolic fitting of the total energy  per MoS$_2$ vs  $c$ and finding the range  where the energy is within 0.01 eV of the minimum. As we lower the doping concentration, lattice parameters necessarily converge towards the pristine value, as the effect of a single Ni atom is diluted among pristine surroundings. 
	
\begin{figure}[h]
			
				\includegraphics[width=0.99\linewidth]{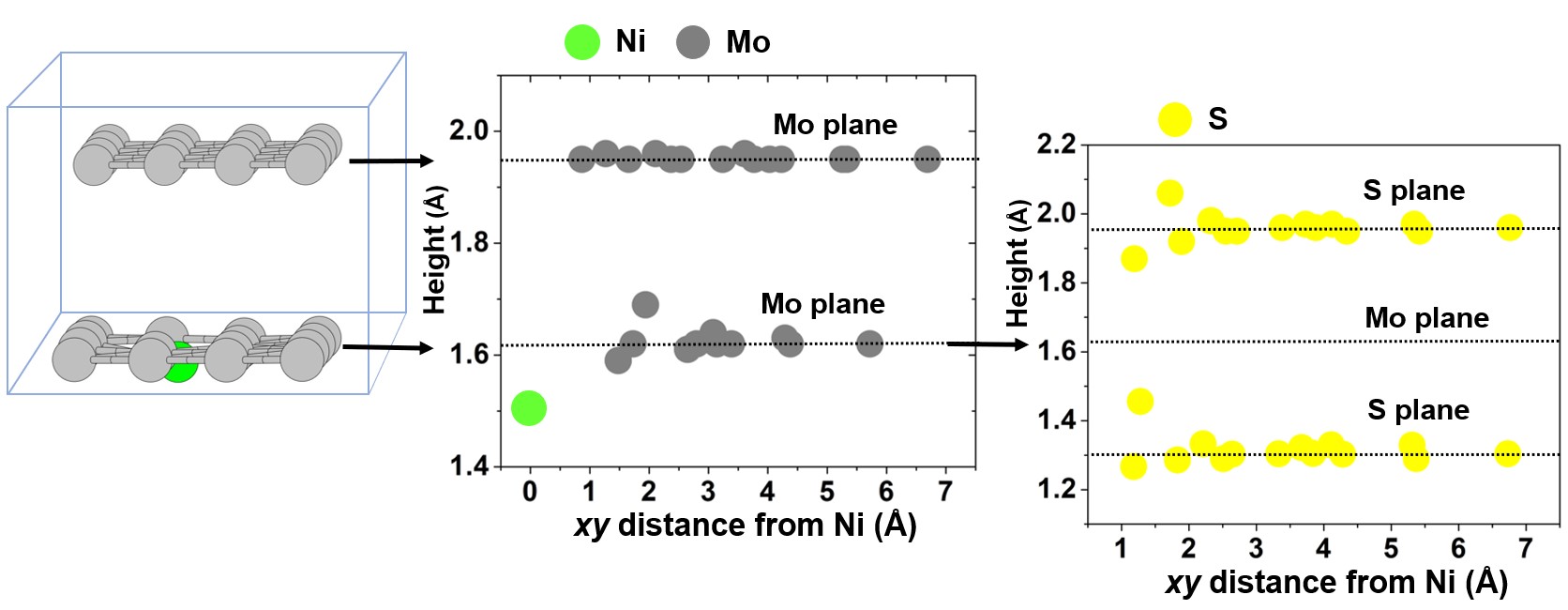}
				\caption{Vertical atomic displacements (from PBE+GD2) in a $4 \times 4 \times 1$ supercell of Mo-substituted 2H-MoS$_2$, with planes shifted vertically to show changes clearly. Blue arrows point to  the plotted heights of the Mo planes.  Center: height of atoms with respect to pristine, indicated by green and purple dotted lines. Right: the two S-planes sandwiching the Mo-plane containing Ni. Most changes in height are observed close to the Ni-atom.}
				\label{fig:plane}
		
		\end{figure}

	We have also analyzed the vertical displacement of atoms in an Mo or S plane. We find rough planes in Mo substitution that could reduce the effective contact area, resulting in a lower coefficient of friction. Fig. \ref{fig:plane} shows the planes of Mo atoms under Mo substitution by Ni in the 2H phase. The Mo plane containing Ni has an uneven surface where the maximum displacement of an Ni atom is 0.24 \rm\AA \space and the maximum displacement of an Mo atom is 0.10 \rm\AA \space with respect to the pristine plane, occurring mostly for nearest neighbors of Ni. The two S planes sandwiching the substituted  Mo plane  also have uneven surfaces: S atoms near to Ni were displaced up to 0.22 \rm\AA \space towards the Mo plane. In S substitution, only the Ni atom was significantly displaced, while the shift was negligible for all atoms in their respective planes. For all substituted structures, the planes of Mo and S  in the unsubstituted layer are smooth and have negligible displacements. Intercalations also showed negligible displacements in Mo or S planes.
	
\begin{table}[htbp]
    \begin{tabular}{|c|r|r|r|r|r|}
    \Xhline{1 pt}
    \textbf{System} & \multicolumn{1}{c|}{\textbf{$c$ (2H)}} & \multicolumn{1}{c|}{\textbf{$a$ (2H)}} & \multicolumn{1}{c|}{\textbf{$c$ (3R)}} & \multicolumn{1}{c|}{\textbf{$a$ (3R)}} & \multicolumn{1}{c|}{\textbf{$a$ (1H)}}  \\
     \Xhline{1 pt}
    Pristine & \multicolumn{1}{r|}{ 12.4 $\pm$ 0.1} & 3.19 & \multicolumn{1}{r|}{18.5 $\pm$ 0.1} & 3.19  & \multicolumn{1}{r|}{3.19}  \\
    \hline
    Mo subs. & \multicolumn{1}{r|}{12.3 $\pm$ 0.1} & 3.20   & \multicolumn{1}{r|}{ 18.4 $\pm$ 0.1} & 3.20   & \multicolumn{1}{r|}{3.21}  \\
    \hline
    S subs. & \multicolumn{1}{r|}{12.2 $\pm$ 0.1} & 3.17  & \multicolumn{1}{r|}{18.3 $\pm$ 0.1} & 3.18  & \multicolumn{1}{r|}{3.15}  \\
    \hline
    Intercal./ & \multicolumn{1}{r|}{12.5 $\pm$ 0.2} & \multicolumn{1}{r|}{3.19} & \multicolumn{1}{r|}{18.7 $\pm$ 0.2} & \multicolumn{1}{r|}{3.20} & \multicolumn{1}{r|}{3.20} \\
    adatom & (tetrahed.) &       & (Mo/S-atop tetrahed.) &       & (Mo-atop)  \\
    \hline
    Intercal./ & \multicolumn{1}{r|}{12.4 $\pm$ 0.2} & \multicolumn{1}{r|}{3.19} & \multicolumn{1}{r|}{18.5 $\pm$ 0.2} & \multicolumn{1}{r|}{3.19} & \multicolumn{1}{r|}{3.19}  \\
    adatom & (octahed.) &       & (trigon. pyram.) &       & (S-atop)  \\
     \Xhline{1 pt}
    \end{tabular}%
   \caption{The lattice parameters $a$ and $c$ (all in \rm\AA)\space from PBE+GD2 for different polytypes and doping cases, in a $2\times2\times1$ supercell.  In all cases $a=b$. The lattice parameters are expressed in terms of the conventional unit cell of the pristine phases. See text for definition of $c$ error bar. Only substitutional doping decreases the $c$-parameter whereas the  intercalations lie mostly  within the error bar of pristine  and the mean values indicate little or no expansion along $c$. Monolayer 1H does not have a $c$-parameter.
   \label{tab:cparameter}}
\end{table}%

	These structural results can be a guide to identifying the doping site in Ni-doped MoS$_2$. Small expansion or unaltered interlayer spacing, which can be detected by X-ray diffraction \cite{Ni_2H_3R} or atomic force microscopy measurements \cite{lieber}, indicates the presence of intercalants. The contraction of interlayer spacing indicates substitutional doping, consistent with recent  experimental work on Ni-doped MoS$_2$: Mosconi {\it et al.} \cite{Ni_2H_3R} found layer contraction in a sample believed to have Mo substitution. Thin films made under an excess of sulfur \cite{Giang_2020}, which favors Mo substitution \cite{enrique}, were found to have a layer contraction increasing with Ni concentration, and also an increasing nanoscale roughness, consistent with our results. Lieber and Kim \cite{lieber}  interpreted lack of  change in the in-plane lattice in atomic force microscopy (AFM) measurements of Ni-doped MoS$_2$ as indicative of Mo substitution, but  our results shows little change in $a$ in any case. Instead, from our results, we propose that Mo or S substitution can be distinguished by observing contracted Ni-S bonds (Mo substitution) or elongated Mo-Ni bonds (S substitution) via transmission electron microscopy (TEM), scanning tunneling microscopy (STM), or extended X-ray absorption fine structure (EXAFS). Thus, these results may be useful to experimentalists to characterize the Ni location in MoS$_2$.

	\subsection{Electronic properties}
\begin{figure}[h]
				\centering{
			\begin{tikzpicture}
			\node [anchor=north west] (imgA) at (-0.15\linewidth,.58\linewidth){\includegraphics[width=0.51\linewidth]{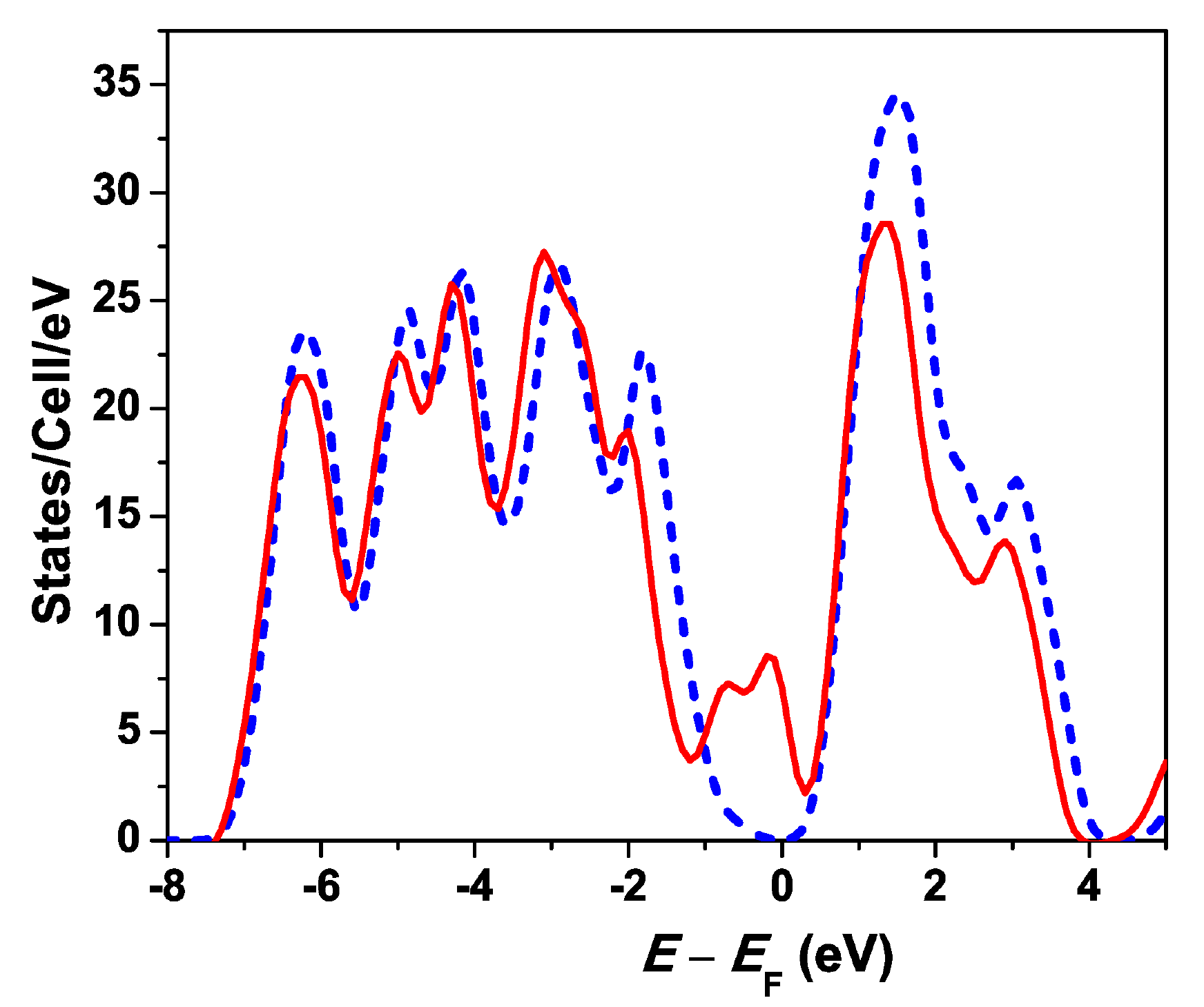}};
            \node [anchor=north west] (imgB) at (0.36\linewidth,.58\linewidth){\includegraphics[width=0.51\linewidth]{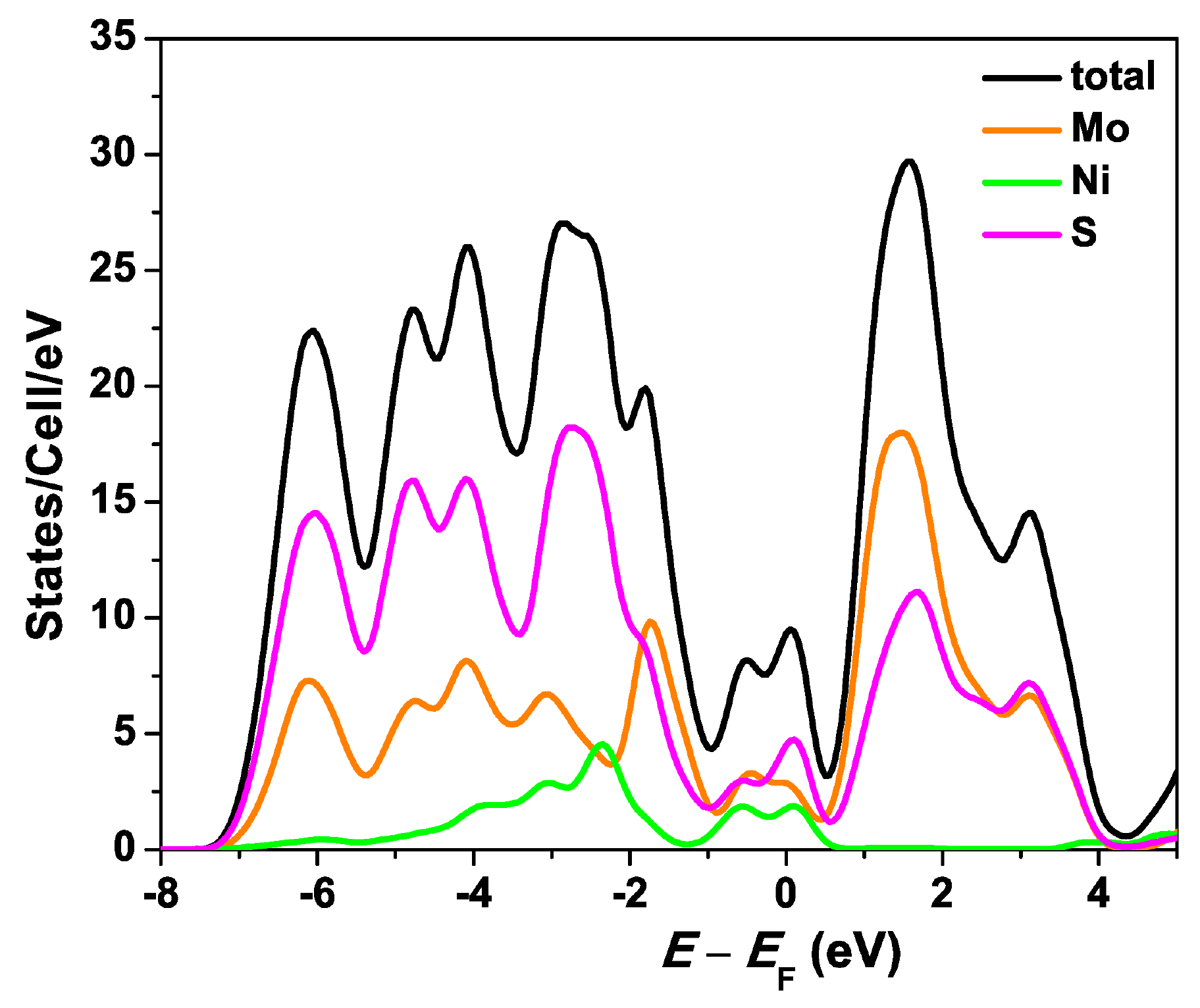}};
          	\node [anchor=north west] (imgA) at (-0.15\linewidth,.08\linewidth){\includegraphics[width=0.51\linewidth]{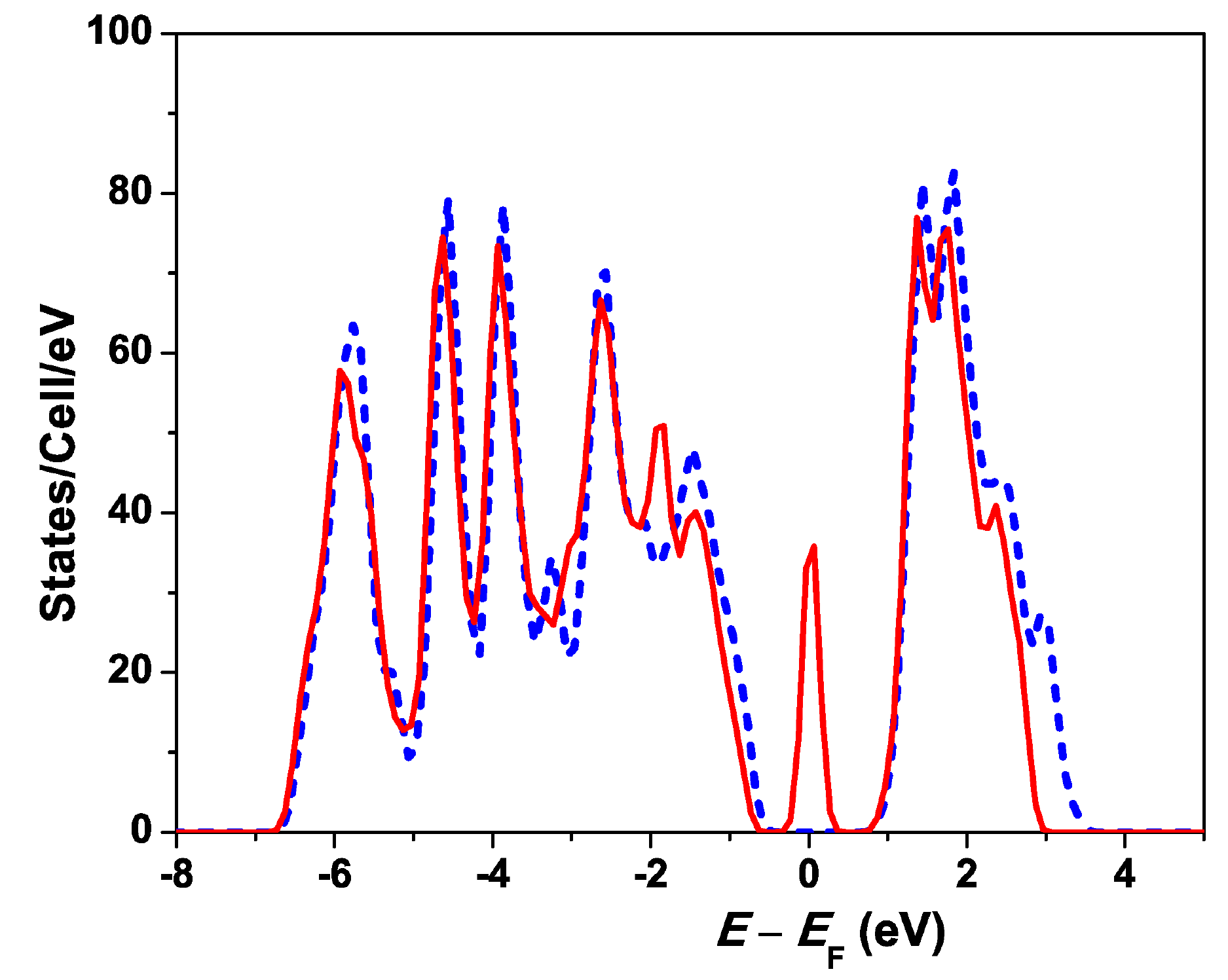}};
            \node [anchor=north west] (imgB) at (0.36\linewidth,.08\linewidth){\includegraphics[width=0.51\linewidth]{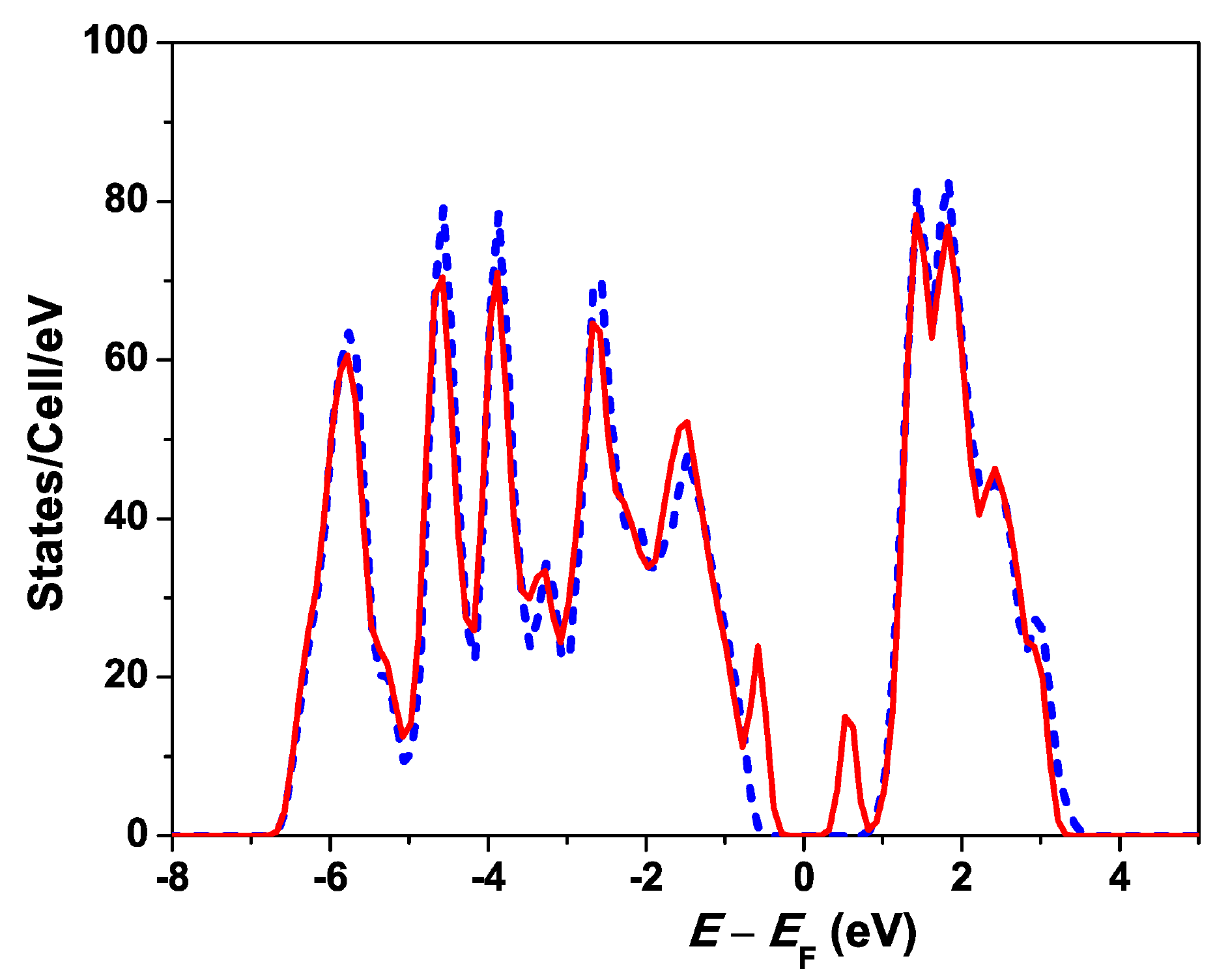}};

            \draw [anchor=north west] (-0.12\linewidth, .62\linewidth) node {(a) {\fontfamily{Arial}\selectfont \textbf{Mo substitution in 2H ($2\times2\times1)$}}};
            \draw [anchor=north west] (0.350\linewidth, .62\linewidth) node {(b) {\fontfamily{Arial}\selectfont \textbf{Partial density of states for (a)}}};
            \draw [anchor=north west] (-0.12\linewidth, .12\linewidth) node {(c) {\fontfamily{Arial}\selectfont \textbf{Mo substitution in 1H $(4\times4\times1)$}}};
            \draw [anchor=north west] (0.350\linewidth, .12\linewidth) node {(d) {\fontfamily{Arial}\selectfont \textbf{S substitution in 1H $(4\times4\times1)$}}};
            \end{tikzpicture}
			}
		
		\caption{Density of states for doped 2H and 1H structures, from PBE+GD2. The blue dotted curves in (a), (c), (d) are the pristine density of states. (b) shows the partial density of states for 2H with Mo substitution.  Around the Fermi level, we find (a) metallic behavior, (c) one in-gap state, and (d) two in-gap states. }
		\label{fig:dos}
		
	\end{figure}
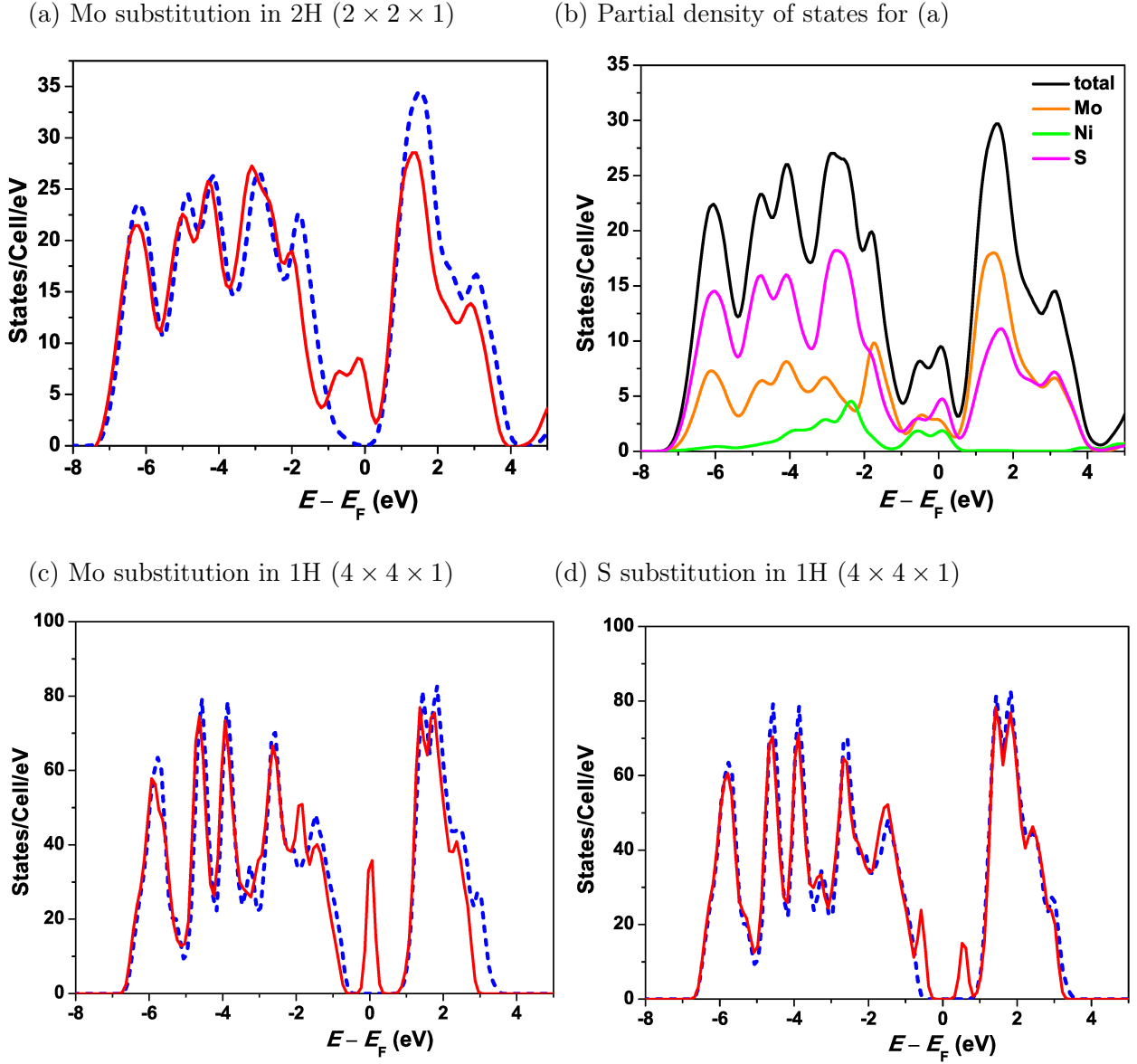

	We studied the electronic properties to find observable signatures of the doping site, as well as to identify effects of Ni-doping that may be useful for electronic applications. When undoped, 1H is a direct-gap semiconductor, and 2H and 3R are indirect-gap semiconductors \cite{lattice,T_3R_bandgap}. The density of states for pristine MoS$_2$ polytypes is shown in Fig. S6 \cite{SM}. We find that Mo substitution makes bulk phases (2H and 3R) metallic, as shown in Fig. \ref{fig:dos} and Fig. S5 \cite{SM}. The partial density of states (PDOS) reveals that such metallic character  arises due to  contributions  at the Fermi level from $d$ orbitals of Mo and Ni and $p$ orbitals from S -- not from Ni alone as one might expect. S substitution and intercalations in 2H and 3R preserved the semiconducting gap. In the 1H monolayer, Mo substitution showed a single in-gap state whereas S substitution showed two in-gap states  at the band edges as shown in Fig. \ref{fig:dos}(c)-(d). Adatom systems are semiconducting; S-atop has a deep in-gap state, and Mo-atop and hollow have an in-gap state close to the valence band (Fig. S4 \cite{SM}).  These in-gap states are due to  $d$-orbitals of Mo and Ni and $p$-orbitals from S, as in 2H substitution. Zhao {\it et al.} \cite{O2_ads} similarly reported two in-gap states at the band edges for S-substituted 1H. All the adatom cases in 1H showed semiconducting behavior. 
        
        These results are relevant for possible applications: the in-gap states could be useful for optoelectronic applications such as quantum emitters \cite{quantum_emitters,emitters_ingap}. The metallic Mo-substituted 2H-MoS$_2$ could be used as an electrode for electrochemical energy storage devices, as semiconducting pristine 2H-MoS$_2$ is not an ideal electrode material \cite{as_electrode_mos2}  and attempts have been made  to enhance electrical conductivity by forming hybrid electrodes \cite{hybrid_electrode,hybrid_electrode2}. The metallic nature for 3R also suggests suitability of Ni substituting Mo as an active site for catalysis, as is well established for 2H \cite{3R_catalysis}. The conductivity of Mo-substituted 2H- or 3R-MoS$_2$ could be identified by transport measurements or STM as a way of determining the doping site, in conjunction with the characteristic structural parameters mentioned in the previous section. Indeed, a recent study \cite{Giang_2020} grew Ni-doped MoS$_2$ under S-rich conditions, which favors Mo substitution \cite{enrique}, and found decreasing resistivity with Ni concentration, consistent with our results.
        
        The oxidation state of Ni, as measured by X-ray photoelectron spectroscopy (XPS), has been used to infer the location of doping by transition metals \cite{doped_apl4} in MoS$_2$. For example, XPS has found Ni$^{0}$ and Ni$^{2+}$ in Ni-doped 2H-MoS$_2$ \cite{Ni_intercalation_oxidation_state}, which could be assumed to represent intercalation and Mo substitution. The relation between oxidation states and local charges in \textit{ab initio} calculations is not simple, but to investigate this issue, we first computed local charges on each atom by Löwdin charges \cite{lowdin} (similar to an integration of the PDOS) and found only small changes in local electron numbers for Mo and S in doped MoS$_2$ compared to pristine. In 2H, Ni has the most partial positive charge in Mo substitution (+0.75$e$), followed by intercalations (+0.52$e$) and S substitution (+0.38$e$) as least positive. We find only a small dependence on supercell sizes, and similar results for 1H as for 2H, which does not suggest detectable differences in XPS. On the other hand, we also tried the more advanced method of Sit \textit{et al.} \cite{accurate_OS} for assigning oxidation states based on $d$-orbital occupations from spin-polarized DFT+$U$ calculations, using $U$ values of 5.5 eV for Ni \cite{magnetization} and 4 eV for Mo \cite{Hub_U_Mo}. We found an oxidation state of Ni$^{4+}$ in Mo substitution at high Ni concentrations in 2H ($>$16\%), 3R ($>$11\%), and 1H ($>$8\%); lower concentrations showed Ni$^{2+}$ in 1H, 2H, and 3R.  S substitution and intercalations/adatoms showed Ni$^{2-}$ in all polytypes. By contrast, the oxidation state for Mo could not be unambiguously assigned in this approach for either pristine or doped systems. These results indicate XPS can help detect Ni-doping sites, and are consistent with the idea of Ni$^{2+}$ for Mo substitution, though we find the unusual Ni$^{2-}$ for intercalation and adatoms instead of Ni$^{0}$ which has been previously expected. We compared DFT+U results for the density of states and did not observe any significant differences with respect to results without U.

        We also checked the presence of magnetic moments. While bulk Ni is ferromagnetic, the magnetism is itinerant, not localized on Ni, and therefore adding Ni dopants does not typically cause magnetism. All Ni-doped 2H and 3R structures are non-magnetic. Total magnetic moments per cell in Mo-substituted 1H were 3.99 $\mu_B$ for a $4\times4\times1$ supercell, in accordance with a previous calculation \cite{magnetization},  and 3.62 $\mu_B$ for a $3\times3\times1$ supercell. No other doped  1H cases showed magnetization.

\subsection{Layer dissociation energy}

 The layer dissociation energy is defined as the energy required to dissociate  MoS$_2$ into separate layers, as used in the work of Mounet \textit{et al.} \cite{Layer_dissociation} to identify  exfoliable compounds from a large database. We study the layer dissociation energy to assess how interlayer interactions change with doping and gain insight into the likelihood of wear in solid lubrication. The layer dissociation energy for pristine 2H- and 3R-MoS$_2$ is calculated as
\begin{equation}
 \Delta E_{\rm pristine} = N E_{\rm 1H-pristine} - E_{\rm bulk-pristine}
\end{equation}
 where $\Delta E_{\rm pristine}$ is the layer dissociation energy for pristine, $N$ is the number of layers in a bulk system, $E_{\rm 1H-pristine}$ is the energy of the monolayer of equivalent size used in $E_{\rm 1H-doped}$, and $E_{\rm bulk-pristine}$ is the energy of the undoped bulk system. Similarly, the layer dissociation energy for the doped system is given by:
\begin{equation}
\Delta E_{\rm doped} = E_{\rm 1H-doped} + (N-1) E_{\rm 1H-pristine} - E_{\rm bulk-doped}
\end{equation}
 where $E_{\rm 1H-doped}$ is the energy of the  doped monolayer of equivalent size and structure as used in $E_{\rm 1H-pristine}$, and $ E_{\rm bulk-doped}$ is the energy of the doped bulk system. To illustrate this, consider the $2\times2\times2$  supercell of 2H with Mo substitution by Ni. Here, $N=4$, $E_{\rm bulk-pristine}$ is the energy of  $2\times2\times2$ undoped MoS$_2$ (Mo$_{16}$S$_{32}$), $E_{\rm 1H-pristine}$ is the energy of $2\times2\times1$ undoped 1H (Mo$_4$S$_8$), $E_{\rm 1H-doped}$ is the energy of  Mo-substituted  $2\times2\times1$  1H  (NiMo$_3$S$_8$), and $E_{\rm bulk-doped}$ is the energy of Mo-substituted  $2\times2\times2$ 2H (NiMo$_{15}$S$_{32}$). Given these quantities, we define a  relative layer dissociation energy  by $\Delta E_{\rm rel} =\Delta E_{\rm doped} - \Delta E_{\rm pristine} $ which accounts for changes  due to doping and is per Ni dopant atom since all  our supercells contain just one Ni atom. The absolute value of the layer dissociation energy in a supercell can be much bigger than the pristine primitive cell's, and so it is more convenient to analyze this relative quantity.

\begin{figure}[h]

		\centering{
			\begin{tikzpicture}
			\node [anchor=north west] (imgA) at (0.0\linewidth,.593\linewidth){\includegraphics[width=0.7\linewidth]{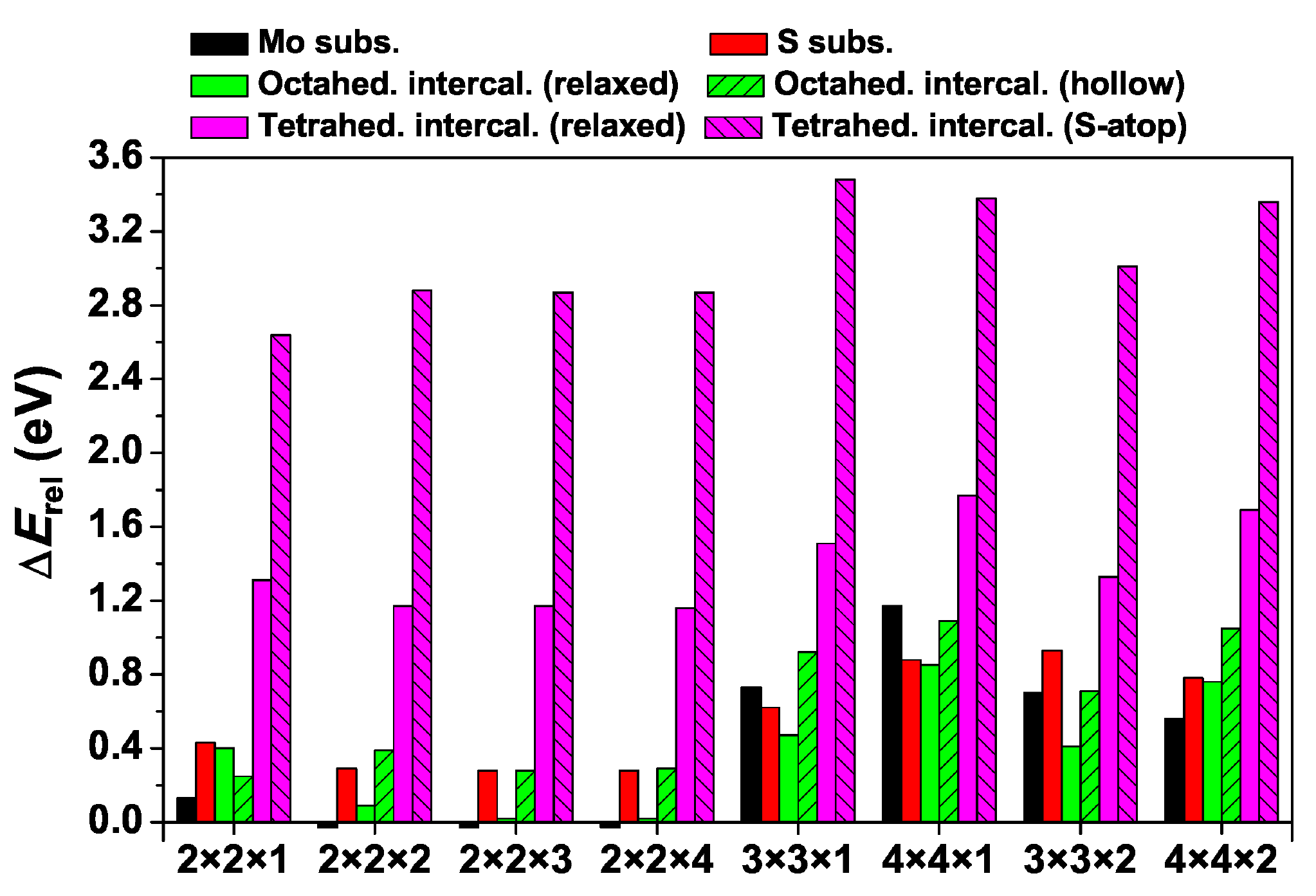}};
            \node [anchor=north west] (imgB) at (0.0\linewidth,.08\linewidth){\includegraphics[width=0.7\linewidth]{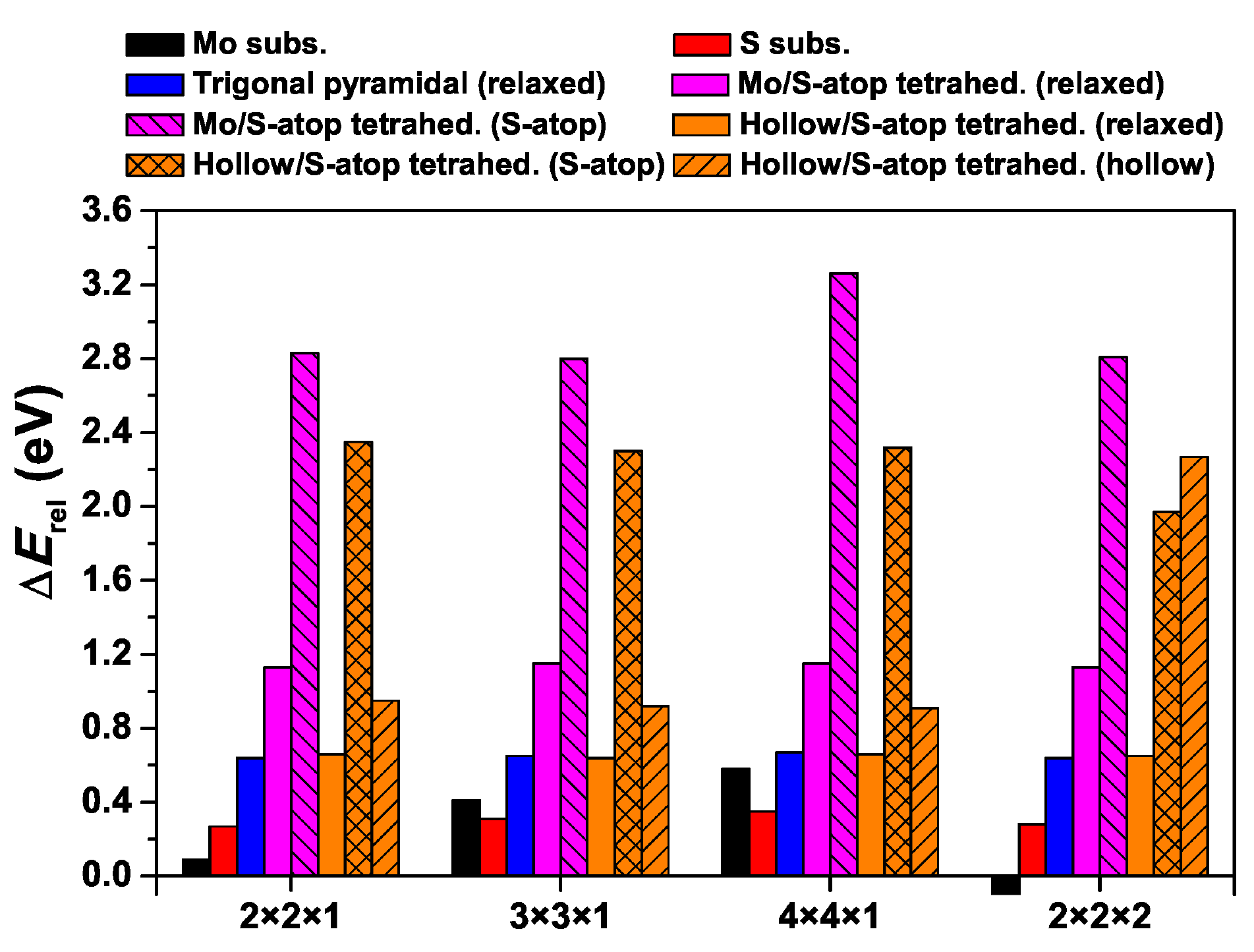}};
          
            \draw [anchor=north west] (0.000\linewidth, .62\linewidth) node {(a)};
            \draw [anchor=north west] (0.0\linewidth, .09\linewidth) node {(b)};
         
            \end{tikzpicture}
			}
		\caption{ (a) Relative layer dissociation energy from PBE+GD2 in (a) 2H-MoS$_2$ and (b) 3R-MoS$_2$, for different doping sites and concentrations, showing a general increase in interlayer binding. We consider both adiabatic and relaxed structures of the doped monolayer.}
		\label{fig:layer_dissociation}
\end{figure}

In intercalation cases, we need to consider whether the dopant will remain in place or relax to a more favorable site, as shown in left side of Fig. \ref{fig:mono_atop}. For example, 2H octahedral intercalation consists of two  hollow sites  with respect to the two layers, and  when the layers are dissociated, the dopant may remain as before in the hollow of one layer (adiabatic case), or may relax to the most favorable site, Mo-atop, depending on the thermal energy available to overcome any barrier. $\Delta E_{\rm rel}$ is always smaller with monolayer relaxation than in the adiabatic case. In tetrahedral intercalation of 2H, Ni remaining on the layer with S-atop has higher  $\Delta E_{\rm rel}$ than remaining on the layer with Mo-atop  since Mo-atop in monolayer is more strongly bound (Section \ref{monolayer}): the energy difference between Mo-atop and S-atop in  1H  is 1.68 eV and the average difference in $\Delta E_{\rm rel}$ in tetrahedral intercalation for Mo-atop and S-atop is 1.70 eV. For comparison, our calculation shows the layer dissociation energy is 0.15 eV per MoS$_2$ unit in the layer (0.27 N m$^{-1}$), which is consistent with the  result  0.22 N m$^{-1}$ from experiment \cite{exp_surface_E}, and  $\sim$0.32 N m$^{-1}$ from a more accurate random-phase
approximation calculation \cite{rpa_surface_E}.   We have considered the different doping cases and found that  nearly all  increase  the binding energy in both 2H and 3R, as shown in Fig. \ref{fig:layer_dissociation}. We also considered higher Ni concentration, $i.e.$ Ni between each layer, and found that $\Delta E_{\rm rel}$ increased, even though doping formation energy per Ni is similar.

 The additional binding between layers is strongest with intercalation while substitution has a lesser effect, consistent with Ni forming  covalent bonds between layers \cite{enrique}. We note that by contrast intercalation with other metals is found to \textit{reduce} interaction between layers, as inferred from Raman spectroscopy for Li \cite{Sekine} and Ag \cite{Sheremetyeva_2021}. In-plane supercell expansion generally increases  $\Delta E_{\rm rel}$, suggesting bilayers will give similar results, whereas supercell expansion out of the plane has little effect.  Ni-doping strengthens the binding between layers which  can explain the experimental report of a reduced low wear rate in Ni-doped MoS$_2$ \cite{low_wear_rate} compared to pristine. The lower wear rate is an important parameter leading to longer life of MoS$_2$ material  for lubrication application \cite{lubricant_by_David}. We may expect  a similar trend in a related quantity, the work of adhesion, which is defined as the energy needed to divide the material in two, whereas the layer dissociation energy involves splitting apart all the layers.  Pastewka {\it et al.} \cite{work_adhesion}  have argued that a larger work of adhesion increases contact area, which would be expected to increase the coefficient of friction, though this may be offset by the increase in atomistic roughness in doped layers. 

Comparing tetrahedral intercalation (the most favorable structure for 2H and 3R), we see 3R has a higher layer dissociation energy in most cases, by 0.04 eV on average. The two structures have similar structural changes around Ni and a similar trend in doping  formation energy. The doping formation energy and  total energy have  only a small difference between 2H and 3R tetrahedral intercalation: 2H is favored by 0.020 eV in doping formation energy and 0.022 eV in total energy. Insofar  as the  layer dissociation energy is determining the wear rate, we do not expect significant differences in wear   between these two phases. Layer sliding simulations for bilayer MoS$_2$ in 2H and 3R-type configurations found that under load 3R has the less corrugated sliding path, both spatially and energetically, suggesting  better lubricant properties \cite{layer_sliding}.  Based on this, Ni-doped 3R may be an attractive target  for further lubrication studies, with experimental tests of coefficient of friction and wear.

		\begin{table}
		\footnotesize
		\begin{tabular}{c| c|c|c|c  }
			\Xhline{1 pt}
	\textbf{Doping site} & \textbf{Lattice in $c$--direction} & \textbf{Bond length} & \textbf{Band gap} &  \textbf{Ni oxidation state}\\
	\Xhline{1 pt}
	Mo subs. & Decreases      &    Decreases (Ni-S)        &    Metallic        &    Ni$^{2+}$, Ni$^{4+}$          \\
	\hline
	S subs. &  Decreases     &  Increases (Ni-Mo)          &   Semiconductor         &  Ni$^{2-}$            \\
	\hline
	Intercal.&  No change/ small increase     &   New bonds           &       Semiconductor     & Ni$^{2-}$              \\
	\Xhline{1 pt}

		\end{tabular}
		\caption{Summary of experimentally observable structural and electronic signatures of different Ni doping sites in bulk 2H- and 3R-MoS$_2$. 1H is similar except: $c$-parameter does not apply, and all sites give semiconductors with in-gap states.
		\label{tab:Detection_techniques}}
	\end{table}

\section{Conclusion}
\label{conclusion}
In this work, we studied MoS$_2$ and its different phases by doping with Ni. We found that the most thermodynamically favorable sites for Ni-doping are Mo-atop adatom in 1H monolayer, tetrahedral intercalation in 2H bulk, and 2H-like intercalation (Mo/S-atop) in 3R bulk. To help guide experimental identification of these sites and substitutional sites, we calculated structural and electronic signatures of the different sites, summarized in Table \ref{tab:Detection_techniques}. In bulk phases, irrespective of doping concentration, we  found Ni bond length contraction for Mo substitution and expansion in S substitution, \textit{vs.} Mo-S bonds.  The bulk lattice parameter out of plane  contracted at high concentration of substitutional Ni but was unaltered or expanded for intercalation. The Mo-S bond lengths away from the dopant were  close to pristine, and lattice parameters converged to pristine at lower doping concentration indicating that Ni-doping has only a local effect. On the other hand, we also found the possibility of a phase change from 2H to 3R with Mo or S substitution. Mo substitution also induced atomic roughening of the Mo and S planes. While most doped structures were semiconducting, we found  that  semiconducting 2H and 3R became metallic with Mo substitution by Ni, of interest for electrochemical or catalytic applications,   whereas substitution and adatoms in 1H are semiconducting with in-gap states of potential interest for quantum emitters. We find oxidation states characteristic of doping site, which can be probed by XPS: Mo substitution has Ni$^{4+}$ or Ni$^{2+}$, and S substitution, intercalations and adatoms have Ni$^{2-}$, for all polytypes.  Lastly, we computed the layer dissociation energy where doped MoS$_2$ shows stronger binding between layers than pristine, particularly for intercalations -- in contrast to findings for other transition metals. In conclusion we present local effects on bonding and structural changes, a possibility of phase change, interesting electronic properties upon Ni-doping and finally strongly increased interlayer interactions. This layer dissociation energy relates to improved lubrication performance lifetime (wear) for doped MoS$_2$. Our results suggest that the Ni-doped 3R polytype may not have significant differences in wear rate than 2H and we propose further exploration of the friction properties of 3R, along with investigation of potential electronic and optical applications of Ni-doped MoS$_2$. Our structural and electronic results can be useful for identification of doping sites in Ni-doped samples.

\begin{acknowledgments}
We acknowledge useful discussions with Ashlie Martini and Mehmet Baykara. This work was supported by  UC Merced start-up funds and  the Merced nAnomaterials Center for Energy and Sensing (MACES), a NASA-funded research and education center, under award NNX15AQ01. This work used computational resources from the Multi-Environment Computer for Exploration and Discovery (MERCED) cluster at UC Merced, funded by National Science Foundation Grant No. ACI-1429783, and the National Energy Research Scientific Computing Center (NERSC), a U.S. Department of Energy Office of Science User Facility operated under Contract No. DE-AC02-05CH11231.
R.K. and E.G. performed calculations, R.K. and D.A.S. wrote the manuscript, and D.A.S. supervised the project.
\end{acknowledgments}


\begin{thebibliography}{79}%
\makeatletter
\providecommand \@ifxundefined [1]{%
 \@ifx{#1\undefined}
}%
\providecommand \@ifnum [1]{%
 \ifnum #1\expandafter \@firstoftwo
 \else \expandafter \@secondoftwo
 \fi
}%
\providecommand \@ifx [1]{%
 \ifx #1\expandafter \@firstoftwo
 \else \expandafter \@secondoftwo
 \fi
}%
\providecommand \natexlab [1]{#1}%
\providecommand \enquote  [1]{``#1''}%
\providecommand \bibnamefont  [1]{#1}%
\providecommand \bibfnamefont [1]{#1}%
\providecommand \citenamefont [1]{#1}%
\providecommand \href@noop [0]{\@secondoftwo}%
\providecommand \href [0]{\begingroup \@sanitize@url \@href}%
\providecommand \@href[1]{\@@startlink{#1}\@@href}%
\providecommand \@@href[1]{\endgroup#1\@@endlink}%
\providecommand \@sanitize@url [0]{\catcode `\\12\catcode `\$12\catcode
  `\&12\catcode `\#12\catcode `\^12\catcode `\_12\catcode `\%12\relax}%
\providecommand \@@startlink[1]{}%
\providecommand \@@endlink[0]{}%
\providecommand \url  [0]{\begingroup\@sanitize@url \@url }%
\providecommand \@url [1]{\endgroup\@href {#1}{\urlprefix }}%
\providecommand \urlprefix  [0]{URL }%
\providecommand \Eprint [0]{\href }%
\providecommand \doibase [0]{https://doi.org/}%
\providecommand \selectlanguage [0]{\@gobble}%
\providecommand \bibinfo  [0]{\@secondoftwo}%
\providecommand \bibfield  [0]{\@secondoftwo}%
\providecommand \translation [1]{[#1]}%
\providecommand \BibitemOpen [0]{}%
\providecommand \bibitemStop [0]{}%
\providecommand \bibitemNoStop [0]{.\EOS\space}%
\providecommand \EOS [0]{\spacefactor3000\relax}%
\providecommand \BibitemShut  [1]{\csname bibitem#1\endcsname}%
\let\auto@bib@innerbib\@empty
\bibitem [{\citenamefont {Splendiani}\ \emph {et~al.}(2010)\citenamefont
  {Splendiani}, \citenamefont {Sun}, \citenamefont {Zhang}, \citenamefont {Li},
  \citenamefont {Kim}, \citenamefont {Chim}, \citenamefont {Galli},\ and\
  \citenamefont {Wang}}]{appl_opticL}%
  \BibitemOpen
  \bibfield  {author} {\bibinfo {author} {\bibfnamefont {A.}~\bibnamefont
  {Splendiani}}, \bibinfo {author} {\bibfnamefont {L.}~\bibnamefont {Sun}},
  \bibinfo {author} {\bibfnamefont {Y.}~\bibnamefont {Zhang}}, \bibinfo
  {author} {\bibfnamefont {T.}~\bibnamefont {Li}}, \bibinfo {author}
  {\bibfnamefont {J.}~\bibnamefont {Kim}}, \bibinfo {author} {\bibfnamefont
  {C.-Y.}\ \bibnamefont {Chim}}, \bibinfo {author} {\bibfnamefont
  {G.}~\bibnamefont {Galli}},\ and\ \bibinfo {author} {\bibfnamefont
  {F.}~\bibnamefont {Wang}},\ }\bibfield  {title} {\bibinfo {title} {Emerging
  photoluminescence in monolayer {MoS}$_2$},\ }\href@noop {} {\bibfield
  {journal} {\bibinfo  {journal} {Nano Lett.}\ }\textbf {\bibinfo {volume}
  {10}},\ \bibinfo {pages} {1271} (\bibinfo {year} {2010})}\BibitemShut
  {NoStop}%
\bibitem [{\citenamefont {Mak}\ \emph {et~al.}(2010)\citenamefont {Mak},
  \citenamefont {Lee}, \citenamefont {Hone}, \citenamefont {Shan},\ and\
  \citenamefont {Heinz}}]{gap_variation}%
  \BibitemOpen
  \bibfield  {author} {\bibinfo {author} {\bibfnamefont {K.~F.}\ \bibnamefont
  {Mak}}, \bibinfo {author} {\bibfnamefont {C.}~\bibnamefont {Lee}}, \bibinfo
  {author} {\bibfnamefont {J.}~\bibnamefont {Hone}}, \bibinfo {author}
  {\bibfnamefont {J.}~\bibnamefont {Shan}},\ and\ \bibinfo {author}
  {\bibfnamefont {T.~F.}\ \bibnamefont {Heinz}},\ }\bibfield  {title} {\bibinfo
  {title} {Atomically thin {MoS}$_2$: A new direct-gap semiconductor},\
  }\href@noop {} {\bibfield  {journal} {\bibinfo  {journal} {Phys. Rev. Lett.}\
  }\textbf {\bibinfo {volume} {105}},\ \bibinfo {pages} {136805} (\bibinfo
  {year} {2010})}\BibitemShut {NoStop}%
\bibitem [{\citenamefont {Li}\ \emph {et~al.}(2008)\citenamefont {Li},
  \citenamefont {Zhou}, \citenamefont {Zhang},\ and\ \citenamefont
  {Chen}}]{appl_magnetic}%
  \BibitemOpen
  \bibfield  {author} {\bibinfo {author} {\bibfnamefont {Y.}~\bibnamefont
  {Li}}, \bibinfo {author} {\bibfnamefont {Z.}~\bibnamefont {Zhou}}, \bibinfo
  {author} {\bibfnamefont {S.}~\bibnamefont {Zhang}},\ and\ \bibinfo {author}
  {\bibfnamefont {Z.}~\bibnamefont {Chen}},\ }\bibfield  {title} {\bibinfo
  {title} {{MoS}$_2$ nanoribbons: High stability and unusual electronic and
  magnetic properties},\ }\href@noop {} {\bibfield  {journal} {\bibinfo
  {journal} {J. Am. Chem. Soc.}\ }\textbf {\bibinfo {volume} {130}},\ \bibinfo
  {pages} {16739} (\bibinfo {year} {2008})}\BibitemShut {NoStop}%
\bibitem [{\citenamefont {Kaasbjerg}\ \emph {et~al.}(2012)\citenamefont
  {Kaasbjerg}, \citenamefont {Thygesen},\ and\ \citenamefont
  {Jacobsen}}]{mobility}%
  \BibitemOpen
  \bibfield  {author} {\bibinfo {author} {\bibfnamefont {K.}~\bibnamefont
  {Kaasbjerg}}, \bibinfo {author} {\bibfnamefont {K.~S.}\ \bibnamefont
  {Thygesen}},\ and\ \bibinfo {author} {\bibfnamefont {K.~W.}\ \bibnamefont
  {Jacobsen}},\ }\bibfield  {title} {\bibinfo {title} {Phonon-limited mobility
  in n-type single-layer {MoS}$_2$ from first principles},\ }\href@noop {}
  {\bibfield  {journal} {\bibinfo  {journal} {Phys. Rev. B}\ }\textbf {\bibinfo
  {volume} {85}},\ \bibinfo {pages} {115317} (\bibinfo {year}
  {2012})}\BibitemShut {NoStop}%
\bibitem [{\citenamefont {Lin}\ \emph {et~al.}(2016)\citenamefont {Lin},
  \citenamefont {Carvalho}, \citenamefont {Kahn}, \citenamefont {Lv},
  \citenamefont {Rao}, \citenamefont {Terrones}, \citenamefont {Pimenta},\ and\
  \citenamefont {Terrones}}]{defect_engg}%
  \BibitemOpen
  \bibfield  {author} {\bibinfo {author} {\bibfnamefont {Z.}~\bibnamefont
  {Lin}}, \bibinfo {author} {\bibfnamefont {B.~R.}\ \bibnamefont {Carvalho}},
  \bibinfo {author} {\bibfnamefont {E.}~\bibnamefont {Kahn}}, \bibinfo {author}
  {\bibfnamefont {R.}~\bibnamefont {Lv}}, \bibinfo {author} {\bibfnamefont
  {R.}~\bibnamefont {Rao}}, \bibinfo {author} {\bibfnamefont {H.}~\bibnamefont
  {Terrones}}, \bibinfo {author} {\bibfnamefont {M.~A.}\ \bibnamefont
  {Pimenta}},\ and\ \bibinfo {author} {\bibfnamefont {M.}~\bibnamefont
  {Terrones}},\ }\bibfield  {title} {\bibinfo {title} {Defect engineering of
  two-dimensional transition metal dichalcogenides},\ }\href@noop {} {\bibfield
   {journal} {\bibinfo  {journal} {2D Mater.}\ }\textbf {\bibinfo {volume}
  {3}},\ \bibinfo {pages} {022002} (\bibinfo {year} {2016})}\BibitemShut
  {NoStop}%
\bibitem [{\citenamefont {Lembke}\ and\ \citenamefont
  {Kis}(2012)}]{high_current}%
  \BibitemOpen
  \bibfield  {author} {\bibinfo {author} {\bibfnamefont {D.}~\bibnamefont
  {Lembke}}\ and\ \bibinfo {author} {\bibfnamefont {A.}~\bibnamefont {Kis}},\
  }\bibfield  {title} {\bibinfo {title} {Breakdown of high-performance
  monolayer {MoS}$_2$ transistors},\ }\href@noop {} {\bibfield  {journal}
  {\bibinfo  {journal} {ACS Nano}\ }\textbf {\bibinfo {volume} {6}},\ \bibinfo
  {pages} {10070} (\bibinfo {year} {2012})}\BibitemShut {NoStop}%
\bibitem [{\citenamefont {Mao}\ \emph {et~al.}(2018)\citenamefont {Mao},
  \citenamefont {Wang}, \citenamefont {Zheng},\ and\ \citenamefont
  {Deng}}]{Mos2_application1}%
  \BibitemOpen
  \bibfield  {author} {\bibinfo {author} {\bibfnamefont {J.}~\bibnamefont
  {Mao}}, \bibinfo {author} {\bibfnamefont {Y.}~\bibnamefont {Wang}}, \bibinfo
  {author} {\bibfnamefont {Z.}~\bibnamefont {Zheng}},\ and\ \bibinfo {author}
  {\bibfnamefont {D.}~\bibnamefont {Deng}},\ }\bibfield  {title} {\bibinfo
  {title} {The rise of two-dimensional {MoS}$_2$ for catalysis},\ }\href@noop
  {} {\bibfield  {journal} {\bibinfo  {journal} {Front. Phys.}\ }\textbf
  {\bibinfo {volume} {13}},\ \bibinfo {pages} {138118} (\bibinfo {year}
  {2018})}\BibitemShut {NoStop}%
\bibitem [{\citenamefont {Radisavljevic}\ \emph {et~al.}(2011)\citenamefont
  {Radisavljevic}, \citenamefont {Radenovic}, \citenamefont {Brivio},
  \citenamefont {Giacometti},\ and\ \citenamefont {Kis}}]{Mos2_application2}%
  \BibitemOpen
  \bibfield  {author} {\bibinfo {author} {\bibfnamefont {B.}~\bibnamefont
  {Radisavljevic}}, \bibinfo {author} {\bibfnamefont {A.}~\bibnamefont
  {Radenovic}}, \bibinfo {author} {\bibfnamefont {J.}~\bibnamefont {Brivio}},
  \bibinfo {author} {\bibfnamefont {V.}~\bibnamefont {Giacometti}},\ and\
  \bibinfo {author} {\bibfnamefont {A.}~\bibnamefont {Kis}},\ }\bibfield
  {title} {\bibinfo {title} {Single-layer {MoS}$_2$ transistors},\ }\href@noop
  {} {\bibfield  {journal} {\bibinfo  {journal} {Nat. Nanotech.}\ }\textbf
  {\bibinfo {volume} {6}},\ \bibinfo {pages} {147–150} (\bibinfo {year}
  {2011})}\BibitemShut {NoStop}%
\bibitem [{\citenamefont {Lopez-Sanchez}\ \emph {et~al.}(2013)\citenamefont
  {Lopez-Sanchez}, \citenamefont {Lembke}, \citenamefont {Kayci}, \citenamefont
  {Radenovic},\ and\ \citenamefont {Kis}}]{Mos2_application3}%
  \BibitemOpen
  \bibfield  {author} {\bibinfo {author} {\bibfnamefont {O.}~\bibnamefont
  {Lopez-Sanchez}}, \bibinfo {author} {\bibfnamefont {D.}~\bibnamefont
  {Lembke}}, \bibinfo {author} {\bibfnamefont {M.}~\bibnamefont {Kayci}},
  \bibinfo {author} {\bibfnamefont {A.}~\bibnamefont {Radenovic}},\ and\
  \bibinfo {author} {\bibfnamefont {A.}~\bibnamefont {Kis}},\ }\bibfield
  {title} {\bibinfo {title} {Ultrasensitive photodetectors based on monolayer
  {MoS}$_2$},\ }\href@noop {} {\bibfield  {journal} {\bibinfo  {journal} {Nat.
  Nanotech.}\ }\textbf {\bibinfo {volume} {8}},\ \bibinfo {pages} {497–501}
  (\bibinfo {year} {2013})}\BibitemShut {NoStop}%
\bibitem [{\citenamefont {Krause}\ \emph {et~al.}(2010)\citenamefont {Krause},
  \citenamefont {Müller}, \citenamefont {Birkmann}, \citenamefont {Böhm},
  \citenamefont {Ebert}, \citenamefont {Grözinger}, \citenamefont {Henning},
  \citenamefont {Hofferbert}, \citenamefont {Huber}, \citenamefont {Lemke},
  \citenamefont {Rohloff}, \citenamefont {Scheithauer}, \citenamefont {Gross},
  \citenamefont {Fischer}, \citenamefont {Luichtel}, \citenamefont {Merkle},
  \citenamefont {Übele}, \citenamefont {Wieland}, \citenamefont {Amiaux},
  \citenamefont {Jager}, \citenamefont {Glauser}, \citenamefont {Parr-Burman},\
  and\ \citenamefont {Sykes}}]{Mos2_application4}%
  \BibitemOpen
  \bibfield  {author} {\bibinfo {author} {\bibfnamefont {O.}~\bibnamefont
  {Krause}}, \bibinfo {author} {\bibfnamefont {F.}~\bibnamefont {Müller}},
  \bibinfo {author} {\bibfnamefont {S.}~\bibnamefont {Birkmann}}, \bibinfo
  {author} {\bibfnamefont {A.}~\bibnamefont {Böhm}}, \bibinfo {author}
  {\bibfnamefont {M.}~\bibnamefont {Ebert}}, \bibinfo {author} {\bibfnamefont
  {U.}~\bibnamefont {Grözinger}}, \bibinfo {author} {\bibfnamefont
  {T.}~\bibnamefont {Henning}}, \bibinfo {author} {\bibfnamefont
  {R.}~\bibnamefont {Hofferbert}}, \bibinfo {author} {\bibfnamefont
  {A.}~\bibnamefont {Huber}}, \bibinfo {author} {\bibfnamefont
  {D.}~\bibnamefont {Lemke}}, \bibinfo {author} {\bibfnamefont {R.-R.}\
  \bibnamefont {Rohloff}}, \bibinfo {author} {\bibfnamefont {S.}~\bibnamefont
  {Scheithauer}}, \bibinfo {author} {\bibfnamefont {T.}~\bibnamefont {Gross}},
  \bibinfo {author} {\bibfnamefont {T.}~\bibnamefont {Fischer}}, \bibinfo
  {author} {\bibfnamefont {G.}~\bibnamefont {Luichtel}}, \bibinfo {author}
  {\bibfnamefont {H.}~\bibnamefont {Merkle}}, \bibinfo {author} {\bibfnamefont
  {M.}~\bibnamefont {Übele}}, \bibinfo {author} {\bibfnamefont {H.-U.}\
  \bibnamefont {Wieland}}, \bibinfo {author} {\bibfnamefont {J.}~\bibnamefont
  {Amiaux}}, \bibinfo {author} {\bibfnamefont {R.}~\bibnamefont {Jager}},
  \bibinfo {author} {\bibfnamefont {A.}~\bibnamefont {Glauser}}, \bibinfo
  {author} {\bibfnamefont {P.}~\bibnamefont {Parr-Burman}},\ and\ \bibinfo
  {author} {\bibfnamefont {J.}~\bibnamefont {Sykes}},\ }\bibfield  {title}
  {\bibinfo {title} {High-precision cryogenic wheel mechanisms of the
  {JWST}/{MIRI} instrument: performance of the flight models},\ }\href@noop {}
  {\bibfield  {journal} {\bibinfo  {journal} {SPIE}\ }\textbf {\bibinfo
  {volume} {7739}},\ \bibinfo {pages} {421 } (\bibinfo {year}
  {2010})}\BibitemShut {NoStop}%
\bibitem [{\citenamefont {Jeong}\ \emph {et~al.}(2018)\citenamefont {Jeong},
  \citenamefont {Kim}, \citenamefont {Hur}, \citenamefont {Han}, \citenamefont
  {Lee},\ and\ \citenamefont {Lee}}]{doped_apl1}%
  \BibitemOpen
  \bibfield  {author} {\bibinfo {author} {\bibfnamefont {G.}~\bibnamefont
  {Jeong}}, \bibinfo {author} {\bibfnamefont {C.~H.}\ \bibnamefont {Kim}},
  \bibinfo {author} {\bibfnamefont {Y.~G.}\ \bibnamefont {Hur}}, \bibinfo
  {author} {\bibfnamefont {G.-H.}\ \bibnamefont {Han}}, \bibinfo {author}
  {\bibfnamefont {S.~H.}\ \bibnamefont {Lee}},\ and\ \bibinfo {author}
  {\bibfnamefont {K.-Y.}\ \bibnamefont {Lee}},\ }\bibfield  {title} {\bibinfo
  {title} {Ni-doped {MoS}$_2$ nanoparticles prepared via core--shell
  nanoclusters and catalytic activity for upgrading heavy oil},\ }\href@noop {}
  {\bibfield  {journal} {\bibinfo  {journal} {Energy Fuels}\ }\textbf {\bibinfo
  {volume} {32}},\ \bibinfo {pages} {9263} (\bibinfo {year}
  {2018})}\BibitemShut {NoStop}%
\bibitem [{\citenamefont {Hakala}\ \emph {et~al.}(2017)\citenamefont {Hakala},
  \citenamefont {Kronberg},\ and\ \citenamefont {Laasonen}}]{doped_apl2}%
  \BibitemOpen
  \bibfield  {author} {\bibinfo {author} {\bibfnamefont {M.}~\bibnamefont
  {Hakala}}, \bibinfo {author} {\bibfnamefont {R.}~\bibnamefont {Kronberg}},\
  and\ \bibinfo {author} {\bibfnamefont {K.}~\bibnamefont {Laasonen}},\
  }\bibfield  {title} {\bibinfo {title} {Hydrogen adsorption on doped {MoS}$_2$
  nanostructures},\ }\href@noop {} {\bibfield  {journal} {\bibinfo  {journal}
  {Sci. Rep.}\ }\textbf {\bibinfo {volume} {7}},\ \bibinfo {pages} {15243}
  (\bibinfo {year} {2017})}\BibitemShut {NoStop}%
\bibitem [{\citenamefont {Ma}\ \emph {et~al.}(2016{\natexlab{a}})\citenamefont
  {Ma}, \citenamefont {Li}, \citenamefont {An}, \citenamefont {Feng},
  \citenamefont {Chi}, \citenamefont {Liu}, \citenamefont {Zhang},\ and\
  \citenamefont {Sun}}]{solar_hydro_pro}%
  \BibitemOpen
  \bibfield  {author} {\bibinfo {author} {\bibfnamefont {X.}~\bibnamefont
  {Ma}}, \bibinfo {author} {\bibfnamefont {J.}~\bibnamefont {Li}}, \bibinfo
  {author} {\bibfnamefont {C.}~\bibnamefont {An}}, \bibinfo {author}
  {\bibfnamefont {J.}~\bibnamefont {Feng}}, \bibinfo {author} {\bibfnamefont
  {Y.}~\bibnamefont {Chi}}, \bibinfo {author} {\bibfnamefont {J.}~\bibnamefont
  {Liu}}, \bibinfo {author} {\bibfnamefont {J.}~\bibnamefont {Zhang}},\ and\
  \bibinfo {author} {\bibfnamefont {Y.}~\bibnamefont {Sun}},\ }\bibfield
  {title} {\bibinfo {title} {Ultrathin {Co(Ni)}-doped {MoS}$_2$ nanosheets as
  catalytic promoters enabling efficient solar hydrogen production},\
  }\href@noop {} {\bibfield  {journal} {\bibinfo  {journal} {Nano Res.}\
  }\textbf {\bibinfo {volume} {9}},\ \bibinfo {pages} {2284–2293} (\bibinfo
  {year} {2016}{\natexlab{a}})}\BibitemShut {NoStop}%
\bibitem [{\citenamefont {Jun-Feng}\ \emph {et~al.}(2012)\citenamefont
  {Jun-Feng}, \citenamefont {Braham},\ and\ \citenamefont
  {Qian-Feng}}]{appl_tribology}%
  \BibitemOpen
  \bibfield  {author} {\bibinfo {author} {\bibfnamefont {Y.}~\bibnamefont
  {Jun-Feng}}, \bibinfo {author} {\bibfnamefont {P.}~\bibnamefont {Braham}},\
  and\ \bibinfo {author} {\bibfnamefont {F.}~\bibnamefont {Qian-Feng}},\
  }\bibfield  {title} {\bibinfo {title} {Tribological properties of transition
  metal di-chalcogenide based lubricant coatings},\ }\href@noop {} {\bibfield
  {journal} {\bibinfo  {journal} {Front. Mater. Sci.}\ }\textbf {\bibinfo
  {volume} {6}},\ \bibinfo {pages} {116} (\bibinfo {year} {2012})}\BibitemShut
  {NoStop}%
\bibitem [{\citenamefont {Ma}\ \emph {et~al.}(2016{\natexlab{b}})\citenamefont
  {Ma}, \citenamefont {Ju}, \citenamefont {Li}, \citenamefont {Zhang},
  \citenamefont {He}, \citenamefont {Ma}, \citenamefont {Lu},\ and\
  \citenamefont {Yang}}]{CO_and_NO_ads}%
  \BibitemOpen
  \bibfield  {author} {\bibinfo {author} {\bibfnamefont {D.}~\bibnamefont
  {Ma}}, \bibinfo {author} {\bibfnamefont {W.}~\bibnamefont {Ju}}, \bibinfo
  {author} {\bibfnamefont {T.}~\bibnamefont {Li}}, \bibinfo {author}
  {\bibfnamefont {X.}~\bibnamefont {Zhang}}, \bibinfo {author} {\bibfnamefont
  {C.}~\bibnamefont {He}}, \bibinfo {author} {\bibfnamefont {B.}~\bibnamefont
  {Ma}}, \bibinfo {author} {\bibfnamefont {Z.}~\bibnamefont {Lu}},\ and\
  \bibinfo {author} {\bibfnamefont {Z.}~\bibnamefont {Yang}},\ }\bibfield
  {title} {\bibinfo {title} {The adsorption of {CO} and {NO} on the {MoS}$_2$
  monolayer doped with {Au}, {Pt}, {Pd}, or {Ni}: A first-principles study},\
  }\href@noop {} {\bibfield  {journal} {\bibinfo  {journal} {Appl. Surf. Sci.}\
  }\textbf {\bibinfo {volume} {383}},\ \bibinfo {pages} {98 } (\bibinfo {year}
  {2016}{\natexlab{b}})}\BibitemShut {NoStop}%
\bibitem [{\citenamefont {Tedstone}\ \emph {et~al.}(2016)\citenamefont
  {Tedstone}, \citenamefont {Lewis},\ and\ \citenamefont
  {O’Brien}}]{doped_apl4}%
  \BibitemOpen
  \bibfield  {author} {\bibinfo {author} {\bibfnamefont {A.~A.}\ \bibnamefont
  {Tedstone}}, \bibinfo {author} {\bibfnamefont {D.~J.}\ \bibnamefont
  {Lewis}},\ and\ \bibinfo {author} {\bibfnamefont {P.}~\bibnamefont
  {O’Brien}},\ }\bibfield  {title} {\bibinfo {title} {Synthesis, properties,
  and applications of transition metal-doped layered transition metal
  dichalcogenides},\ }\href@noop {} {\bibfield  {journal} {\bibinfo  {journal}
  {Chem. Mater.}\ }\textbf {\bibinfo {volume} {28}},\ \bibinfo {pages} {1965}
  (\bibinfo {year} {2016})}\BibitemShut {NoStop}%
\bibitem [{\citenamefont {Guerrero}\ \emph {et~al.}(2021)\citenamefont
  {Guerrero}, \citenamefont {Karkee},\ and\ \citenamefont {Strubbe}}]{enrique}%
  \BibitemOpen
  \bibfield  {author} {\bibinfo {author} {\bibfnamefont {E.}~\bibnamefont
  {Guerrero}}, \bibinfo {author} {\bibfnamefont {R.}~\bibnamefont {Karkee}},\
  and\ \bibinfo {author} {\bibfnamefont {D.~A.}\ \bibnamefont {Strubbe}},\
  }\bibfield  {title} {\bibinfo {title} {Phase stability and {Raman/IR}
  signatures of {Ni}--doped {MoS}$_2$ from {DFT} studies},\ }\href@noop {}
  {\bibfield  {journal} {\bibinfo  {journal} {J. Phys. Chem. C}\ }
  \textbf {\bibinfo {volume} {125}},\ \bibinfo {pages} {13401}
  (\bibinfo {year} {2021})}\BibitemShut {NoStop}%
\bibitem [{\citenamefont {Savage}(1948)}]{graphite}%
  \BibitemOpen
  \bibfield  {author} {\bibinfo {author} {\bibfnamefont {R.~H.}\ \bibnamefont
  {Savage}},\ }\bibfield  {title} {\bibinfo {title} {Graphite lubrication},\
  }\href@noop {} {\bibfield  {journal} {\bibinfo  {journal} {J. Appl. Phys.}\
  }\textbf {\bibinfo {volume} {19}},\ \bibinfo {pages} {1} (\bibinfo {year}
  {1948})}\BibitemShut {NoStop}%
\bibitem [{\citenamefont {Khare}\ and\ \citenamefont
  {Burris}(2013)}]{lubrication_in_mos2_in_ambient}%
  \BibitemOpen
  \bibfield  {author} {\bibinfo {author} {\bibfnamefont {H.~S.}\ \bibnamefont
  {Khare}}\ and\ \bibinfo {author} {\bibfnamefont {D.~L.}\ \bibnamefont
  {Burris}},\ }\bibfield  {title} {\bibinfo {title} {The effects of
  environmental water and oxygen on the temperature-dependent friction of
  sputtered molybdenum disulfide},\ }\href@noop {} {\bibfield  {journal}
  {\bibinfo  {journal} {Tribology Lett.}\ }\textbf {\bibinfo {volume} {52}},\
  \bibinfo {pages} {485–493} (\bibinfo {year} {2013})}\BibitemShut {NoStop}%
\bibitem [{\citenamefont {Lince}(2020)}]{lubricant_new}%
  \BibitemOpen
  \bibfield  {author} {\bibinfo {author} {\bibfnamefont {J.~R.}\ \bibnamefont
  {Lince}},\ }\bibfield  {title} {\bibinfo {title} {Effective application of
  solid lubricants in spacecraft mechanisms},\ }\href@noop {} {\bibfield
  {journal} {\bibinfo  {journal} {Lubricants}\ }\textbf {\bibinfo {volume}
  {8}},\ \bibinfo {pages} {74} (\bibinfo {year} {2020})}\BibitemShut {NoStop}%
\bibitem [{\citenamefont {Stupp}(1981)}]{mos2_experiment}%
  \BibitemOpen
  \bibfield  {author} {\bibinfo {author} {\bibfnamefont {B.}~\bibnamefont
  {Stupp}},\ }\bibfield  {title} {\bibinfo {title} {Synergistic effects of
  metals co-sputtered with {MoS}$_2$},\ }\href@noop {} {\bibfield  {journal}
  {\bibinfo  {journal} {Thin Solid Films}\ }\textbf {\bibinfo {volume} {84}},\
  \bibinfo {pages} {256} (\bibinfo {year} {1981})}\BibitemShut {NoStop}%
\bibitem [{\citenamefont {Pope}\ \emph {et~al.}(1990)\citenamefont {Pope},
  \citenamefont {Jervis},\ and\ \citenamefont {Nastasi}}]{mos2_experiment_2}%
  \BibitemOpen
  \bibfield  {author} {\bibinfo {author} {\bibfnamefont {L.~E.}\ \bibnamefont
  {Pope}}, \bibinfo {author} {\bibfnamefont {T.~R.}\ \bibnamefont {Jervis}},\
  and\ \bibinfo {author} {\bibfnamefont {M.}~\bibnamefont {Nastasi}},\
  }\bibfield  {title} {\bibinfo {title} {Effects of laser processing and doping
  on the lubrication and chemical properties of thin {MoS}$_2$ films},\
  }\href@noop {} {\bibfield  {journal} {\bibinfo  {journal} {Surf. Coat.
  Technol.}\ }\textbf {\bibinfo {volume} {42}},\ \bibinfo {pages} {217 }
  (\bibinfo {year} {1990})}\BibitemShut {NoStop}%
\bibitem [{\citenamefont {Wells}(1943)}]{Ni_abundance}%
  \BibitemOpen
  \bibfield  {author} {\bibinfo {author} {\bibfnamefont {R.~C.}\ \bibnamefont
  {Wells}},\ }\bibfield  {title} {\bibinfo {title} {Relative abundance of
  nickel in the {Earth}'s crust},\ }\href@noop {} {\bibfield  {journal}
  {\bibinfo  {journal} {U.S. Dept. of the Interior Professional Paper 205-A}\
  ,\ \bibinfo {pages} {1}} (\bibinfo {year} {1943})},\ \bibinfo {note}
  {https://pubs.usgs.gov/pp/0205a/report.pdf}\BibitemShut {NoStop}%
\bibitem [{\citenamefont {Schulz}\ \emph {et~al.}(2017)\citenamefont {Schulz},
  \citenamefont {Piatak},\ and\ \citenamefont {Papp}}]{Ta_abundance}%
  \BibitemOpen
  \bibfield  {author} {\bibinfo {author} {\bibfnamefont {K.~J.}\ \bibnamefont
  {Schulz}}, \bibinfo {author} {\bibfnamefont {N.~M.}\ \bibnamefont {Piatak}},\
  and\ \bibinfo {author} {\bibfnamefont {J.~F.}\ \bibnamefont {Papp}},\
  }\bibfield  {title} {\bibinfo {title} {Niobium and tantalum, chap. {M} of
  {Schulz}, {K.J.}, {DeYoung, J.H., Jr.}, {Seal, R.R., II}, and {Bradley,
  D.C.}, eds., {Critical} mineral resources of the {United States}—{Economic}
  and environmental geology and prospects for future supply},\ }\href@noop {}
  {\bibfield  {journal} {\bibinfo  {journal} {U.S. Geological Survey
  Professional Paper 1802}\ ,\ \bibinfo {pages} {M1–M34}} (\bibinfo {year}
  {2017})},\ \bibinfo {note} {https://doi.org/10.3133/pp1802M}\BibitemShut
  {NoStop}%
\bibitem [{\citenamefont {Zhao}\ \emph {et~al.}(2017)\citenamefont {Zhao},
  \citenamefont {Liu}, \citenamefont {Cheng}, \citenamefont {Li}, \citenamefont
  {Qi}, \citenamefont {Chen},\ and\ \citenamefont {Tang}}]{O2_ads}%
  \BibitemOpen
  \bibfield  {author} {\bibinfo {author} {\bibfnamefont {B.}~\bibnamefont
  {Zhao}}, \bibinfo {author} {\bibfnamefont {L.~L.}\ \bibnamefont {Liu}},
  \bibinfo {author} {\bibfnamefont {G.~D.}\ \bibnamefont {Cheng}}, \bibinfo
  {author} {\bibfnamefont {T.}~\bibnamefont {Li}}, \bibinfo {author}
  {\bibfnamefont {N.}~\bibnamefont {Qi}}, \bibinfo {author} {\bibfnamefont
  {Z.~Q.}\ \bibnamefont {Chen}},\ and\ \bibinfo {author} {\bibfnamefont
  {Z.}~\bibnamefont {Tang}},\ }\bibfield  {title} {\bibinfo {title}
  {Interaction of {O}$_2$ with monolayer {MoS}$_2$: Effect of doping and
  hydrogenation},\ }\href@noop {} {\bibfield  {journal} {\bibinfo  {journal}
  {Mater. Des.}\ }\textbf {\bibinfo {volume} {383}},\ \bibinfo {pages} {1 }
  (\bibinfo {year} {2017})}\BibitemShut {NoStop}%
\bibitem [{\citenamefont {Vazirisereshk}\ \emph {et~al.}(2019)\citenamefont
  {Vazirisereshk}, \citenamefont {Martini}, \citenamefont {Strubbe},\ and\
  \citenamefont {Baykara}}]{lubricant_by_David}%
  \BibitemOpen
  \bibfield  {author} {\bibinfo {author} {\bibfnamefont {M.~R.}\ \bibnamefont
  {Vazirisereshk}}, \bibinfo {author} {\bibfnamefont {A.}~\bibnamefont
  {Martini}}, \bibinfo {author} {\bibfnamefont {D.~A.}\ \bibnamefont
  {Strubbe}},\ and\ \bibinfo {author} {\bibfnamefont {M.~Z.}\ \bibnamefont
  {Baykara}},\ }\bibfield  {title} {\bibinfo {title} {Solid lubrication with
  {MoS}$_2$: A review},\ }\href@noop {} {\bibfield  {journal} {\bibinfo
  {journal} {Lubricants}\ }\textbf {\bibinfo {volume} {7}},\ \bibinfo {pages}
  {57} (\bibinfo {year} {2019})}\BibitemShut {NoStop}%
\bibitem [{\citenamefont {Renevier}\ \emph {et~al.}(2012)\citenamefont
  {Renevier}, \citenamefont {Fox}, \citenamefont {Teer},\ and\ \citenamefont
  {Hampshire}}]{fric_proposed_mechanism1}%
  \BibitemOpen
  \bibfield  {author} {\bibinfo {author} {\bibfnamefont {N.~M.}\ \bibnamefont
  {Renevier}}, \bibinfo {author} {\bibfnamefont {V.~C.}\ \bibnamefont {Fox}},
  \bibinfo {author} {\bibfnamefont {D.~G.}\ \bibnamefont {Teer}},\ and\
  \bibinfo {author} {\bibfnamefont {J.}~\bibnamefont {Hampshire}},\ }\bibfield
  {title} {\bibinfo {title} {Coating characteristics and tribological
  properties of sputter-deposited {MoS}$_2$/metal composite coatings deposited
  by closed field unbalanced magnetron sputter ion plating},\ }\href@noop {}
  {\bibfield  {journal} {\bibinfo  {journal} {Surf. Coat. Technol.}\ }\textbf
  {\bibinfo {volume} {127}},\ \bibinfo {pages} {24} (\bibinfo {year}
  {2012})}\BibitemShut {NoStop}%
\bibitem [{\citenamefont {Teer}(2001)}]{fric_proposed_mechanism2}%
  \BibitemOpen
  \bibfield  {author} {\bibinfo {author} {\bibfnamefont {D.~G.}\ \bibnamefont
  {Teer}},\ }\bibfield  {title} {\bibinfo {title} {New solid lubricant
  coatings},\ }\href@noop {} {\bibfield  {journal} {\bibinfo  {journal} {Wear}\
  }\textbf {\bibinfo {volume} {251}},\ \bibinfo {pages} {1068} (\bibinfo {year}
  {2001})}\BibitemShut {NoStop}%
\bibitem [{\citenamefont {Vellore}\ \emph {et~al.}(2020)\citenamefont
  {Vellore}, \citenamefont {Garcia}, \citenamefont {Johnson},\ and\
  \citenamefont {Martini}}]{low_wear_rate}%
  \BibitemOpen
  \bibfield  {author} {\bibinfo {author} {\bibfnamefont {A.}~\bibnamefont
  {Vellore}}, \bibinfo {author} {\bibfnamefont {S.~R.}\ \bibnamefont {Garcia}},
  \bibinfo {author} {\bibfnamefont {D.}~\bibnamefont {Johnson}},\ and\ \bibinfo
  {author} {\bibfnamefont {A.}~\bibnamefont {Martini}},\ }\bibfield  {title}
  {\bibinfo {title} {Ni-doped {MoS}$_2$ dry film lubricant life},\ }\href@noop
  {} {\bibfield  {journal} {\bibinfo  {journal} {Adv. Mater. Interfaces}\
  }\textbf {\bibinfo {volume} {7}},\ \bibinfo {pages} {2001109} (\bibinfo
  {year} {2020})}\BibitemShut {NoStop}%
\bibitem [{\citenamefont {Benavente}\ \emph {et~al.}(2002)\citenamefont
  {Benavente}, \citenamefont {Santa-Ana}, \citenamefont {Mendizábal},\ and\
  \citenamefont {González}}]{phase_change_Li}%
  \BibitemOpen
  \bibfield  {author} {\bibinfo {author} {\bibfnamefont {E.}~\bibnamefont
  {Benavente}}, \bibinfo {author} {\bibfnamefont {M.}~\bibnamefont
  {Santa-Ana}}, \bibinfo {author} {\bibfnamefont {F.}~\bibnamefont
  {Mendizábal}},\ and\ \bibinfo {author} {\bibfnamefont {G.}~\bibnamefont
  {González}},\ }\bibfield  {title} {\bibinfo {title} {Intercalation chemistry
  of molybdenum disulfide},\ }\href@noop {} {\bibfield  {journal} {\bibinfo
  {journal} {Coord. Chem. Rev.}\ }\textbf {\bibinfo {volume} {224}},\ \bibinfo
  {pages} {87 } (\bibinfo {year} {2002})}\BibitemShut {NoStop}%
\bibitem [{\citenamefont {Suh}\ \emph {et~al.}(2018)\citenamefont {Suh},
  \citenamefont {Tan}, \citenamefont {Zhao}, \citenamefont {Park},
  \citenamefont {Lin}, \citenamefont {Park}, \citenamefont {Kim}, \citenamefont
  {Jin}, \citenamefont {Saigal}, \citenamefont {Ghosh}, \citenamefont {Wong},
  \citenamefont {Chen}, \citenamefont {Wang}, \citenamefont {Walukiewicz},
  \citenamefont {Eda},\ and\ \citenamefont {Wu}}]{phase_change_Nb}%
  \BibitemOpen
  \bibfield  {author} {\bibinfo {author} {\bibfnamefont {J.}~\bibnamefont
  {Suh}}, \bibinfo {author} {\bibfnamefont {T.~L.}\ \bibnamefont {Tan}},
  \bibinfo {author} {\bibfnamefont {W.}~\bibnamefont {Zhao}}, \bibinfo {author}
  {\bibfnamefont {J.}~\bibnamefont {Park}}, \bibinfo {author} {\bibfnamefont
  {D.-Y.}\ \bibnamefont {Lin}}, \bibinfo {author} {\bibfnamefont {T.-E.}\
  \bibnamefont {Park}}, \bibinfo {author} {\bibfnamefont {J.}~\bibnamefont
  {Kim}}, \bibinfo {author} {\bibfnamefont {C.}~\bibnamefont {Jin}}, \bibinfo
  {author} {\bibfnamefont {N.}~\bibnamefont {Saigal}}, \bibinfo {author}
  {\bibfnamefont {S.}~\bibnamefont {Ghosh}}, \bibinfo {author} {\bibfnamefont
  {Z.~M.}\ \bibnamefont {Wong}}, \bibinfo {author} {\bibfnamefont
  {Y.}~\bibnamefont {Chen}}, \bibinfo {author} {\bibfnamefont {F.}~\bibnamefont
  {Wang}}, \bibinfo {author} {\bibfnamefont {W.}~\bibnamefont {Walukiewicz}},
  \bibinfo {author} {\bibfnamefont {G.}~\bibnamefont {Eda}},\ and\ \bibinfo
  {author} {\bibfnamefont {J.}~\bibnamefont {Wu}},\ }\bibfield  {title}
  {\bibinfo {title} {Reconfiguring crystal and electronic structures of
  {MoS}$_2$ by substitutional doping},\ }\href@noop {} {\bibfield  {journal}
  {\bibinfo  {journal} {Nat. Commun.}\ }\textbf {\bibinfo {volume} {9}},\
  \bibinfo {pages} {199} (\bibinfo {year} {2018})}\BibitemShut {NoStop}%
\bibitem [{\citenamefont {Giannozzi}\ \emph {et~al.}(2009)\citenamefont
  {Giannozzi}, \citenamefont {Baroni}, \citenamefont {Bonini}, \citenamefont
  {Calandra}, \citenamefont {Car}, \citenamefont {Cavazzoni}, \citenamefont
  {Ceresoli}, \citenamefont {Chiarotti}, \citenamefont {Cococcioni},
  \citenamefont {Dabo}, \citenamefont {{Dal Corso}}, \citenamefont {{de
  Gironcoli}}, \citenamefont {Fabris}, \citenamefont {Fratesi}, \citenamefont
  {Gebauer}, \citenamefont {Gerstmann}, \citenamefont {Gougoussis},
  \citenamefont {Kokalj}, \citenamefont {Lazzeri}, \citenamefont
  {Martin-Samos}, \citenamefont {Marzari}, \citenamefont {Mauri}, \citenamefont
  {Mazzarello}, \citenamefont {Paolini}, \citenamefont {Pasquarello},
  \citenamefont {Paulatto}, \citenamefont {Sbraccia}, \citenamefont {Scandolo},
  \citenamefont {Sclauzero}, \citenamefont {Seitsonen}, \citenamefont
  {Smogunov}, \citenamefont {Umari},\ and\ \citenamefont
  {Wentzcovitch}}]{QE-2009}%
  \BibitemOpen
  \bibfield  {author} {\bibinfo {author} {\bibfnamefont {P.}~\bibnamefont
  {Giannozzi}}, \bibinfo {author} {\bibfnamefont {S.}~\bibnamefont {Baroni}},
  \bibinfo {author} {\bibfnamefont {N.}~\bibnamefont {Bonini}}, \bibinfo
  {author} {\bibfnamefont {M.}~\bibnamefont {Calandra}}, \bibinfo {author}
  {\bibfnamefont {R.}~\bibnamefont {Car}}, \bibinfo {author} {\bibfnamefont
  {C.}~\bibnamefont {Cavazzoni}}, \bibinfo {author} {\bibfnamefont
  {D.}~\bibnamefont {Ceresoli}}, \bibinfo {author} {\bibfnamefont {G.~L.}\
  \bibnamefont {Chiarotti}}, \bibinfo {author} {\bibfnamefont {M.}~\bibnamefont
  {Cococcioni}}, \bibinfo {author} {\bibfnamefont {I.}~\bibnamefont {Dabo}},
  \bibinfo {author} {\bibfnamefont {A.}~\bibnamefont {{Dal Corso}}}, \bibinfo
  {author} {\bibfnamefont {S.}~\bibnamefont {{de Gironcoli}}}, \bibinfo
  {author} {\bibfnamefont {S.}~\bibnamefont {Fabris}}, \bibinfo {author}
  {\bibfnamefont {G.}~\bibnamefont {Fratesi}}, \bibinfo {author} {\bibfnamefont
  {R.}~\bibnamefont {Gebauer}}, \bibinfo {author} {\bibfnamefont
  {U.}~\bibnamefont {Gerstmann}}, \bibinfo {author} {\bibfnamefont
  {C.}~\bibnamefont {Gougoussis}}, \bibinfo {author} {\bibfnamefont
  {A.}~\bibnamefont {Kokalj}}, \bibinfo {author} {\bibfnamefont
  {M.}~\bibnamefont {Lazzeri}}, \bibinfo {author} {\bibfnamefont
  {L.}~\bibnamefont {Martin-Samos}}, \bibinfo {author} {\bibfnamefont
  {N.}~\bibnamefont {Marzari}}, \bibinfo {author} {\bibfnamefont
  {F.}~\bibnamefont {Mauri}}, \bibinfo {author} {\bibfnamefont
  {R.}~\bibnamefont {Mazzarello}}, \bibinfo {author} {\bibfnamefont
  {S.}~\bibnamefont {Paolini}}, \bibinfo {author} {\bibfnamefont
  {A.}~\bibnamefont {Pasquarello}}, \bibinfo {author} {\bibfnamefont
  {L.}~\bibnamefont {Paulatto}}, \bibinfo {author} {\bibfnamefont
  {C.}~\bibnamefont {Sbraccia}}, \bibinfo {author} {\bibfnamefont
  {S.}~\bibnamefont {Scandolo}}, \bibinfo {author} {\bibfnamefont
  {G.}~\bibnamefont {Sclauzero}}, \bibinfo {author} {\bibfnamefont {A.~P.}\
  \bibnamefont {Seitsonen}}, \bibinfo {author} {\bibfnamefont {A.}~\bibnamefont
  {Smogunov}}, \bibinfo {author} {\bibfnamefont {P.}~\bibnamefont {Umari}},\
  and\ \bibinfo {author} {\bibfnamefont {R.~M.}\ \bibnamefont {Wentzcovitch}},\
  }\bibfield  {title} {\bibinfo {title} {{QUANTUM ESPRESSO}: a modular and
  open-source software project for quantum simulations of materials},\
  }\href@noop {} {\bibfield  {journal} {\bibinfo  {journal} {J. Phys.: Condens.
  Matter}\ }\textbf {\bibinfo {volume} {21}} (\bibinfo {year}
  {2009})}\BibitemShut {NoStop}%
\bibitem [{\citenamefont {Giannozzi}\ \emph {et~al.}(2017)\citenamefont
  {Giannozzi}, \citenamefont {Andreussi}, \citenamefont {Brumme}, \citenamefont
  {Bunau}, \citenamefont {{Buongiorno Nardelli}}, \citenamefont {Calandra},
  \citenamefont {Car}, \citenamefont {Cavazzoni}, \citenamefont {Ceresoli},
  \citenamefont {Cococcioni}, \citenamefont {Colonna}, \citenamefont
  {Carnimeo}, \citenamefont {{Dal Corso}}, \citenamefont {{de Gironcoli}},
  \citenamefont {Delugas}, \citenamefont {DiStasio}, \citenamefont {Ferretti},
  \citenamefont {Floris}, \citenamefont {Fratesi}, \citenamefont {Fugallo},
  \citenamefont {Gebauer}, \citenamefont {Gerstmann}, \citenamefont {Giustino},
  \citenamefont {Gorni}, \citenamefont {Jia}, \citenamefont {Kawamura},
  \citenamefont {Ko}, \citenamefont {Kokalj}, \citenamefont
  {Kü{\c{c}}ükbenli}, \citenamefont {Lazzeri}, \citenamefont {Marsili},
  \citenamefont {Marzari}, \citenamefont {Mauri}, \citenamefont {Nguyen},
  \citenamefont {Nguyen}, \citenamefont {{Otero-de-la-Roza}}, \citenamefont
  {Paulatto}, \citenamefont {Ponc{\'{e}}}, \citenamefont {Rocca}, \citenamefont
  {Sabatini}, \citenamefont {Santra}, \citenamefont {Schlipf}, \citenamefont
  {Seitsonen}, \citenamefont {Smogunov}, \citenamefont {Timrov}, \citenamefont
  {Thonhauser}, \citenamefont {Umari}, \citenamefont {Vast}, \citenamefont
  {Wu},\ and\ \citenamefont {Baroni}}]{QE_2}%
  \BibitemOpen
  \bibfield  {author} {\bibinfo {author} {\bibfnamefont {P.}~\bibnamefont
  {Giannozzi}}, \bibinfo {author} {\bibfnamefont {O.}~\bibnamefont
  {Andreussi}}, \bibinfo {author} {\bibfnamefont {T.}~\bibnamefont {Brumme}},
  \bibinfo {author} {\bibfnamefont {O.}~\bibnamefont {Bunau}}, \bibinfo
  {author} {\bibfnamefont {M.}~\bibnamefont {{Buongiorno Nardelli}}}, \bibinfo
  {author} {\bibfnamefont {M.}~\bibnamefont {Calandra}}, \bibinfo {author}
  {\bibfnamefont {R.}~\bibnamefont {Car}}, \bibinfo {author} {\bibfnamefont
  {C.}~\bibnamefont {Cavazzoni}}, \bibinfo {author} {\bibfnamefont
  {D.}~\bibnamefont {Ceresoli}}, \bibinfo {author} {\bibfnamefont
  {M.}~\bibnamefont {Cococcioni}}, \bibinfo {author} {\bibfnamefont
  {N.}~\bibnamefont {Colonna}}, \bibinfo {author} {\bibfnamefont
  {I.}~\bibnamefont {Carnimeo}}, \bibinfo {author} {\bibfnamefont
  {A.}~\bibnamefont {{Dal Corso}}}, \bibinfo {author} {\bibfnamefont
  {S.}~\bibnamefont {{de Gironcoli}}}, \bibinfo {author} {\bibfnamefont
  {P.}~\bibnamefont {Delugas}}, \bibinfo {author} {\bibfnamefont {R.~A.}\
  \bibnamefont {DiStasio}}, \bibinfo {author} {\bibfnamefont {A.}~\bibnamefont
  {Ferretti}}, \bibinfo {author} {\bibfnamefont {A.}~\bibnamefont {Floris}},
  \bibinfo {author} {\bibfnamefont {G.}~\bibnamefont {Fratesi}}, \bibinfo
  {author} {\bibfnamefont {G.}~\bibnamefont {Fugallo}}, \bibinfo {author}
  {\bibfnamefont {R.}~\bibnamefont {Gebauer}}, \bibinfo {author} {\bibfnamefont
  {U.}~\bibnamefont {Gerstmann}}, \bibinfo {author} {\bibfnamefont
  {F.}~\bibnamefont {Giustino}}, \bibinfo {author} {\bibfnamefont
  {T.}~\bibnamefont {Gorni}}, \bibinfo {author} {\bibfnamefont
  {J.}~\bibnamefont {Jia}}, \bibinfo {author} {\bibfnamefont {M.}~\bibnamefont
  {Kawamura}}, \bibinfo {author} {\bibfnamefont {H.-Y.}\ \bibnamefont {Ko}},
  \bibinfo {author} {\bibfnamefont {A.}~\bibnamefont {Kokalj}}, \bibinfo
  {author} {\bibfnamefont {E.}~\bibnamefont {Kü{\c{c}}ükbenli}}, \bibinfo
  {author} {\bibfnamefont {M.}~\bibnamefont {Lazzeri}}, \bibinfo {author}
  {\bibfnamefont {M.}~\bibnamefont {Marsili}}, \bibinfo {author} {\bibfnamefont
  {N.}~\bibnamefont {Marzari}}, \bibinfo {author} {\bibfnamefont
  {F.}~\bibnamefont {Mauri}}, \bibinfo {author} {\bibfnamefont {N.~L.}\
  \bibnamefont {Nguyen}}, \bibinfo {author} {\bibfnamefont {H.-V.}\
  \bibnamefont {Nguyen}}, \bibinfo {author} {\bibfnamefont {A.}~\bibnamefont
  {{Otero-de-la-Roza}}}, \bibinfo {author} {\bibfnamefont {L.}~\bibnamefont
  {Paulatto}}, \bibinfo {author} {\bibfnamefont {S.}~\bibnamefont
  {Ponc{\'{e}}}}, \bibinfo {author} {\bibfnamefont {D.}~\bibnamefont {Rocca}},
  \bibinfo {author} {\bibfnamefont {R.}~\bibnamefont {Sabatini}}, \bibinfo
  {author} {\bibfnamefont {B.}~\bibnamefont {Santra}}, \bibinfo {author}
  {\bibfnamefont {M.}~\bibnamefont {Schlipf}}, \bibinfo {author} {\bibfnamefont
  {A.~P.}\ \bibnamefont {Seitsonen}}, \bibinfo {author} {\bibfnamefont
  {A.}~\bibnamefont {Smogunov}}, \bibinfo {author} {\bibfnamefont
  {I.}~\bibnamefont {Timrov}}, \bibinfo {author} {\bibfnamefont
  {T.}~\bibnamefont {Thonhauser}}, \bibinfo {author} {\bibfnamefont
  {P.}~\bibnamefont {Umari}}, \bibinfo {author} {\bibfnamefont
  {N.}~\bibnamefont {Vast}}, \bibinfo {author} {\bibfnamefont {X.}~\bibnamefont
  {Wu}},\ and\ \bibinfo {author} {\bibfnamefont {S.}~\bibnamefont {Baroni}},\
  }\bibfield  {title} {\bibinfo {title} {Advanced capabilities for materials
  modelling with {Quantum ESPRESSO}},\ }\href@noop {} {\bibfield  {journal}
  {\bibinfo  {journal} {J. Phys.: Condens. Matter}\ }\textbf {\bibinfo {volume}
  {29}},\ \bibinfo {pages} {465901} (\bibinfo {year} {2017})}\BibitemShut
  {NoStop}%
\bibitem [{\citenamefont {Perdew}\ \emph {et~al.}(1996)\citenamefont {Perdew},
  \citenamefont {Burke},\ and\ \citenamefont {Ernzerhof}}]{original_pbe}%
  \BibitemOpen
  \bibfield  {author} {\bibinfo {author} {\bibfnamefont {J.~P.}\ \bibnamefont
  {Perdew}}, \bibinfo {author} {\bibfnamefont {K.}~\bibnamefont {Burke}},\ and\
  \bibinfo {author} {\bibfnamefont {M.}~\bibnamefont {Ernzerhof}},\ }\bibfield
  {title} {\bibinfo {title} {Generalized gradient approximation made simple},\
  }\href@noop {} {\bibfield  {journal} {\bibinfo  {journal} {Phys. Rev. Lett.}\
  }\textbf {\bibinfo {volume} {77}},\ \bibinfo {pages} {3865} (\bibinfo {year}
  {1996})}\BibitemShut {NoStop}%
\bibitem [{\citenamefont {Perdew}\ and\ \citenamefont
  {Wang}(1992)}]{lda_perdew}%
  \BibitemOpen
  \bibfield  {author} {\bibinfo {author} {\bibfnamefont {J.~P.}\ \bibnamefont
  {Perdew}}\ and\ \bibinfo {author} {\bibfnamefont {Y.}~\bibnamefont {Wang}},\
  }\bibfield  {title} {\bibinfo {title} {Accurate and simple analytic
  representation of the electron--gas correlation energy},\ }\href@noop {}
  {\bibfield  {journal} {\bibinfo  {journal} {J. Phys. Chem. C}\ }\textbf
  {\bibinfo {volume} {45}},\ \bibinfo {pages} {13244} (\bibinfo {year}
  {1992})}\BibitemShut {NoStop}%
\bibitem [{\citenamefont {Grimme}(2006)}]{Grimme}%
  \BibitemOpen
  \bibfield  {author} {\bibinfo {author} {\bibfnamefont {S.}~\bibnamefont
  {Grimme}},\ }\bibfield  {title} {\bibinfo {title} {Semiempirical {GGA}-type
  density functional constructed with a long-range dispersion correction},\
  }\href@noop {} {\bibfield  {journal} {\bibinfo  {journal} {J. Comput. Chem.}\
  }\textbf {\bibinfo {volume} {27}},\ \bibinfo {pages} {1787–1799} (\bibinfo
  {year} {2006})}\BibitemShut {NoStop}%
\bibitem [{\citenamefont {Peelaers}\ and\ \citenamefont {{Van de
  Walle}}(2014)}]{Mos2_exp_fit}%
  \BibitemOpen
  \bibfield  {author} {\bibinfo {author} {\bibfnamefont {H.}~\bibnamefont
  {Peelaers}}\ and\ \bibinfo {author} {\bibfnamefont {C.~G.}\ \bibnamefont
  {{Van de Walle}}},\ }\bibfield  {title} {\bibinfo {title} {Elastic constants
  and pressure-induced effects in {MoS}$_2$},\ }\href@noop {} {\bibfield
  {journal} {\bibinfo  {journal} {J. Phys. Chem. C}\ }\textbf {\bibinfo
  {volume} {118}},\ \bibinfo {pages} {12073} (\bibinfo {year}
  {2014})}\BibitemShut {NoStop}%
\bibitem [{\citenamefont {Andersson}(2013)}]{GD2_metal}%
  \BibitemOpen
  \bibfield  {author} {\bibinfo {author} {\bibfnamefont {M.~P.}\ \bibnamefont
  {Andersson}},\ }\bibfield  {title} {\bibinfo {title} {Density functional
  theory with modified dispersion correction for metals applied to
  self-assembled monolayers of thiols on {Au}(111)},\ }\href@noop {} {\bibfield
   {journal} {\bibinfo  {journal} {J. Theor. Chem.}\ }\textbf {\bibinfo
  {volume} {2013}},\ \bibinfo {pages} {2356} (\bibinfo {year}
  {2013})}\BibitemShut {NoStop}%
\bibitem [{\citenamefont {Hamann}(2013)}]{ONCV}%
  \BibitemOpen
  \bibfield  {author} {\bibinfo {author} {\bibfnamefont {D.~R.}\ \bibnamefont
  {Hamann}},\ }\bibfield  {title} {\bibinfo {title} {Optimized norm-conserving
  {Vanderbilt} pseudopotentials},\ }\href@noop {} {\bibfield  {journal}
  {\bibinfo  {journal} {Phys. Rev. B}\ }\textbf {\bibinfo {volume} {88}},\
  \bibinfo {pages} {085117} (\bibinfo {year} {2013})}\BibitemShut {NoStop}%
\bibitem [{\citenamefont {Schlipf}\ and\ \citenamefont {Gygi}(2015)}]{Gygi}%
  \BibitemOpen
  \bibfield  {author} {\bibinfo {author} {\bibfnamefont {M.}~\bibnamefont
  {Schlipf}}\ and\ \bibinfo {author} {\bibfnamefont {F.}~\bibnamefont {Gygi}},\
  }\bibfield  {title} {\bibinfo {title} {Optimization algorithm for the
  generation of {ONCV} pseudopotentials},\ }\href@noop {} {\bibfield  {journal}
  {\bibinfo  {journal} {Comput. Phys. Commun.}\ }\textbf {\bibinfo {volume}
  {196}},\ \bibinfo {pages} {36} (\bibinfo {year} {2015})}\BibitemShut
  {NoStop}%
\bibitem [{web(ials)}]{web_sg15}%
  \BibitemOpen
  \href@noop {} {\bibfield  {journal} {\bibinfo  {journal} {web}\ } (\bibinfo
  {year} {http://www.quantum-simulation.org/potentials})}\BibitemShut {NoStop}%
\bibitem [{web(oorg)}]{web_pseudojo}%
  \BibitemOpen
  \href@noop {} {\bibfield  {journal} {\bibinfo  {journal} {web}\ } (\bibinfo
  {year} {http://www.pseudo-dojo.org})}\BibitemShut {NoStop}%
\bibitem [{\citenamefont {Komsa}\ and\ \citenamefont
  {Krasheninnikov}(2015)}]{lattice}%
  \BibitemOpen
  \bibfield  {author} {\bibinfo {author} {\bibfnamefont {H.-P.}\ \bibnamefont
  {Komsa}}\ and\ \bibinfo {author} {\bibfnamefont {A.~V.}\ \bibnamefont
  {Krasheninnikov}},\ }\bibfield  {title} {\bibinfo {title} {Native defects in
  bulk and monolayer {MoS}$_2$ from first principles},\ }\href
  {https://doi.org/10.1103/physrevb.91.125304} {\bibfield  {journal} {\bibinfo
  {journal} {Phys. Rev. B}\ }\textbf {\bibinfo {volume} {91}},\ \bibinfo
  {pages} {125304} (\bibinfo {year} {2015})}\BibitemShut {NoStop}%
\bibitem [{\citenamefont {Bengtsson}(1999)}]{dipole_corr}%
  \BibitemOpen
  \bibfield  {author} {\bibinfo {author} {\bibfnamefont {L.}~\bibnamefont
  {Bengtsson}},\ }\bibfield  {title} {\bibinfo {title} {Dipole correction for
  surface supercell calculations},\ }\href@noop {} {\bibfield  {journal}
  {\bibinfo  {journal} {Phys. Rev. B}\ }\textbf {\bibinfo {volume} {59}},\
  \bibinfo {pages} {12301} (\bibinfo {year} {1999})}\BibitemShut {NoStop}%
\bibitem [{\citenamefont {McKeehan}(1923)}]{lattice_exp}%
  \BibitemOpen
  \bibfield  {author} {\bibinfo {author} {\bibfnamefont {L.~W.}\ \bibnamefont
  {McKeehan}},\ }\bibfield  {title} {\bibinfo {title} {The crystal structure of
  iron-nickel alloys},\ }\href@noop {} {\bibfield  {journal} {\bibinfo
  {journal} {Phys. Rev.}\ }\textbf {\bibinfo {volume} {21}},\ \bibinfo {pages}
  {402} (\bibinfo {year} {1923})}\BibitemShut {NoStop}%
\bibitem [{web(7869)}]{web_S}%
  \BibitemOpen
  \href@noop {} {\bibfield  {journal} {\bibinfo  {journal} {web}\ } (\bibinfo
  {year} {https://materialsproject.org/materials/mp-557869/})}\BibitemShut
  {NoStop}%
\bibitem [{\citenamefont {Bo}\ \emph {et~al.}(2013)\citenamefont {Bo},
  \citenamefont {Lian},\ and\ \citenamefont {Jun}}]{2H&3R_agreement}%
  \BibitemOpen
  \bibfield  {author} {\bibinfo {author} {\bibfnamefont {C.~X.}\ \bibnamefont
  {Bo}}, \bibinfo {author} {\bibfnamefont {C.~Z.}\ \bibnamefont {Lian}},\ and\
  \bibinfo {author} {\bibfnamefont {L.}~\bibnamefont {Jun}},\ }\bibfield
  {title} {\bibinfo {title} {Critical electronic structures controlling phase
  transitions induced by lithium ion intercalation in molybdenum disulphide},\
  }\href@noop {} {\bibfield  {journal} {\bibinfo  {journal} {Chin. Sci. Bull.}\
  }\textbf {\bibinfo {volume} {58}},\ \bibinfo {pages} {1632} (\bibinfo {year}
  {2013})}\BibitemShut {NoStop}%
\bibitem [{SM()}]{SM}%
  \BibitemOpen
  \href@noop {} {}\bibinfo {note} {See Supplemental Material at [URL will be
  inserted by publisher] for doping formation energy tables, density of states
  for Ni-doped and pristine polytypes, and relaxed atomic coordinates for
  Ni-doped MoS$_2$ polytypes, pristine polytypes, and 3R primitive unit
  cell.}\BibitemShut {Stop}%
\bibitem [{\citenamefont {Ivanovskaya}\ \emph {et~al.}(2008)\citenamefont
  {Ivanovskaya}, \citenamefont {Zobelli}, \citenamefont {Gloter}, \citenamefont
  {Brun}, \citenamefont {Serin},\ and\ \citenamefont
  {Colliex}}]{E_form_bulk_ref1}%
  \BibitemOpen
  \bibfield  {author} {\bibinfo {author} {\bibfnamefont {V.~V.}\ \bibnamefont
  {Ivanovskaya}}, \bibinfo {author} {\bibfnamefont {A.}~\bibnamefont
  {Zobelli}}, \bibinfo {author} {\bibfnamefont {A.}~\bibnamefont {Gloter}},
  \bibinfo {author} {\bibfnamefont {N.}~\bibnamefont {Brun}}, \bibinfo {author}
  {\bibfnamefont {V.}~\bibnamefont {Serin}},\ and\ \bibinfo {author}
  {\bibfnamefont {C.}~\bibnamefont {Colliex}},\ }\bibfield  {title} {\bibinfo
  {title} {Ab initio study of bilateral doping within the {MoS}$_2$-{NbS}$_2$
  system},\ }\href@noop {} {\bibfield  {journal} {\bibinfo  {journal} {Phys.
  Rev. B}\ }\textbf {\bibinfo {volume} {78}},\ \bibinfo {pages} {134104}
  (\bibinfo {year} {2008})}\BibitemShut {NoStop}%
\bibitem [{\citenamefont {Dolui}\ \emph {et~al.}(2013)\citenamefont {Dolui},
  \citenamefont {Rungger}, \citenamefont {{Das Pemmaraju}},\ and\ \citenamefont
  {Sanvito}}]{E_form_bulk_ref2}%
  \BibitemOpen
  \bibfield  {author} {\bibinfo {author} {\bibfnamefont {K.}~\bibnamefont
  {Dolui}}, \bibinfo {author} {\bibfnamefont {I.}~\bibnamefont {Rungger}},
  \bibinfo {author} {\bibfnamefont {C.}~\bibnamefont {{Das Pemmaraju}}},\ and\
  \bibinfo {author} {\bibfnamefont {S.}~\bibnamefont {Sanvito}},\ }\bibfield
  {title} {\bibinfo {title} {Possible doping strategies for {MoS}$_2$
  monolayers: An ab initio study},\ }\href@noop {} {\bibfield  {journal}
  {\bibinfo  {journal} {Phys. Rev. B}\ }\textbf {\bibinfo {volume} {88}},\
  \bibinfo {pages} {075420} (\bibinfo {year} {2013})}\BibitemShut {NoStop}%
\bibitem [{\citenamefont {Zhao}\ \emph {et~al.}(2016)\citenamefont {Zhao},
  \citenamefont {Jin}, \citenamefont {Wu},\ and\ \citenamefont
  {Ji}}]{Mo-atop1}%
  \BibitemOpen
  \bibfield  {author} {\bibinfo {author} {\bibfnamefont {C.}~\bibnamefont
  {Zhao}}, \bibinfo {author} {\bibfnamefont {C.}~\bibnamefont {Jin}}, \bibinfo
  {author} {\bibfnamefont {J.}~\bibnamefont {Wu}},\ and\ \bibinfo {author}
  {\bibfnamefont {W.}~\bibnamefont {Ji}},\ }\bibfield  {title} {\bibinfo
  {title} {Magnetism in molybdenum disulphide monolayer with sulfur substituted
  by 3d transition metals},\ }\href@noop {} {\bibfield  {journal} {\bibinfo
  {journal} {J. Appl. Phys.}\ }\textbf {\bibinfo {volume} {120}},\ \bibinfo
  {pages} {144305} (\bibinfo {year} {2016})}\BibitemShut {NoStop}%
\bibitem [{\citenamefont {Makov}\ and\ \citenamefont
  {Payne}(1995)}]{extrapolation}%
  \BibitemOpen
  \bibfield  {author} {\bibinfo {author} {\bibfnamefont {G.}~\bibnamefont
  {Makov}}\ and\ \bibinfo {author} {\bibfnamefont {M.~C.}\ \bibnamefont
  {Payne}},\ }\bibfield  {title} {\bibinfo {title} {Periodic boundary
  conditions in ab initio calculations},\ }\href@noop {} {\bibfield  {journal}
  {\bibinfo  {journal} {Phys. Rev. B}\ }\textbf {\bibinfo {volume} {51}},\
  \bibinfo {pages} {4014} (\bibinfo {year} {1995})}\BibitemShut {NoStop}%
\bibitem [{\citenamefont {Mosconi}\ \emph {et~al.}(2019)\citenamefont
  {Mosconi}, \citenamefont {Till}, \citenamefont {Calvillo}, \citenamefont
  {Kosmala}, \citenamefont {Garoli}, \citenamefont {Debellis}, \citenamefont
  {Martucci}, \citenamefont {Agnoli},\ and\ \citenamefont
  {Granozzi}}]{Ni_2H_3R}%
  \BibitemOpen
  \bibfield  {author} {\bibinfo {author} {\bibfnamefont {D.}~\bibnamefont
  {Mosconi}}, \bibinfo {author} {\bibfnamefont {P.}~\bibnamefont {Till}},
  \bibinfo {author} {\bibfnamefont {L.}~\bibnamefont {Calvillo}}, \bibinfo
  {author} {\bibfnamefont {T.}~\bibnamefont {Kosmala}}, \bibinfo {author}
  {\bibfnamefont {D.}~\bibnamefont {Garoli}}, \bibinfo {author} {\bibfnamefont
  {D.}~\bibnamefont {Debellis}}, \bibinfo {author} {\bibfnamefont
  {A.}~\bibnamefont {Martucci}}, \bibinfo {author} {\bibfnamefont
  {S.}~\bibnamefont {Agnoli}},\ and\ \bibinfo {author} {\bibfnamefont
  {G.}~\bibnamefont {Granozzi}},\ }\bibfield  {title} {\bibinfo {title} {Effect
  of {Ni} doping on the {MoS}$_2$ structure and its hydrogen evolution activity
  in acid and alkaline electrolytes},\ }\href@noop {} {\bibfield  {journal}
  {\bibinfo  {journal} {Surfaces}\ }\textbf {\bibinfo {volume} {2}},\ \bibinfo
  {pages} {531–545} (\bibinfo {year} {2019})}\BibitemShut {NoStop}%
\bibitem [{\citenamefont {Cordero}\ \emph {et~al.}(2008)\citenamefont
  {Cordero}, \citenamefont {Gómez}, \citenamefont {Platero-Prats},
  \citenamefont {Revés}, \citenamefont {Echeverría}, \citenamefont
  {Cremades}, \citenamefont {Barragán},\ and\ \citenamefont
  {Alvarez}}]{covalent_radii}%
  \BibitemOpen
  \bibfield  {author} {\bibinfo {author} {\bibfnamefont {B.}~\bibnamefont
  {Cordero}}, \bibinfo {author} {\bibfnamefont {V.}~\bibnamefont {Gómez}},
  \bibinfo {author} {\bibfnamefont {A.~E.}\ \bibnamefont {Platero-Prats}},
  \bibinfo {author} {\bibfnamefont {M.}~\bibnamefont {Revés}}, \bibinfo
  {author} {\bibfnamefont {J.}~\bibnamefont {Echeverría}}, \bibinfo {author}
  {\bibfnamefont {E.}~\bibnamefont {Cremades}}, \bibinfo {author}
  {\bibfnamefont {F.}~\bibnamefont {Barragán}},\ and\ \bibinfo {author}
  {\bibfnamefont {S.}~\bibnamefont {Alvarez}},\ }\bibfield  {title} {\bibinfo
  {title} {Covalent radii revisited},\ }\href@noop {} {\bibfield  {journal}
  {\bibinfo  {journal} {Dalton Trans.}\ }\textbf {\bibinfo {volume} {21}},\
  \bibinfo {pages} {2832} (\bibinfo {year} {2008})}\BibitemShut {NoStop}%
\bibitem [{\citenamefont {Aquilanti}\ \emph {et~al.}(2000)\citenamefont
  {Aquilanti}, \citenamefont {Ascenzi}, \citenamefont {Braca}, \citenamefont
  {Cappelletti},\ and\ \citenamefont {Pirani}}]{Van_radii_S}%
  \BibitemOpen
  \bibfield  {author} {\bibinfo {author} {\bibfnamefont {V.}~\bibnamefont
  {Aquilanti}}, \bibinfo {author} {\bibfnamefont {D.}~\bibnamefont {Ascenzi}},
  \bibinfo {author} {\bibfnamefont {E.}~\bibnamefont {Braca}}, \bibinfo
  {author} {\bibfnamefont {D.}~\bibnamefont {Cappelletti}},\ and\ \bibinfo
  {author} {\bibfnamefont {F.}~\bibnamefont {Pirani}},\ }\bibfield  {title}
  {\bibinfo {title} {Production, characterization and scattering of a sulfur
  atom beam : Interatomic potentials for the rare-gas sulfides, {RS} ({R} =
  {Ne}, {Ar}, {Kr}, {Xe})},\ }\href@noop {} {\bibfield  {journal} {\bibinfo
  {journal} {Phys. Chem. Chem. Phys.}\ }\textbf {\bibinfo {volume} {2}},\
  \bibinfo {pages} {4081} (\bibinfo {year} {2000})}\BibitemShut {NoStop}%
\bibitem [{\citenamefont {Batsanov}(2000)}]{Van_radii_MoNi}%
  \BibitemOpen
  \bibfield  {author} {\bibinfo {author} {\bibfnamefont {S.}~\bibnamefont
  {Batsanov}},\ }\bibfield  {title} {\bibinfo {title} {Intramolecular contact
  radii close to the van der {Waals} radii},\ }\href@noop {} {\bibfield
  {journal} {\bibinfo  {journal} {Zh. Neorg. Khim.}\ }\textbf {\bibinfo
  {volume} {45}},\ \bibinfo {pages} {992–996} (\bibinfo {year}
  {2000})}\BibitemShut {NoStop}%
\bibitem [{\citenamefont {Deng}\ \emph {et~al.}(2015)\citenamefont {Deng},
  \citenamefont {Li}, \citenamefont {Xiao}, \citenamefont {Tu}, \citenamefont
  {Deng}, \citenamefont {Yang}, \citenamefont {Tian}, \citenamefont {Li},
  \citenamefont {Rena},\ and\ \citenamefont {Bao}}]{four_coodination}%
  \BibitemOpen
  \bibfield  {author} {\bibinfo {author} {\bibfnamefont {J.}~\bibnamefont
  {Deng}}, \bibinfo {author} {\bibfnamefont {H.}~\bibnamefont {Li}}, \bibinfo
  {author} {\bibfnamefont {J.}~\bibnamefont {Xiao}}, \bibinfo {author}
  {\bibfnamefont {Y.}~\bibnamefont {Tu}}, \bibinfo {author} {\bibfnamefont
  {D.}~\bibnamefont {Deng}}, \bibinfo {author} {\bibfnamefont {H.}~\bibnamefont
  {Yang}}, \bibinfo {author} {\bibfnamefont {H.}~\bibnamefont {Tian}}, \bibinfo
  {author} {\bibfnamefont {J.}~\bibnamefont {Li}}, \bibinfo {author}
  {\bibfnamefont {P.}~\bibnamefont {Rena}},\ and\ \bibinfo {author}
  {\bibfnamefont {X.}~\bibnamefont {Bao}},\ }\bibfield  {title} {\bibinfo
  {title} {Triggering the electrocatalytic hydrogen evolution activity of the
  inert two-dimensional {MoS}$_2$ surface via single-atom metal doping},\
  }\href@noop {} {\bibfield  {journal} {\bibinfo  {journal} {Energy Environ.
  Sci.}\ }\textbf {\bibinfo {volume} {8}},\ \bibinfo {pages} {1594} (\bibinfo
  {year} {2015})}\BibitemShut {NoStop}%
\bibitem [{\citenamefont {Xie}\ \emph {et~al.}(2013)\citenamefont {Xie},
  \citenamefont {Zhang}, \citenamefont {Li}, \citenamefont {Grote},
  \citenamefont {Zhang}, \citenamefont {Zhang}, \citenamefont {Wang},
  \citenamefont {Lei}, \citenamefont {Pan},\ and\ \citenamefont
  {Xie}}]{Intercalation_inc_layers}%
  \BibitemOpen
  \bibfield  {author} {\bibinfo {author} {\bibfnamefont {J.}~\bibnamefont
  {Xie}}, \bibinfo {author} {\bibfnamefont {J.}~\bibnamefont {Zhang}}, \bibinfo
  {author} {\bibfnamefont {S.}~\bibnamefont {Li}}, \bibinfo {author}
  {\bibfnamefont {F.}~\bibnamefont {Grote}}, \bibinfo {author} {\bibfnamefont
  {X.}~\bibnamefont {Zhang}}, \bibinfo {author} {\bibfnamefont
  {H.}~\bibnamefont {Zhang}}, \bibinfo {author} {\bibfnamefont
  {R.}~\bibnamefont {Wang}}, \bibinfo {author} {\bibfnamefont {Y.}~\bibnamefont
  {Lei}}, \bibinfo {author} {\bibfnamefont {B.}~\bibnamefont {Pan}},\ and\
  \bibinfo {author} {\bibfnamefont {Y.}~\bibnamefont {Xie}},\ }\bibfield
  {title} {\bibinfo {title} {Controllable disorder engineering in
  oxygen-incorporated {MoS}$_2$ ultrathin nanosheets for efficient hydrogen
  evolution},\ }\href@noop {} {\bibfield  {journal} {\bibinfo  {journal} {J.
  Am. Chem. Soc.}\ }\textbf {\bibinfo {volume} {135}},\ \bibinfo {pages}
  {17881} (\bibinfo {year} {2013})}\BibitemShut {NoStop}%
\bibitem [{\citenamefont {Lieber}\ and\ \citenamefont {Kim}(1991)}]{lieber}%
  \BibitemOpen
  \bibfield  {author} {\bibinfo {author} {\bibfnamefont {C.~M.}\ \bibnamefont
  {Lieber}}\ and\ \bibinfo {author} {\bibfnamefont {Y.}~\bibnamefont {Kim}},\
  }\bibfield  {title} {\bibinfo {title} {Characterization of the structural,
  electronic and tribological properties of metal dichalcogenides by scanning
  probe microscopies},\ }\href@noop {} {\bibfield  {journal} {\bibinfo
  {journal} {Thin Solid Films}\ }\textbf {\bibinfo {volume} {206}},\ \bibinfo
  {pages} {355} (\bibinfo {year} {1991})}\BibitemShut {NoStop}%
\bibitem [{\citenamefont {Giang}\ \emph {et~al.}(2020)\citenamefont {Giang},
  \citenamefont {Adil},\ and\ \citenamefont {Suni}}]{Giang_2020}%
  \BibitemOpen
  \bibfield  {author} {\bibinfo {author} {\bibfnamefont {H.}~\bibnamefont
  {Giang}}, \bibinfo {author} {\bibfnamefont {O.}~\bibnamefont {Adil}},\ and\
  \bibinfo {author} {\bibfnamefont {I.~I.}\ \bibnamefont {Suni}},\ }\bibfield
  {title} {\bibinfo {title} {Electrodeposition of {Ni}-doped {MoS}$_2$ thin
  films},\ }\href {https://doi.org/10.1149/1945-7111/ab8ce0} {\bibfield
  {journal} {\bibinfo  {journal} {J. Electrochem. Soc.}\ }\textbf {\bibinfo
  {volume} {167}},\ \bibinfo {pages} {082512} (\bibinfo {year}
  {2020})}\BibitemShut {NoStop}%
\bibitem [{\citenamefont {Coutinho}\ \emph {et~al.}(2017)\citenamefont
  {Coutinho}, \citenamefont {Tavares}, \citenamefont {Barboza}, \citenamefont
  {Frazão}, \citenamefont {Moreira},\ and\ \citenamefont
  {Azevedo}}]{T_3R_bandgap}%
  \BibitemOpen
  \bibfield  {author} {\bibinfo {author} {\bibfnamefont {S.}~\bibnamefont
  {Coutinho}}, \bibinfo {author} {\bibfnamefont {M.}~\bibnamefont {Tavares}},
  \bibinfo {author} {\bibfnamefont {C.}~\bibnamefont {Barboza}}, \bibinfo
  {author} {\bibfnamefont {N.}~\bibnamefont {Frazão}}, \bibinfo {author}
  {\bibfnamefont {E.}~\bibnamefont {Moreira}},\ and\ \bibinfo {author}
  {\bibfnamefont {D.~L.}\ \bibnamefont {Azevedo}},\ }\bibfield  {title}
  {\bibinfo {title} {{3R} and {2H} polytypes of {MoS}$_2$: {DFT} and {DFPT}
  calculations of structural, optoelectronic, vibrational and thermodynamic
  properties},\ }\href@noop {} {\bibfield  {journal} {\bibinfo  {journal} {J.
  Phys. Chem. Solids}\ }\textbf {\bibinfo {volume} {111}},\ \bibinfo {pages}
  {25 } (\bibinfo {year} {2017})}\BibitemShut {NoStop}%
\bibitem [{\citenamefont {Peyskens}\ \emph {et~al.}(2019)\citenamefont
  {Peyskens}, \citenamefont {Chakraborty}, \citenamefont {Muneeb},
  \citenamefont {Thourhout},\ and\ \citenamefont {Englund}}]{quantum_emitters}%
  \BibitemOpen
  \bibfield  {author} {\bibinfo {author} {\bibfnamefont {F.}~\bibnamefont
  {Peyskens}}, \bibinfo {author} {\bibfnamefont {C.}~\bibnamefont
  {Chakraborty}}, \bibinfo {author} {\bibfnamefont {M.}~\bibnamefont {Muneeb}},
  \bibinfo {author} {\bibfnamefont {D.~V.}\ \bibnamefont {Thourhout}},\ and\
  \bibinfo {author} {\bibfnamefont {D.}~\bibnamefont {Englund}},\ }\bibfield
  {title} {\bibinfo {title} {Integration of single photon emitters in {2D}
  layered materials with a silicon nitride photonic chip},\ }\href@noop {}
  {\bibfield  {journal} {\bibinfo  {journal} {Nat. Commun.}\ }\textbf {\bibinfo
  {volume} {10}},\ \bibinfo {pages} {4435} (\bibinfo {year}
  {2019})}\BibitemShut {NoStop}%
\bibitem [{\citenamefont {Schuler}\ \emph {et~al.}(2019)\citenamefont
  {Schuler}, \citenamefont {Qiu}, \citenamefont {Refaely-Abramson},
  \citenamefont {Kastl}, \citenamefont {Chen}, \citenamefont {Barja},
  \citenamefont {Koch}, \citenamefont {Ogletree}, \citenamefont {Aloni},
  \citenamefont {Schwartzberg}, \citenamefont {Neaton}, \citenamefont {Louie},\
  and\ \citenamefont {Weber-Bargioni}}]{emitters_ingap}%
  \BibitemOpen
  \bibfield  {author} {\bibinfo {author} {\bibfnamefont {B.}~\bibnamefont
  {Schuler}}, \bibinfo {author} {\bibfnamefont {D.~Y.}\ \bibnamefont {Qiu}},
  \bibinfo {author} {\bibfnamefont {S.}~\bibnamefont {Refaely-Abramson}},
  \bibinfo {author} {\bibfnamefont {C.}~\bibnamefont {Kastl}}, \bibinfo
  {author} {\bibfnamefont {C.~T.}\ \bibnamefont {Chen}}, \bibinfo {author}
  {\bibfnamefont {S.}~\bibnamefont {Barja}}, \bibinfo {author} {\bibfnamefont
  {R.~J.}\ \bibnamefont {Koch}}, \bibinfo {author} {\bibfnamefont {D.~F.}\
  \bibnamefont {Ogletree}}, \bibinfo {author} {\bibfnamefont {S.}~\bibnamefont
  {Aloni}}, \bibinfo {author} {\bibfnamefont {A.~M.}\ \bibnamefont
  {Schwartzberg}}, \bibinfo {author} {\bibfnamefont {J.~B.}\ \bibnamefont
  {Neaton}}, \bibinfo {author} {\bibfnamefont {S.~G.}\ \bibnamefont {Louie}},\
  and\ \bibinfo {author} {\bibfnamefont {A.}~\bibnamefont {Weber-Bargioni}},\
  }\bibfield  {title} {\bibinfo {title} {Large spin-orbit splitting of deep
  in-gap defect states of engineered sulfur vacancies in monolayer {WS}$_2$},\
  }\href@noop {} {\bibfield  {journal} {\bibinfo  {journal} {Phys. Rev. Lett.}\
  }\textbf {\bibinfo {volume} {123}},\ \bibinfo {pages} {076801} (\bibinfo
  {year} {2019})}\BibitemShut {NoStop}%
\bibitem [{\citenamefont {Acerce}\ \emph {et~al.}(2015)\citenamefont {Acerce},
  \citenamefont {Voiry},\ and\ \citenamefont {Chhowalla}}]{as_electrode_mos2}%
  \BibitemOpen
  \bibfield  {author} {\bibinfo {author} {\bibfnamefont {M.}~\bibnamefont
  {Acerce}}, \bibinfo {author} {\bibfnamefont {D.}~\bibnamefont {Voiry}},\ and\
  \bibinfo {author} {\bibfnamefont {M.}~\bibnamefont {Chhowalla}},\ }\bibfield
  {title} {\bibinfo {title} {Metallic {1T} phase {MoS}$_2$ nanosheets as
  supercapacitor electrode materials},\ }\href@noop {} {\bibfield  {journal}
  {\bibinfo  {journal} {Nat. Nanotech.}\ }\textbf {\bibinfo {volume} {10}},\
  \bibinfo {pages} {313} (\bibinfo {year} {2015})}\BibitemShut {NoStop}%
\bibitem [{\citenamefont {da~Silveira~Firmiano}\ \emph
  {et~al.}(2014)\citenamefont {da~Silveira~Firmiano}, \citenamefont {Rabelo},
  \citenamefont {Dalmaschio}, \citenamefont {Pinheiro}, \citenamefont
  {Pereira}, \citenamefont {Schreiner},\ and\ \citenamefont
  {Leite}}]{hybrid_electrode}%
  \BibitemOpen
  \bibfield  {author} {\bibinfo {author} {\bibfnamefont {E.~G.}\ \bibnamefont
  {da~Silveira~Firmiano}}, \bibinfo {author} {\bibfnamefont {A.~C.}\
  \bibnamefont {Rabelo}}, \bibinfo {author} {\bibfnamefont {C.~J.}\
  \bibnamefont {Dalmaschio}}, \bibinfo {author} {\bibfnamefont {A.~N.}\
  \bibnamefont {Pinheiro}}, \bibinfo {author} {\bibfnamefont {E.~C.}\
  \bibnamefont {Pereira}}, \bibinfo {author} {\bibfnamefont {W.~H.}\
  \bibnamefont {Schreiner}},\ and\ \bibinfo {author} {\bibfnamefont {E.~R.}\
  \bibnamefont {Leite}},\ }\bibfield  {title} {\bibinfo {title} {Supercapacitor
  electrodes obtained by directly bonding {2D} {MoS}$_2$ on reduced graphene
  oxide},\ }\href@noop {} {\bibfield  {journal} {\bibinfo  {journal} {Adv.
  Energy Mater.}\ }\textbf {\bibinfo {volume} {4}},\ \bibinfo {pages} {1301380}
  (\bibinfo {year} {2014})}\BibitemShut {NoStop}%
\bibitem [{\citenamefont {Huang}\ \emph {et~al.}(2013)\citenamefont {Huang},
  \citenamefont {Wang}, \citenamefont {Liu}, \citenamefont {Wang},
  \citenamefont {Liu},\ and\ \citenamefont {Wang}}]{hybrid_electrode2}%
  \BibitemOpen
  \bibfield  {author} {\bibinfo {author} {\bibfnamefont {K.-J.}\ \bibnamefont
  {Huang}}, \bibinfo {author} {\bibfnamefont {L.}~\bibnamefont {Wang}},
  \bibinfo {author} {\bibfnamefont {Y.-J.}\ \bibnamefont {Liu}}, \bibinfo
  {author} {\bibfnamefont {H.-B.}\ \bibnamefont {Wang}}, \bibinfo {author}
  {\bibfnamefont {Y.-M.}\ \bibnamefont {Liu}},\ and\ \bibinfo {author}
  {\bibfnamefont {L.-L.}\ \bibnamefont {Wang}},\ }\bibfield  {title} {\bibinfo
  {title} {Synthesis of polyaniline/2-dimensional graphene analog {MoS}$_2$
  composites for high-performance supercapacitor},\ }\href@noop {} {\bibfield
  {journal} {\bibinfo  {journal} {Electrochim. Acta}\ }\textbf {\bibinfo
  {volume} {109}},\ \bibinfo {pages} {587} (\bibinfo {year}
  {2013})}\BibitemShut {NoStop}%
\bibitem [{\citenamefont {Gómez‐Balderas}\ \emph {et~al.}(2000)\citenamefont
  {Gómez‐Balderas}, \citenamefont {Martínez‐Magadán}, \citenamefont
  {Santamaria},\ and\ \citenamefont {Amador}}]{3R_catalysis}%
  \BibitemOpen
  \bibfield  {author} {\bibinfo {author} {\bibfnamefont {R.}~\bibnamefont
  {Gómez‐Balderas}}, \bibinfo {author} {\bibfnamefont {J.~M.}\ \bibnamefont
  {Martínez‐Magadán}}, \bibinfo {author} {\bibfnamefont {R.}~\bibnamefont
  {Santamaria}},\ and\ \bibinfo {author} {\bibfnamefont {C.}~\bibnamefont
  {Amador}},\ }\bibfield  {title} {\bibinfo {title} {Promotional effect of {Co}
  or {Ni} impurity in the catalytic activity of {MoS}$_2$: An electronic
  structure study},\ }\href@noop {} {\bibfield  {journal} {\bibinfo  {journal}
  {Int. J. Quantum Chem.}\ }\textbf {\bibinfo {volume} {80}},\ \bibinfo {pages}
  {406–415} (\bibinfo {year} {2000})}\BibitemShut {NoStop}%
\bibitem [{\citenamefont {Kondekar}\ \emph {et~al.}(2019)\citenamefont
  {Kondekar}, \citenamefont {Boebinger}, \citenamefont {Tian}, \citenamefont
  {Kirmani},\ and\ \citenamefont
  {McDowell}}]{Ni_intercalation_oxidation_state}%
  \BibitemOpen
  \bibfield  {author} {\bibinfo {author} {\bibfnamefont {N.}~\bibnamefont
  {Kondekar}}, \bibinfo {author} {\bibfnamefont {M.~G.}\ \bibnamefont
  {Boebinger}}, \bibinfo {author} {\bibfnamefont {M.}~\bibnamefont {Tian}},
  \bibinfo {author} {\bibfnamefont {M.~H.}\ \bibnamefont {Kirmani}},\ and\
  \bibinfo {author} {\bibfnamefont {M.~T.}\ \bibnamefont {McDowell}},\
  }\bibfield  {title} {\bibinfo {title} {The effect of nickel on {MoS}$_2$
  growth revealed with \emph{in situ} transmission electron microscopy},\
  }\href@noop {} {\bibfield  {journal} {\bibinfo  {journal} {ACS Nano}\
  }\textbf {\bibinfo {volume} {13}},\ \bibinfo {pages} {7117} (\bibinfo {year}
  {2019})}\BibitemShut {NoStop}%
\bibitem [{\citenamefont {Löwdin}(1950)}]{lowdin}%
  \BibitemOpen
  \bibfield  {author} {\bibinfo {author} {\bibfnamefont {P.}~\bibnamefont
  {Löwdin}},\ }\bibfield  {title} {\bibinfo {title} {On the
  non‐orthogonality problem connected with the use of atomic wave functions
  in the theory of molecules and crystals},\ }\href@noop {} {\bibfield
  {journal} {\bibinfo  {journal} {J. Chem. Phys.}\ }\textbf {\bibinfo {volume}
  {18}},\ \bibinfo {pages} {365} (\bibinfo {year} {1950})}\BibitemShut
  {NoStop}%
\bibitem [{\citenamefont {Sit}\ \emph {et~al.}(2011)\citenamefont {Sit},
  \citenamefont {Car}, \citenamefont {Cohen},\ and\ \citenamefont
  {Selloni}}]{accurate_OS}%
  \BibitemOpen
  \bibfield  {author} {\bibinfo {author} {\bibfnamefont {P.~H.-L.}\
  \bibnamefont {Sit}}, \bibinfo {author} {\bibfnamefont {R.}~\bibnamefont
  {Car}}, \bibinfo {author} {\bibfnamefont {M.~H.}\ \bibnamefont {Cohen}},\
  and\ \bibinfo {author} {\bibfnamefont {A.}~\bibnamefont {Selloni}},\
  }\bibfield  {title} {\bibinfo {title} {Simple, unambiguous theoretical
  approach to oxidation state determination via first-principles
  calculations},\ }\href@noop {} {\bibfield  {journal} {\bibinfo  {journal}
  {Inorg. Chem.}\ }\textbf {\bibinfo {volume} {50}},\ \bibinfo {pages}
  {10259–10267} (\bibinfo {year} {2011})}\BibitemShut {NoStop}%
\bibitem [{\citenamefont {Zhu}\ \emph {et~al.}(2018)\citenamefont {Zhu},
  \citenamefont {Liang}, \citenamefont {Qin}, \citenamefont {Deng},\ and\
  \citenamefont {Bi}}]{magnetization}%
  \BibitemOpen
  \bibfield  {author} {\bibinfo {author} {\bibfnamefont {Y.}~\bibnamefont
  {Zhu}}, \bibinfo {author} {\bibfnamefont {X.}~\bibnamefont {Liang}}, \bibinfo
  {author} {\bibfnamefont {J.}~\bibnamefont {Qin}}, \bibinfo {author}
  {\bibfnamefont {L.}~\bibnamefont {Deng}},\ and\ \bibinfo {author}
  {\bibfnamefont {L.}~\bibnamefont {Bi}},\ }\bibfield  {title} {\bibinfo
  {title} {Strain tunable magnetic properties of 3d transition-metal ion doped
  monolayer {MoS}$_2$: A first-principles study},\ }\href@noop {} {\bibfield
  {journal} {\bibinfo  {journal} {Int. J. Quantum Chem.}\ }\textbf {\bibinfo
  {volume} {8}},\ \bibinfo {pages} {055917} (\bibinfo {year}
  {2018})}\BibitemShut {NoStop}%
\bibitem [{\citenamefont {Wu}\ \emph {et~al.}(2018)\citenamefont {Wu},
  \citenamefont {Yao}, \citenamefont {Hao}, \citenamefont {Dong}, \citenamefont
  {Cheng}, \citenamefont {Liu}, \citenamefont {Lu}, \citenamefont {Wang},
  \citenamefont {Cho},\ and\ \citenamefont {Wang}}]{Hub_U_Mo}%
  \BibitemOpen
  \bibfield  {author} {\bibinfo {author} {\bibfnamefont {M.}~\bibnamefont
  {Wu}}, \bibinfo {author} {\bibfnamefont {X.}~\bibnamefont {Yao}}, \bibinfo
  {author} {\bibfnamefont {Y.}~\bibnamefont {Hao}}, \bibinfo {author}
  {\bibfnamefont {H.}~\bibnamefont {Dong}}, \bibinfo {author} {\bibfnamefont
  {Y.}~\bibnamefont {Cheng}}, \bibinfo {author} {\bibfnamefont
  {H.}~\bibnamefont {Liu}}, \bibinfo {author} {\bibfnamefont {F.}~\bibnamefont
  {Lu}}, \bibinfo {author} {\bibfnamefont {W.}~\bibnamefont {Wang}}, \bibinfo
  {author} {\bibfnamefont {K.}~\bibnamefont {Cho}},\ and\ \bibinfo {author}
  {\bibfnamefont {W.-H.}\ \bibnamefont {Wang}},\ }\bibfield  {title} {\bibinfo
  {title} {Electronic structures, magnetic properties and band alignments of 3d
  transition metal atoms doped monolayer {MoS}$_2$},\ }\href@noop {} {\bibfield
   {journal} {\bibinfo  {journal} {Phys. Lett. A}\ }\textbf {\bibinfo {volume}
  {382}},\ \bibinfo {pages} {111} (\bibinfo {year} {2018})}\BibitemShut
  {NoStop}%
\bibitem [{\citenamefont {Mounet}\ \emph {et~al.}(2018)\citenamefont {Mounet},
  \citenamefont {Gibertini}, \citenamefont {Schwaller}, \citenamefont {Campi},
  \citenamefont {Merkys}, \citenamefont {Marrazzo}, \citenamefont {Sohier},
  \citenamefont {Castelli}, \citenamefont {Cepellotti}, \citenamefont {Pizzi},\
  and\ \citenamefont {Marzari}}]{Layer_dissociation}%
  \BibitemOpen
  \bibfield  {author} {\bibinfo {author} {\bibfnamefont {N.}~\bibnamefont
  {Mounet}}, \bibinfo {author} {\bibfnamefont {M.}~\bibnamefont {Gibertini}},
  \bibinfo {author} {\bibfnamefont {P.}~\bibnamefont {Schwaller}}, \bibinfo
  {author} {\bibfnamefont {D.}~\bibnamefont {Campi}}, \bibinfo {author}
  {\bibfnamefont {A.}~\bibnamefont {Merkys}}, \bibinfo {author} {\bibfnamefont
  {A.}~\bibnamefont {Marrazzo}}, \bibinfo {author} {\bibfnamefont
  {T.}~\bibnamefont {Sohier}}, \bibinfo {author} {\bibfnamefont {I.~E.}\
  \bibnamefont {Castelli}}, \bibinfo {author} {\bibfnamefont {A.}~\bibnamefont
  {Cepellotti}}, \bibinfo {author} {\bibfnamefont {G.}~\bibnamefont {Pizzi}},\
  and\ \bibinfo {author} {\bibfnamefont {N.}~\bibnamefont {Marzari}},\
  }\bibfield  {title} {\bibinfo {title} {Two-dimensional materials from
  high-throughput computational exfoliation of experimentally known
  compounds},\ }\href@noop {} {\bibfield  {journal} {\bibinfo  {journal} {Nat.
  Nanotech.}\ }\textbf {\bibinfo {volume} {13}},\ \bibinfo {pages} {246–252}
  (\bibinfo {year} {2018})}\BibitemShut {NoStop}%
\bibitem [{\citenamefont {Tang}\ \emph {et~al.}(2014)\citenamefont {Tang},
  \citenamefont {Kvashnin}, \citenamefont {Najmaei}, \citenamefont {Bando},
  \citenamefont {Kimoto}, \citenamefont {Koskinen}, \citenamefont {Ajayan},
  \citenamefont {Yakobson}, \citenamefont {Sorokin}, \citenamefont {Lou},\ and\
  \citenamefont {Golberg}}]{exp_surface_E}%
  \BibitemOpen
  \bibfield  {author} {\bibinfo {author} {\bibfnamefont {D.-M.}\ \bibnamefont
  {Tang}}, \bibinfo {author} {\bibfnamefont {D.~G.}\ \bibnamefont {Kvashnin}},
  \bibinfo {author} {\bibfnamefont {S.}~\bibnamefont {Najmaei}}, \bibinfo
  {author} {\bibfnamefont {Y.}~\bibnamefont {Bando}}, \bibinfo {author}
  {\bibfnamefont {K.}~\bibnamefont {Kimoto}}, \bibinfo {author} {\bibfnamefont
  {P.}~\bibnamefont {Koskinen}}, \bibinfo {author} {\bibfnamefont {P.~M.}\
  \bibnamefont {Ajayan}}, \bibinfo {author} {\bibfnamefont {B.~I.}\
  \bibnamefont {Yakobson}}, \bibinfo {author} {\bibfnamefont {P.~B.}\
  \bibnamefont {Sorokin}}, \bibinfo {author} {\bibfnamefont {J.}~\bibnamefont
  {Lou}},\ and\ \bibinfo {author} {\bibfnamefont {D.}~\bibnamefont {Golberg}},\
  }\bibfield  {title} {\bibinfo {title} {Nanomechanical cleavage of molybdenum
  disulphide atomic layers},\ }\href@noop {} {\bibfield  {journal} {\bibinfo
  {journal} {Nat. Commun.}\ }\textbf {\bibinfo {volume} {4}},\ \bibinfo {pages}
  {3631} (\bibinfo {year} {2014})}\BibitemShut {NoStop}%
\bibitem [{\citenamefont {Bj\"{o}rkman}\ \emph {et~al.}(2012)\citenamefont
  {Bj\"{o}rkman}, \citenamefont {Gulans}, \citenamefont {Krasheninnikov},\ and\
  \citenamefont {Nieminen}}]{rpa_surface_E}%
  \BibitemOpen
  \bibfield  {author} {\bibinfo {author} {\bibfnamefont {T.}~\bibnamefont
  {Bj\"{o}rkman}}, \bibinfo {author} {\bibfnamefont {A.}~\bibnamefont
  {Gulans}}, \bibinfo {author} {\bibfnamefont {A.~V.}\ \bibnamefont
  {Krasheninnikov}},\ and\ \bibinfo {author} {\bibfnamefont {R.~M.}\
  \bibnamefont {Nieminen}},\ }\bibfield  {title} {\bibinfo {title} {Van der
  {Waals} bonding in layered compounds from advanced density-functional
  first-principles calculations},\ }\href@noop {} {\bibfield  {journal}
  {\bibinfo  {journal} {Phys. Rev. Lett.}\ }\textbf {\bibinfo {volume} {108}},\
  \bibinfo {pages} {235502} (\bibinfo {year} {2012})}\BibitemShut {NoStop}%
\bibitem [{\citenamefont {Sekine}\ \emph {et~al.}(1989)\citenamefont {Sekine},
  \citenamefont {C.}, \citenamefont {Samaras}, \citenamefont {Jouanne},\ and\
  \citenamefont {Balkanski}}]{Sekine}%
  \BibitemOpen
  \bibfield  {author} {\bibinfo {author} {\bibfnamefont {T.}~\bibnamefont
  {Sekine}}, \bibinfo {author} {\bibfnamefont {J.}~\bibnamefont {C.}}, \bibinfo
  {author} {\bibfnamefont {I.}~\bibnamefont {Samaras}}, \bibinfo {author}
  {\bibfnamefont {M.}~\bibnamefont {Jouanne}},\ and\ \bibinfo {author}
  {\bibfnamefont {M.}~\bibnamefont {Balkanski}},\ }\bibfield  {title} {\bibinfo
  {title} {Vibrational modifications on lithium intercalation in {MoS}$_2$},\
  }\href {https://doi.org/10.1016/0921-5107(89)90195-5} {\bibfield  {journal}
  {\bibinfo  {journal} {Mat. Sci. Eng., B}\ }\textbf {\bibinfo {volume} {3}},\
  \bibinfo {pages} {153} (\bibinfo {year} {1989})}\BibitemShut {NoStop}%
\bibitem [{\citenamefont {Sheremetyeva}\ \emph {et~al.}(2021)\citenamefont
  {Sheremetyeva}, \citenamefont {Niedzielski}, \citenamefont {Tristant},
  \citenamefont {Liang}, \citenamefont {Kerstetter}, \citenamefont {Mohney},\
  and\ \citenamefont {Meunier}}]{Sheremetyeva_2021}%
  \BibitemOpen
  \bibfield  {author} {\bibinfo {author} {\bibfnamefont {N.}~\bibnamefont
  {Sheremetyeva}}, \bibinfo {author} {\bibfnamefont {D.}~\bibnamefont
  {Niedzielski}}, \bibinfo {author} {\bibfnamefont {D.}~\bibnamefont
  {Tristant}}, \bibinfo {author} {\bibfnamefont {L.}~\bibnamefont {Liang}},
  \bibinfo {author} {\bibfnamefont {L.~E.}\ \bibnamefont {Kerstetter}},
  \bibinfo {author} {\bibfnamefont {S.~E.}\ \bibnamefont {Mohney}},\ and\
  \bibinfo {author} {\bibfnamefont {V.}~\bibnamefont {Meunier}},\ }\bibfield
  {title} {\bibinfo {title} {Low-frequency {Raman} signature of
  {Ag}-intercalated few-layer {MoS}$_2$},\ }\href
  {https://doi.org/10.1088/2053-1583/abdbcc} {\bibfield  {journal} {\bibinfo
  {journal} {2D Mater.}\ }\textbf {\bibinfo {volume} {8}},\ \bibinfo {pages}
  {025031} (\bibinfo {year} {2021})}\BibitemShut {NoStop}%
\bibitem [{\citenamefont {Pastewka}\ and\ \citenamefont
  {Robbins}(2014)}]{work_adhesion}%
  \BibitemOpen
  \bibfield  {author} {\bibinfo {author} {\bibfnamefont {L.}~\bibnamefont
  {Pastewka}}\ and\ \bibinfo {author} {\bibfnamefont {M.~O.}\ \bibnamefont
  {Robbins}},\ }\bibfield  {title} {\bibinfo {title} {Contact between rough
  surfaces and a criterion for macroscopic adhesion},\ }\href@noop {}
  {\bibfield  {journal} {\bibinfo  {journal} {Proc. Nat. Acad. Sci.}\ }\textbf
  {\bibinfo {volume} {111}},\ \bibinfo {pages} {3298–3303} (\bibinfo {year}
  {2014})}\BibitemShut {NoStop}%
\bibitem [{\citenamefont {Levita}\ \emph {et~al.}(2014)\citenamefont {Levita},
  \citenamefont {Cavaleiro}, \citenamefont {Molinari}, \citenamefont {Polcar},\
  and\ \citenamefont {Righi}}]{layer_sliding}%
  \BibitemOpen
  \bibfield  {author} {\bibinfo {author} {\bibfnamefont {G.}~\bibnamefont
  {Levita}}, \bibinfo {author} {\bibfnamefont {A.}~\bibnamefont {Cavaleiro}},
  \bibinfo {author} {\bibfnamefont {E.}~\bibnamefont {Molinari}}, \bibinfo
  {author} {\bibfnamefont {T.}~\bibnamefont {Polcar}},\ and\ \bibinfo {author}
  {\bibfnamefont {M.~C.}\ \bibnamefont {Righi}},\ }\bibfield  {title} {\bibinfo
  {title} {Sliding properties of {MoS}$_2$ layers: Load and interlayer
  orientation effects},\ }\href@noop {} {\bibfield  {journal} {\bibinfo
  {journal} {J. Phys. Chem. C}\ }\textbf {\bibinfo {volume} {118}},\ \bibinfo
  {pages} {13809} (\bibinfo {year} {2014})}\BibitemShut {NoStop}%
\end{thebibliography}

%

\end{document}